\documentclass[11pt]{article}

\usepackage[final]{acl}
\usepackage{amsmath}
\usepackage{adjustbox}
\usepackage{times}
\usepackage{latexsym}
\usepackage{booktabs}
\usepackage[T1]{fontenc}
\usepackage{enumitem}
\usepackage[utf8]{inputenc}

\usepackage{microtype}

\usepackage{inconsolata}

\usepackage{graphicx}

\usepackage{subcaption}

\usepackage{tcolorbox}
\usepackage{xcolor}
\newtcolorbox{promptbox}[1][]{
  colback=gray!5!white, 
  colframe=gray!75!black, 
  title=\textbf{Prompt}, 
  fonttitle=\bfseries,
  left=5pt, right=5pt, top=5pt, bottom=5pt,
  boxrule=0.5pt,
  sharp corners, 
  #1
}

\title{Dashboard2Code: Evaluating Multimodal Models on Reconstructing Interactive Dashboards}

\author{
Tianhao Niu\thanks{Equal contribution.}
\quad
Ziyu Han\footnotemark[1]
\quad 
Qiguang Chen \\
\textbf{Shiqi Zhou}
\quad
\textbf{Baocai Shan}
\quad
\textbf{Hengjie Fang}
\quad
\textbf{Qingfu Zhu}
\quad
\textbf{Wanxiang Che}\thanks{Corresponding author.} \\
 Research Center for Social Computing and Interactive Robotics \\
 Harbin Institute of Technology, China \\
\texttt{\{thniu,zyhan,car\}@ir.hit.edu.cn}
}

\begin{document}
\maketitle
\begin{abstract}
Automatic data visualization generation has advanced rapidly with multi-modal large language models, yet existing efforts largely focus on static charts and overlook the interactive dashboards commonly used for real-world data exploration. We introduce Dashboard2Code, a novel task that requires a model to proactively explore an interactive dashboard, acquire and integrate feedback from its own interactions (e.g., clicking and filtering), and generate code that reproduces the target dashboard.
To support comprehensive evaluation, we present DashboardMimic, the first Plotly+Dash benchmark for Dashboard2Code, comprising 180 carefully designed and manually verified dashboard–code pairs spanning three difficulty levels and covering eight common real-world interaction patterns. We further propose an automated evaluation framework tailored to dashboards that combines code semantic analysis with dynamic interaction-based testing to assess visual and interaction consistency, showing strong agreement with human judgments.
Experiments across a range of open- and closed-source multi-modal models reveal that even the strongest systems struggle on high-complexity dashboards and that a substantial performance gap remains between open-source and closed-source models on the Dashboard2Code task.
\footnote{\ \ \href{https://github.com/nth2000/Dashboard2Code}{Code and data}}

\end{abstract}

\section{Introduction}

Automatic data visualization understanding and generation are important tasks in artificial intelligence. With the rapid development of multi-modal large language models, tasks for understanding data visualizations (e.g., chart question answering \citep{masry2022chartqabenchmarkquestionanswering,masry2025chartqaprodiversechallengingbenchmark}, chart summarization \citep{choi2025endtoendchartsummarizationvisual,wang2024charxivchartinggapsrealistic}, chart parsing \citep{xu2025chartmoemixturediverselyaligned}) and tasks for generating visualizations (e.g., text-to-chart \citep{rahman-etal-2025-text2vis,ni-etal-2025-viscoder} and chart-to-code \citep{yang2025chartmimic,wu-etal-2025-plot2code,zhao-etal-2025-chartcoder,tang2025chartscodehierarchicalbenchmark}) show strong results. However, these tasks mainly focus on understanding and generating static visualizations. In practical scenarios, datasets are often large and structurally complex, making interactive dashboards a common form of visualization for in-depth exploration. As interactive interfaces, dashboards enable users to uncover more complex data patterns and logical relations through actions such as clicking and filtering.

\begin{figure}[t]
  \centering
  \includegraphics[width=1.0\columnwidth]{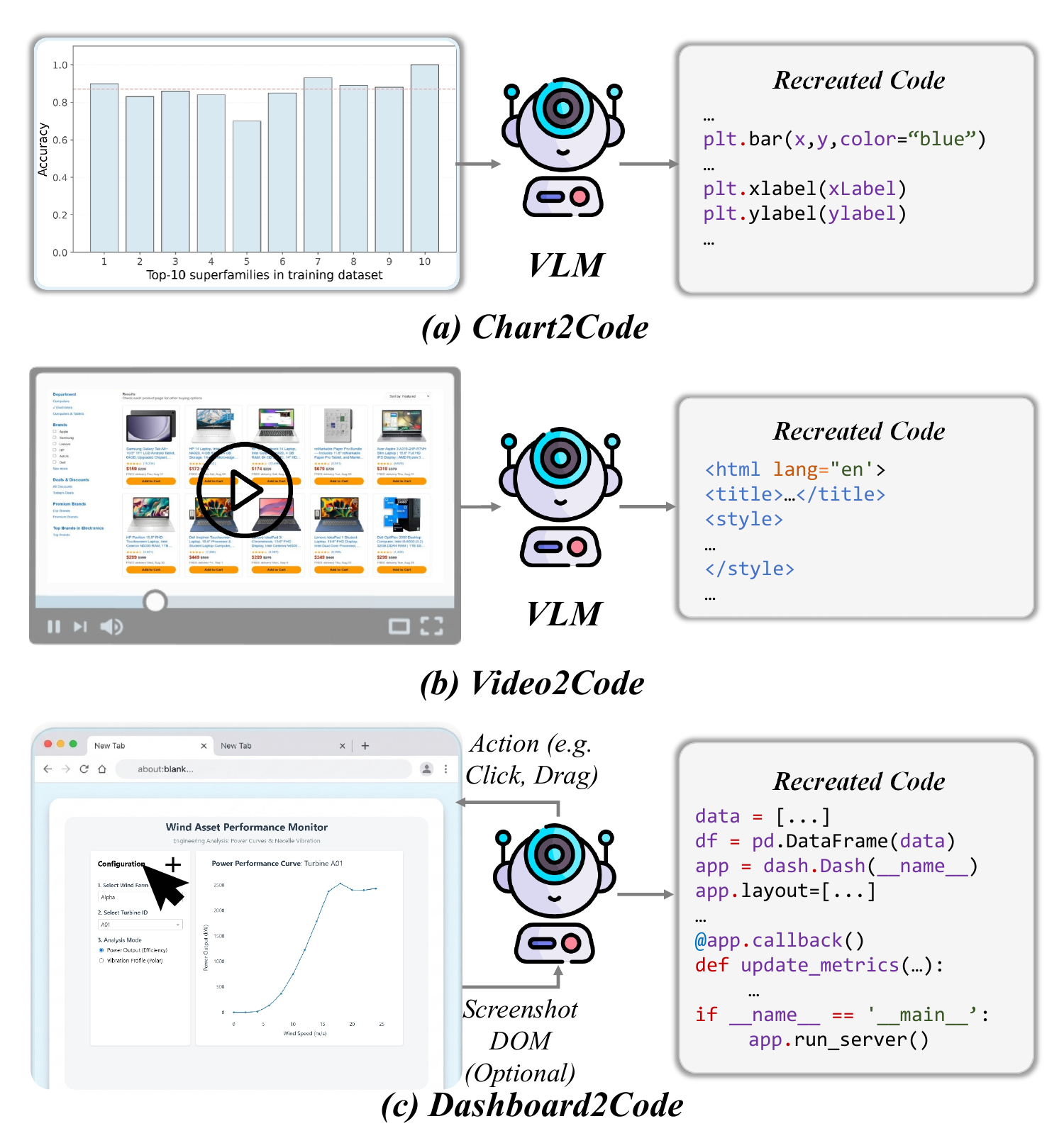}
  \caption {Comparison of Dashboard2Code task with existing tasks. Dashboard2Code requires the model to \emph{\textbf{proactively explore a dashboard, collect feedback during exploration, and integrate this feedback with the interaction history}} to generate code that faithfully reproduces both the visual appearance and interaction logic of the target dashboard.}
  \label {fig:experiment_analysis_low_level}
\end{figure}

With the rapid development of multi-modal large language models \citep{wang2025internvl35advancingopensourcemultimodal,openai2024gpt4technicalreport,gemmateam2025gemma3technicalreport}, several important research tasks have emerged on automating interactive user interfaces. 

One line of research systematically evaluates multi-modal GUI agents on perception, action, and reasoning over interactive interfaces. For instance, OS-World \citep{xie2024osworld} introduces a benchmark that evaluates multi-modal agents in real operating system environments on open-ended computer tasks involving diverse applications, interfaces, and workflows. In the data visualization domain, DashboardQA \citep{kartha2025dashboardqabenchmarkingmultimodalagents} presents the dashboard question answering benchmark that assesses a GUI agent’s ability to interact with a dashboard in response to a given question and return the correct final answer. However, these research tasks typically provide explicit goal instructions, making it difficult to assess whether GUI Agents can \textbf{actively fully} explore, understand, and reason over dynamic interactive interfaces and then recreate it.

Another line of research focuses on reconstructing dynamic interfaces and their interaction logic from trajectories or videos through code. 
For instance, Interaction2Code evaluates whether multi-modal large language models can reconstruct code from a user interaction sequence \citep{xiao2025interaction2codebenchmarkingmllmbasedinteractive}. IWR-Bench \citep{chen2025iwrbenchlvlmsreconstructinteractive} evaluates whether multi-modal large language models can recover an interactive web page from a user interaction video and generate the corresponding code. These studies extend visualization understanding toward code generation for dynamic interactive interfaces. However, these tasks still rely on human-annotated interaction trajectories for evaluation. Such annotations increase cost and do not capture the model’s ability to proactively explore, integrate, and use diverse external feedback obtained during its own exploration.
In addition, these tasks are mainly limited to general web pages and do not consider the interaction logic specific to interactive dashboards. Furthermore, their evaluation metrics are generic and are not tailored for dashboards.

\begin{figure*}[t]
  \centering
  \includegraphics[width=2.0\columnwidth]{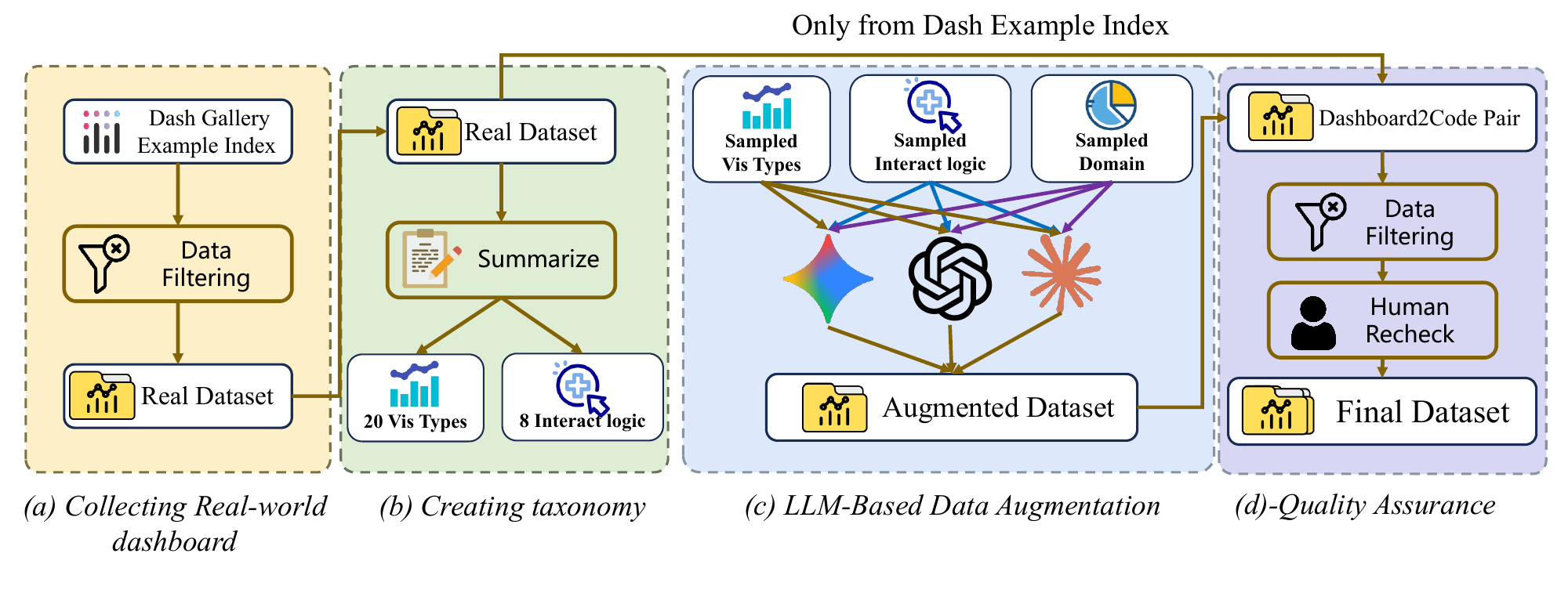}
  \caption {Benchmark construction pipeline.}
  \label {fig:text_anomize_figure}
\end{figure*}



In light of these observations, we introduce the Dashboard2Code task. Unlike previous settings that rely on static snapshots \citep{yang2025chartmimic,si-etal-2025-design2code} or annotated interaction image traces to generate code \citep{chen2025iwrbenchlvlmsreconstructinteractive,xiao2025interaction2codebenchmarkingmllmbasedinteractive} and thus leave models in a \emph{passive} role, Dashboard2Code requires the model to \emph{\textbf{proactively explore a dashboard, collect feedback during exploration, and integrate this feedback with the interaction history}} to generate code that faithfully reproduces both the visual appearance and interaction logic of the target dashboard.
This formulation positions the task as an \emph{active} interaction problem: the model must navigate the interface, interpret dynamic feedback, and reason over a sequence of interactions to make proper action decisions or recover the underlying implementation. 
As a result, Dashboard2Code presents a stronger end‑to‑end challenge for multi‑modal models.

To comprehensively evaluate the Dashboard2Code task, we introduce DashboardMimic, the first Plotly+Dash-based benchmark that covers diverse visualization types and interaction logic. It contains 180 carefully designed dashboard–code pairs across three difficulty levels and includes eight common real-world dashboard interaction patterns, with all annotations manually verified. We further propose an automated evaluation framework tailored to Dashboard2Code, which combines code semantic evaluation and dynamic interaction evaluation to measure interaction-logic consistency and visual consistency, and shows strong agreement with human judgments (Pearson correlation = 0.78).

We evaluate a range of multi-modal models, including both open- and closed-source systems. The strongest model, Gemini 3 Pro, achieves 79.4, but it reaches only 64.2 on Complexity L3, which better reflects real-world complexity. Furthermore, our results reveal a substantial performance gap between open-source and closed-source models on the Dashboard2Code task.

Our contributions are as follows:
\begin{itemize}
    \item We propose the novel Dashboard2Code task, which requires the model to proactively explore a dashboard, obtain feedback, and integrate this feedback with the interaction history to generate code that reproduces the target dashboard.
    \item We construct a Plotly+Dash based Dashboard2Code benchmark DashboardMimic, which provides broad coverage of dashboard visualization types and interaction logic.\footnote{We compare our benchmark with existing benchmarks in the appendix~\ref{sec:comparison_with_existing_benchmarks}. We discuss more related works in appendix~\ref{sec:related work}}.
    \item We propose an automated, multi-aspect evaluation framework for Dashboard2Code and validate its consistency with human evaluation.

\end{itemize}

\section{Task Formulation}

In contrast to chart-to-code and video-to-code settings, Dashboard2Code requires the GUI agent to actively interact with the target dashboard at runtime and infer its structure, component semantics, and interaction logic from iterative feedback. The goal of the task is to generate code that reproduces the dashboard with high fidelity in both functional behavior and visual appearance.

Formally, given a target dashboard $D$ and its initial view $V_0 = V_{\text{initial}}$, the agent interacts with the environment for a variable number of steps. At step $t=1,2,\dots$, the agent receives feedback $o_{t-1}$ and selects an action $a_t \in \mathcal{S}$ from an action space $\mathcal{S}$ (e.g., click, hover, scroll, drag), or chooses a special \textsc{Stop} action to terminate interaction. After executing $a_t$, the environment returns new feedback
\begin{equation}
o_t = \mathrm{Feedback}(D, a_{1:t}),
\end{equation}
where the feedback can take two forms: (i) \emph{visual-only} feedback $o_t = V_t$, or (ii) \emph{visual+DOM} feedback $o_t = (V_t, G_t)$, with $G_t$ denoting the DOM tree (or an equivalent structured UI representation) at step $t$. In both cases, $V_t$ is the rendered view after applying actions $a_{1:t}$:
\begin{equation}
V_t = \mathrm{Render}(D, a_{1:t}).
\end{equation}
When DOM feedback is available, $G_t$ denotes the corresponding DOM tree (or an equivalent structured UI representation) at step $t$, obtained after executing the same action history:
\begin{equation}
G_t = \mathrm{DOM}(D, a_{1:t}).
\end{equation}

The agent terminates at time $\tau$ when it selects \textsc{Stop} and outputs code $C$, which is used to construct a reproduced dashboard $\hat{D} = \mathrm{Build}(C)$. The objective is to maximize the end-to-end reproduction quality:
\begin{equation}
C^* = \arg\max_C \ \mathcal{M}\!\big(D, \mathrm{Build}(C)\big),
\end{equation}

Notably, the goal of this task is to produce code that reproduces the target dashboard and does not enforce a strict two-stage ``explore $\rightarrow$ generate'' pipeline. For instance, the model may interleave exploration and code generation, continuously integrating its interaction history and feedback to draft and iteratively refine the code throughout the interaction process. The already drafted code may be a replacement for some screenshot history.

\section{Benchmark Construction}
\subsection{Overview}
To comprehensively evaluate models for dynamic dashboard recreation, we construct DashboardMimic benchmark based on the Python Plotly+Dash framework. The benchmark consists of 58 carefully curated real-world dashboard-code pairs and 122 high-quality synthetic dashboards enhanced by large language models. It covers 20 common visualization types and 8 callback-logic patterns spanning multiple levels of interaction complexity.

\subsection{Collecting and Filtering Real-world Data}

We first collect a large set of dashboard-code pairs from the open-source community site
\texttt{dash-example-index.herokuapp}, which provides minimal examples covering diverse Dash chart
types and interaction logic. To ensure data quality, we apply the filtering principle defined in appendix~\ref{sec:benchmark_filtering_principle}.
After filtering, we retain \textbf{58} high-quality real-world dashboards as seed data.

In addition, we select \textbf{50} dashboards from \texttt{DashGallery} to summarize
visualization types, interaction logic and interactive components in dashboards along with the 58 collected seed data. \footnote{We do not directly include \texttt{DashGallery} data in our
benchmark as many dashboards there rely heavily on external data sources and additional libraries, which can
unfairly lower code-generation scores when the required dependencies or domain knowledge are missing.}

\subsection{Creating Taxonomy of Interaction Logic}

Interaction logic is a key property of dashboards. Based on the callback-topology graphs derived from \textbf{108} real-world dashboard examples, we categorize callback logic into three difficulty levels with eight subtypes. Detailed definitions are shown in the appendix~\ref{sec:interaction_logic_detail}. And dashboard examples of each subtype are shown in the appendix~\ref{sec:benchmark_cases}.

\subsection{LLM-Based Data Augmentation}

To further enrich interaction logic and visualization types, increase domain diversity, scale up the benchmark, we use LLMs to augment the benchmark based on the taxonomy in \S3.3, together with the visualization types and interaction components observed from 108 real-world dashboards. 

We develop a semi-random sampling-then-generating pipeline that increases benchmark diversity while keeping dashboards semantically and functionally coherent. The pipeline first samples a high-level specification, including a domain, an interaction-complexity level in \S3.3, a UI style (native vs. Bootstrap), and candidate types of UI components (3 basic and 4 advanced) and visualization types (5 in total). In the generation stage, an LLM is conditioned on this specification to write a grounded user story and then implement the dashboard in Plotly+Dash; Notably, the LLM is allowed to select an appropriate subset of the sampled components and visualization types rather than being forced to use all of them. Prompt details are shown in appendix~\ref{sec:llm_aug_detail}.

We use multiple state-of-the-art LLM families (Gemini, GPT, and Claude) as generators. 
Employing multiple generators helps mitigate potential bias introduced by any single model. However, LLM-based synthesis is inherently imperfect. We therefore initially generate around 500 samples in total and then do the quality control in the next selection.

\subsection{Quality Assurance}
This section outlines the rigorous quality assurance protocols employed to guarantee the high standard of dashboard-code pairs.

\textbf{\emph{Data filtering:}}
We manually filter the synthetic dashboards according to the criteria defined in the appendix, retaining only samples that satisfy all requirements. After this selection process, we obtain 122 synthetic dashboards, which we merge with 58 real-world seed examples to form a benchmark of 180 dashboard–code pairs.

\textbf{\emph{Human recheck:}}
As an additional quality-control step, two annotators independently review all 180 dashboard–code pairs to verify compliance with the same criteria defined in the appendix.

We further discuss the Data Contamination mitigation in appendix~\ref{sec:decontamination}.

\subsection{Benchmark Statistics}

The DashboardMimic benchmark contains 180 Plotly+Dash dashboard–code pairs. Among them, 58 are collected from real-world dash-example-index and 122 are synthesized with LLMs. All samples undergo strict manual screening and filtering to ensure data quality. We show more statistics related to UI component, Visualization type diversity and Interaction logic diversity in appendix \ref{sec:benchmark_statistics}.

At this stage, we limit Dashboard2Code to Plotly+Dash. This is a deliberate choice. Plotly is an open-source Python library for building interactive charts, and Dash is an open-source Python framework for building data apps with common dashboard parts such as charts, tables, controls, and callback-based interactions. Both are released under the MIT license, which makes our benchmark easier to reproduce and extend.  In addition, since both Plotly and Dash use Python, this setting helps reduce the effect of weaker code generation in other programming languages on the final evaluation. We therefore use Plotly+Dash as a controlled and practical first step for studying multi-modal agents on dashboard programming. Although this setting does not cover all dashboard frameworks, we hope DashboardMimic can provide a solid starting point for future work on more frameworks.\footnote{We discuss more about other dashboard frameworks in appendix~\ref{sec:discuss_other_framework}}

\section{Evaluation Framework}

\subsection{Overview}

This section describes our evaluation framework for the Dashboard2Code task. The framework is fully automated and combines \emph{Code Semantics Evaluation} with \emph{Dynamic Interaction Evaluation}. Specifically, \emph{Code Semantics Evaluation} compares the ground-truth code (\textsc{GT}) and the generated code (\textsc{Gen}) via static properties without executing either program. The \emph{Dynamic Interaction Evaluation} is inspired by \textsc{IWR-Bench}: a strong GUI agent executes a set of predefined tasks on the generated dashboard and compares outcomes with those on the reference dashboard.

In contrast to prior settings, our dynamic evaluation metrics are designed to better reflect the properties of Dashboard2Code. 
\textbf{(1)} Dashboards are designed for interactive data visualization. Therefore, evaluation should jointly assess both \emph{visual fidelity}---covering visualization type, style, and data---and \emph{interaction-logic correctness}.
\textbf{(2)} In Plotly+Dash dashboards, the front-end exposes structured attribute dictionaries for each visualization object (e.g., data, style, text, and visualization type). Comparing these attribute dictionaries provides a low-cost and less subjective alternative to using LLMs as judges. \textbf{\emph{Empirically, this metric correlates closely with human judgments and performs comparably to LLM-as-judge in evaluating visual consistency.}} Detailed results are provided in the appendix~\ref{sec:eval_framework_detail}.

Overall, the final metric produced by our framework shows strong agreement with human evaluation on the Dashboard2Code task (Pearson correlation $=0.78$). Details of metric validation and quality control are provided in the appendix.

\subsection{Code Semantic Evaluation}
Code Semantics Evaluation compares the static properties of the ground-truth code (\textbf{GT}) and the generated code (\textbf{Gen}) without involving any runtime interaction. It includes two metrics.
\textbf{(1) Key Component Coverage (F1):} We extract component class names from \texttt{app.layout} in both \textbf{GT} and \textbf{Gen}, and compute an F1 score to measure whether the model selects the correct component types.
\textbf{(2) LLM-Semantic-Eval:} We use Gemini-3-Flash to score the semantic consistency between \textbf{GT} and \textbf{Gen} code. In the prompt, we instruct Gemini-3-Flash to focus on the \emph{equivalence of the intended behavior and logic} rather than low-level implementation details. The evaluation prompt is shown in the appendix~\ref{sec:related_prompt_eval}.

\subsection{Dynamic Interaction Evaluation}
\subsubsection{Overall Workflow}
Dynamic Interaction Evaluation aims to directly assess the consistency between the generated dashboard and the reference dashboard in both \emph{visual appearance} and \emph{interactive behavior}. This complements code-semantics evaluation by reducing cases where different implementations lead to the same rendered results.

Inspired by \textsc{IWR-Bench}, we annotate a set of test tasks for each dashboard in our benchmark (e.g., ``Click the `Update' button'' or ``Select \texttt{OptionA} in the dropdown''). We introduce an automated evaluation agent based on an advanced multi-modal LLM. For each task, the agent receives the DOM tree and the task instruction, and executes the task on both the reference and generated dashboards from their initial states. After execution, we score the outcomes using the metrics defined in \S\ref{sec:eval_metric_dynamic}. The dashboard-level score is computed as the average over all its tasks. If the generated dashboard raises an error at any step of a task (e.g., an expected target element is missing), we assign a score of 0 to that task.

We also define the \textbf{Task Execution Rate} metric, which refers to the proportion of tasks successfully executed by the evaluation agent across the entire benchmark, relative to the total number of tasks.

\subsubsection{Evaluation Task Annotation}

To evaluate the consistency of interactive behavior, we manually annotate a set of deterministic test tasks for each dashboard. Further annotation principles, quality assurance details and examples are in the appendix~\ref{sec:eval_task_anno_detail}.

\subsubsection{Evaluation Metric}
\label{sec:eval_metric_dynamic}
\paragraph{Figure Similarity.}
After each task is executed, we extract the internal Plotly figure JSON specifications from the frontend for both the generated dashboard and the reference dashboard, and perform a fine-grained comparison along four dimensions Style, Data, Type and Text and do weighted average as final score.
Details are in the appendix~\ref{sec:fig_sim_detail}.
In particular, Figure Similarity correlates closely with human judgments as an LLM-judge-free metric, as shown in the appendix~\ref{sec:human_correlation}.

\paragraph{LLM-Visual-Eval.}
After each task is executed, we use \textsc{Gemini-3-Flash} to score the visual similarity between the generated dashboard screenshot and the reference dashboard screenshot. The full prompt is provided in the appendix~\ref{sec:related_prompt_eval}.

\paragraph{LLM-Task-Behavior-Eval.}
For each task, we feed \textsc{Gemini-3-Flash} four screenshots: the reference dashboard before execution and after execution, and the generated dashboard before execution and after execution. \textsc{Gemini-3-Flash} is instructed to focus on whether the \emph{responses caused by the interaction} are consistent across the two dashboards, and outputs a behavior-consistency score. The full prompt is provided in the appendix~\ref{sec:related_prompt_eval}.

\subsection{Final Evaluation Metric}
\label{sec:final_eval_metric}
The final evaluation score is a weighted combination of the above metrics. 
\begin{equation}
\begin{aligned}
\mathrm{Score} \;=\;& 0.1\,\mathrm{KCC} + 0.3\,\mathrm{LLM}_{\text{sem}} + 0.2\,\mathrm{FigSim} \\
&+ 0.2\,\mathrm{LLM}_{\text{vis}} + 0.2\,\mathrm{LLM}_{\text{beh}}.
\end{aligned}
\end{equation}
This aggregated score achieves a Pearson correlation of $0.78$ with human evaluation. We further validate the robustness of LLM-as-judge choice. Details of metric validation, quality control and weight determination are provided in the appendix~\ref{sec:eval_framework_detail}. In our evaluation, if the generated code fails to execute, the corresponding dashboard receives a score of 0.

\begin{table*}[t]
\small
\centering
\begin{adjustbox}{max width=2.0\columnwidth}
\begin{tabular}{lcccccccc}
\toprule
\textbf{Model} & \textbf{Code Exec.} & \textbf{Task Exec.} & \textbf{Comp. Cov.} & \textbf{LLM Sem.} & \textbf{Fig Sim} & \textbf{LLM Vis.} & \textbf{LLM Behav.} & \textbf{Total} \\
\midrule
\multicolumn{9}{c}{\emph{Closed-source (with DOM)}} \\
\midrule
Gemini 3 Pro & 97.8 & 94.2 & 91.6 & 81.9 & 80.9 & 70.3 & 77.1 & 79.4 \\
GPT-5.1 & 82.2 & 76.9 & 73.4 & 61.4 & 61.8 & 62.0 & 50.9 & 60.7 \\
Claude Sonnet 4.5 & 78.9 & 71.2 & 69.3 & 48.3 & 53.5 & 50.6 & 41.2 & 50.5 \\
\midrule
\multicolumn{9}{c}{\emph{Closed-source (w/o DOM)}} \\
\midrule
Gemini 3 Pro & 99.4 & 93.2 & 90.1 & 74.1 & 78.0 & 72.1 & 69.6 & 75.2 \\
GPT-5.1 & 78.3 & 63.0 & 68.7 & 53.0 & 51.0 & 49.2 & 45.6 & 51.9 \\
Claude Sonnet 4.5 & 83.3 & 59.8 & 70.3 & 42.9 & 48.6 & 43.0 & 30.2 & 44.3 \\
\midrule
\multicolumn{9}{c}{\emph{Open-source (with DOM)}} \\
\midrule
Qwen3-VL-30B-A3B-Instruct & 63.3 & 12.8 & 50.5 & 30.8 & 29.1 & 12.4 & 12.5 & 25.1 \\
InternVL3.5-8B & 5.6 & 1.1 & 4.8 & 1.7 & 2.8 & 1.1 & 1.1 & 2.0 \\
Qwen3-VL-8B-Instruct & 2.8 & 1.8 & 1.9 & 1.2 & 1.4 & 0.9 & 0.8 & 1.2 \\
\midrule
\multicolumn{9}{c}{\emph{Open-source (w/o DOM)}} \\
\midrule
Qwen3-VL-30B-A3B-Instruct & 60.0 & 12.2 & 46.5 & 27.8 & 28.1 & 12.2 & 12.2 & 23.5 \\
InternVL3.5-8B & 2.2 & 0.6 & 1.5 & 0.7 & 1.2 & 0.6 & 0.6 & 0.8 \\
Qwen3-VL-8B-Instruct & 1.1 & 1.1 & 0.8 & 0.6 & 0.6 & 0.3 & 0.3 & 0.5 \\
\bottomrule
\end{tabular}
\end{adjustbox}
\caption{Overall results on DashboardMimic under settings with and without DOM information.}
\label{tab:main_results}
\end{table*}

\section{Experiments}

\subsection{Experimental Setup}
In this section, we detail the implementation of the Dashboard2Code interactive environment, the agent’s interaction mechanism and context-management strategy, and the base model configurations evaluated in our experiments.

\subsubsection{Interaction Environment}
To simulate realistic web interactions, we establish a lightweight environment using Selenium to drive Google Chrome, with the viewport resolution fixed at \(1920 \times 1080\) for consistent visual observations. Beyond visual feedback, the environment can optionally provide the DOM tree, which undergoes a strict pruning and abstraction strategy to optimize context length. We adopt an MM-React-style framework where the model generates its reasoning and executes actions—such as standard keyboard and mouse operations—within the environment. Full implementation details are provided in the Appendix~\ref{sec:interaction_environment_detail}.

\subsubsection{Baseline Models}

We selected the most advanced proprietary and open-weight multimodal models (LMMs) for evaluation:
For Proprietary Models, we evaluate OpenAI's \texttt{GPT-5.1} \citep{openai2024gpt4technicalreport}, Google's \texttt{Gemini 3 Pro}, and Anthropic's \texttt{Claude-Sonnet-4.5}.
For Open-weight models, we select the \texttt{Qwen3-VL} \citep{bai2025qwen3vltechnicalreport} series (including the 30B-A3B and 8B versions) and the \texttt{InternVL3.5-8B}  \citep{wang2025internvl35advancingopensourcemultimodal}.

\subsubsection{Evaluation Setup}

We use an MM-React-style \citep{yang2023mmreactpromptingchatgptmultimodal} framework for evaluation. At each step, the multimodal model receives the current dashboard screenshot, optional DOM tree, historical screenshots, and actions as input. The model outputs its reasoning and selects an action from its action space.
We further use a resolution compression strategy to reduce overhead. Detailed evaluation setup and examples is shown in the appendix~\ref{sec:eval_setup_detail}.

\subsection{Main Results}

Table~\ref{tab:main_results} presents the results of Dashboard2Code task in DashboardMimic benchmark. We have the following observations.

\paragraph{Dashboard2Code is a challenging task for all models.}  
As shown in the results from Table~\ref{tab:main_results}, the Dashboard2Code task presents a significant challenge for all types of models. Even the best-performing proprietary model, such as Gemini 3 Pro, still has substantial room for improvement. In particular, its performance on the more complex Level 3 tasks, which involve advanced callback logic, is only 64.2 points, highlighting the difficulty that even the strongest current models face when tasked with generating interactive code in scenarios that approach real-world complexity. Moreover, we observe that performance drops dramatically when DOM information is excluded, indicating that models still have significant limitations in GUI visual grounding on dashboards.

\paragraph{Large Performance Gap between Proprietary and Open-Source Models.}  
There is a notable performance gap between proprietary and open-source models. Proprietary models, such as Gemini 3 Pro and GPT-5.1, consistently outperform the current best open-source multimodal models, such as Qwen3-VL and InternVL3.5. This disparity is especially pronounced when DOM support is unavailable, with open-source models showing a significant drop in performance, despite their strong capabilities in GUI interaction and visual reasoning. This suggests that open-source models are still far behind proprietary models in terms of the active exploration, integration of long-context information, visual logic reasoning, and code generation required for Dashboard2Code tasks.

\begin{table}[t]
\centering
\normalsize
\begin{adjustbox}{max width=1.0\columnwidth}
\begin{tabular}{lccc}
\toprule
\textbf{Model} & \textbf{Avg. Steps} & \textbf{Ineff. Exp. (\%)} & \textbf{UI Expl. Ratio} \\
\midrule
Gemini 3 Pro with DOM    & 3.31 & 2.37\%  & 64.58 \\
Gemini 3 Pro w/o DOM     & 3.90 & 40.15\% & 20.61 \\
GPT-5.1 with DOM         & 4.25 & 4.96\%  & 45.06 \\
GPT-5.1 w/o DOM          & 7.18 & 56.96\% & 23.00 \\
\bottomrule
\end{tabular}
\end{adjustbox}
\caption{Exploration statistics under settings with and without DOM feedback. Here, \textbf{Avg. Steps} denotes \textbf{Avg. Exploration Steps}, i.e., the average number of interaction steps taken on the dashboard during exploration; \textbf{Ineff. Exp. (\%)} denotes \textbf{Ineffective Exploration (\%)}, i.e., the percentage of interactions for which no visual changes were observed in the screenshots before and after the interaction; and \textbf{UI Expl. Ratio} denotes \textbf{UI Component Exploration Ratio}, i.e., the ratio of explored UI components that successfully triggered callbacks to the total number of components in the dashboard.}
\label{tab:exploration_stats}
\end{table}
\begin{figure}[t]
      \centering
      \includegraphics[width=1.0\columnwidth]{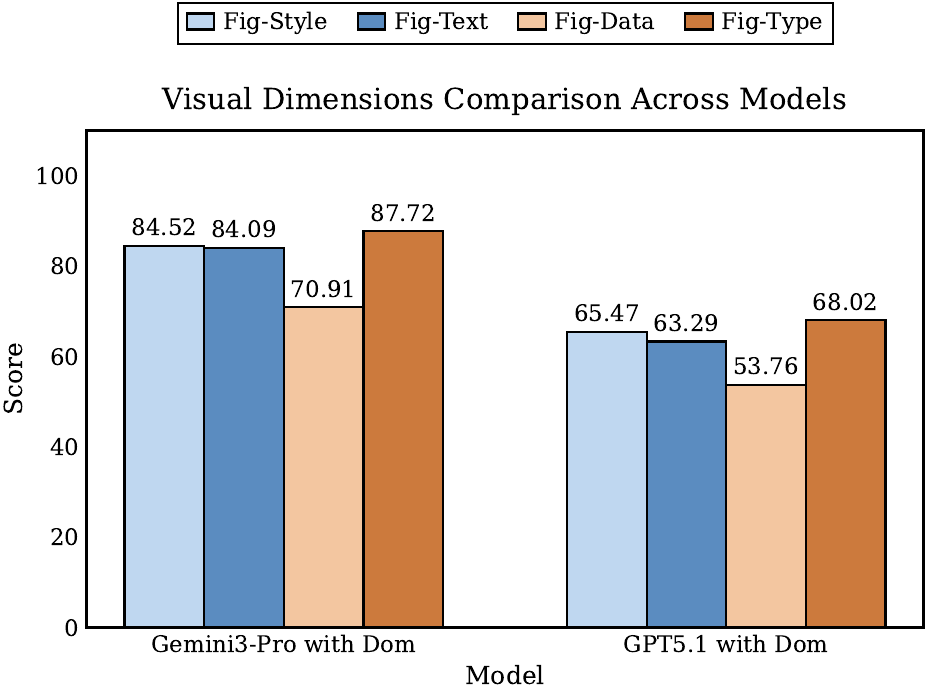}
      \caption {Break down analysis of different visual dimensions.}
      \label {fig:experiment_analysis_low_level}
\end{figure}

\begin{figure}[t]
  \centering
  \includegraphics[width=0.9\columnwidth]{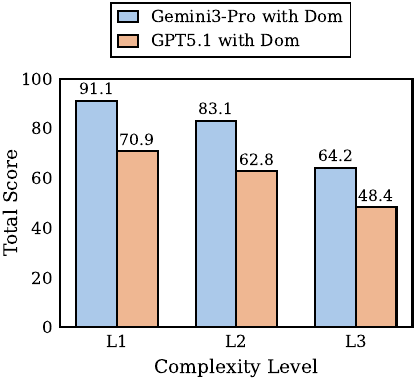}
  \caption {Break down analysis of different interaction logic complexity.}
  \label {fig:experiment_analysis_complexity}
\end{figure}

\subsection{Analysis}
This section conducts a detailed analysis of the model behavior and performance in Dashboard2Code task.

\subsubsection{Breakdown Analysis}

\paragraph{There Is Significant Room for Improvement in Model Exploration Performance.}  
As shown in Table~\ref{tab:exploration_stats}, even the state-of-the-art (SOTA) models achieve only a 64.58\% exploration ratio of interactive components in the dashboard. This highlights the substantial room for improvement in the model's ability to explore and make decisions based on historical visual feedback within the Dashboard2Code task. Furthermore, incorporating the DOM tree, as opposed to relying solely on screenshot-based exploration, significantly reduces redundant exploration and improves the completeness of the exploration process.

\paragraph{Model Struggles to Reproduce Dashboard Visual Details.}

We analyze the model’s performance in reproducing the visual consistency of different dimensions. As shown in Figure~\ref{fig:experiment_analysis_low_level}, the model performs best in reproducing the \texttt{Type}, followed by \texttt{Style} and \texttt{Text}, with the weakest performance in \texttt{Data}. We attribute this to the fact that the first three are frequently changing attributes in dashboards, combined with the model's insufficient perception of fine-grained visual details.

\paragraph{Model Struggles to Reproduce Complex Interaction Logic.}  
We analyze the model’s performance across different levels of callback logic complexity. As shown in Figure~\ref{fig:experiment_analysis_complexity}, the model's overall performance declines significantly as the complexity of the callback logic increases. This indicates that the model still faces considerable difficulty in exploring and reproducing complex logical interactions.
\begin{table}[t]
\centering
\small
\begin{tabular}{l l c}
\toprule
\textbf{Exploration Agent} & \textbf{Coding Agent} & \textbf{Total Score} \\
\midrule
Gemini-3-Pro      & Gemini-3-Pro         & 75.8 (+0.6) \\
Gemini-3-Pro      & Claude-4.5-Sonnet    & 53.4 \\
Claude-4.5-Sonnet & Claude-4.5-Sonnet    & 49.2 (+4.9) \\
\bottomrule
\end{tabular}
\caption{Results under role-separated inference. The numbers in parentheses indicate the gain over the corresponding end-to-end baseline.}
\label{tab:role_separation_results}
\end{table}
\begin{table}[t]
\normalsize
\centering
\begin{adjustbox}{max width=1.0\columnwidth}
\begin{tabular}{lcc}
\toprule
\textbf{Model} & \textbf{LLM Sem.} \\
\midrule
Gemini 3 Pro w/o DOM & 74.11 \\
Gemini 3 Pro w/o DOM + anonymize text & 54.10 \\
Gemini 3 Pro with DOM & 81.89 \\
Gemini 3 Pro with DOM + anonymize text & 60.30 \\
\bottomrule
\end{tabular}
\end{adjustbox}
\caption{LLM Semantic scores for Gemini under different DOM and text anonymization settings. Note that we calculate all the generated code with the ground truth code regardless of the execution error in the anonymization setting which makes the metric in anonymization setting an upperbound.}
\label{tab:anonymization_results}
\end{table}
\label{sec:text_robustness}
\subsubsection{Text Perturbation Robustness Analysis}
\paragraph{Model Performance Sensitivity to Text Perturbation.}
We observed that for some samples, the target components and logic of callback functions can be directly inferred from the relationships between control text, titles, and descriptive labels in the interface. To test whether the model primarily relies on these text cues instead of understanding callback semantics through the interaction process, we created a counterfactual setup: without altering the layout, component types, or the semantic content of the rest of the code, we anonymized the visible text elements in the benchmark dashboards (e.g., control labels, titles) by replacing them with meaningless labels such as "Option1", "Option2", and "Title1" in the \texttt{app.layout}. The details of this anonymization are provided in the appendix~\ref{sec:text_anoy}. Note that we only
modify the textual element in app.layout
and keep all the others the same. After the process,
we manually recheck each modified code to ensure
that only the text in app.layout has changed. The experimental results in Table~\ref{tab:anonymization_results} show that even for state-of-the-art models, anonymization significantly reduces callback semantic replication quality. This phenomenon suggests that, in some cases, the model may rely heavily on text cues for heuristic inference rather than robustly recovering the interaction logic. Further details and examples are shown in appendix~\ref{sec:text_anoy}.

\subsubsection{Agent Role Separation Analysis}
In our main experiments, we employ the same agent to perform both interactive exploration and code generation. However, in practical deployment, these two responsibilities may be handled by different expert foundation models. To examine this setting, we conduct an additional experiment in which inference is decomposed into two distinct phases. In the exploration phase, an Exploration Agent interacts with the dashboard and produces an exploration trajectory, including screenshots, intermediate reasoning, and action history. In the generation phase, a Coding Agent takes this trajectory as input and generates the final code. We evaluate several cross-model role configurations using Gemini 3 Pro and Claude-4.5-Sonnet. The results are reported in Table~\ref{tab:role_separation_results}. As shown in Table~\ref{tab:role_separation_results}, role separation leads to modest improvements over the corresponding end-to-end baselines for individual models (e.g., Claude improves from 44.3 to 49.2). Nevertheless, the strongest overall performance still requires a model that is simultaneously effective at dynamic exploration and visually grounded code generation.
\begin{figure*}
    \centering
    \includegraphics[width=0.85\textwidth]{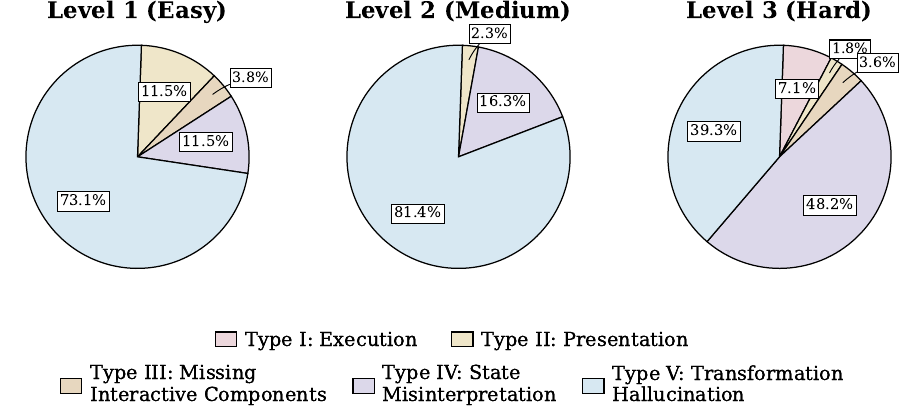}
    \caption{Error distribution of Gemini 3 Pro in different complexity levels.}
    \label{fig:error_distribution}
\end{figure*}
\subsubsection{Error Analysis}

To provide a comprehensive understanding of LLM limitations in the Dashboard2Code task, we conducted a fine-grained manual inspection of failure cases. 
We identify five common error types, progressing from surface-level implementation to deep semantic reasoning. We show examples of each error type in appendix~\ref{sec:error_case_sota}.

\paragraph{Execution dimension - Execution Failure (Syntax \& Runtime Errors)}
The most fundamental failure occurs when the generated code cannot be rendered. This includes invalid syntax, hallucinated dependencies, or deprecated API calls.

\paragraph{Presentation dimension - Presentation Mismatch (Styling \& Layout Deviations)}
The code is executable and logically valid but fails to match the target's aesthetic specifications. Examples including incorrect layout arrangements (e.g., vertical vs. horizontal stacking), mismatched color palettes.

\paragraph{Logical dimension - Missing Interactive Components}
The model misses the specific interactive components in the dashboard or fails to replicate the interactive affordance of specific elements, for instance, the model failed to identify hidden controls during the exploration process, resulting in the inability to replicate them.

\paragraph{Logical dimension - State Misinterpretation (Broken Dependency)}
The model identifies components but misinterprets the causal graph between inputs and outputs. For instance, treating coupled variables as independent or wiring triggers to incorrect outputs.

\paragraph{Logical dimension - Transformation Hallucination}
The model successfully replicates the interaction skeleton (i.e., the causal graph) but fails to reproduce the exact data or visual transformation logic. For instance, when the progress bar is moved to the right, the gold data increases, but the generated dashboard data decreases.

The statistical analysis results are illustrated in the Figure~\ref{fig:error_distribution}. It can be observed that in L1 and L2, the main issue with the model lies in Transformation Hallucination, suggesting that the model is generally capable of understanding the causal relationships between simpler components. However, it still struggles to capture more fine-grained logical changes. In contrast, for L3, the model continues to face challenges in capturing dependencies between components in most cases. Additionally, an analysis of the Missing Interactive Components category revealed that the primary issue stems from the model's inability to effectively explore interactive components that are not explicitly presented on the initial screenshot.

\section{Conclusion}

We propose the Dashboard2Code task, which requires models to explore interactive dashboards, integrate feedback, and generate code that reproduces both visual and interaction logic. We introduce the DashboardMimic benchmark, the first Plotly+Dash-based Dashboard2Code benchmark covering diverse visualization types and interaction patterns. Experimental results reveal that Dashboard2Code poses a strong challenge for multi-modal models.

\section*{Limitations}

Our current benchmark, DashboardMimic, is limited to Plotly+Dash-based dashboards, which may not fully capture the diversity of interactive dashboard frameworks in real-world applications. Additionally, the use of LLMs as judges introduces significant evaluation costs, as it requires large-scale model inference and human evaluation for validation. Future work can explore expanding the benchmark to include other popular dashboard frameworks and investigate more cost-efficient evaluation methods.

\section*{Acknowledgments}

We gratefully acknowledge the support of the
National Natural Science Foundation of China
(NSFC) via grant 62236004 and 62476073.


\bibliography{custom}

\appendix

\section{Appendix}
\label{sec:appendix}

\subsection{Related Work}
\label{sec:related work}
\paragraph{Interactive GUI Agents}

Recent research on interactive GUI agents has focused on enabling AI systems to perceive and act within graphical user interfaces in a human‑like manner — OSWorld \citep{xie2024osworld} offers a scalable, real computer environment with diverse interactive tasks to benchmark multi-modal agents on open‑ended workflows, DashboardQA \citep{kartha2025dashboardqabenchmarkingmultimodalagents} evaluates agents’ ability to interact with and reason over real interactive dashboards, and MMBench‑GUI  \citep{wang2025mmbenchguihierarchicalmultiplatformevaluation} provides a hierarchical, multi‑platform benchmark covering interface understanding, element grounding, task automation. However, these research tasks typically provide explicit goal instructions, making it difficult to assess whether GUI Agents can \textbf{actively fully} explore, understand, and reason over dynamic interactive interfaces and then recreate it.

\paragraph{Multi-modal Code generation}

Multi-modal code generation refers to the task where a model takes inputs from multiple modalities and ultimately generates code. Early works such as Pix2Struct \citep{lee2023pix2structscreenshotparsingpretraining}, Design2Code \citep{si-etal-2025-design2code}, and Web2Code \citep{yun2024web2codelargescalewebpagetocodedataset} focus on evaluating the capability to generate code from static front‑end representations. More recently, multi-modal code generation has been expanded to multiple domains, including image‑to‑CAD, chart‑to‑code \citep{yang2025chartmimic,tang2025chartscodehierarchicalbenchmark,niu-etal-2025-chart2code53,zhao-etal-2025-chartcoder}, slide‑to‑code \citep{tang-etal-2025-slidecoder}, and paper‑to‑poster \citep{pang2025paper2postermultimodalposterautomation} generation. Despite significant progress, these tasks primarily concentrate on static inputs and overlook the interactive characteristics required to construct real‑world applications. To address this limitation, Interaction2Code \citep{xiao2025interaction2codebenchmarkingmllmbasedinteractive} proposes to recover code from pre-interaction and post‑interaction screenshots, and IWR-Bench \citep{chen2025iwrbenchlvlmsreconstructinteractive} introduces recovering web pages from human‑recorded interaction trajectories. However, these tasks still rely on manually annotated trajectories, which increases annotation cost. Moreover, existing evaluations do not examine the model’s ability to actively explore and obtain external feedback.

Concurrently, WebVIA \citep{xu2025webviawebbasedvisionlanguageagentic} presents an agent model in which an exploration agent is responsible for exploring web content and a code‑generation agent generates code based on the exploration results. Our work differs from WebVIA in the following aspects: (1) In terms of task definition, our task’s ultimate goal is to replicate dashboards and does not emphasize a separation between exploration and code generation, whereas WebVIA still evaluates exploration and code generation separately, and its code generation evaluation depends on human‑selected screenshots; (2) Our domain focuses more specifically on dashboards and the diverse callback logical structures within dashboards; (3) We design a comprehensive set of automated evaluation metrics that are more tailored to the Dashboard2Code task.

\subsection{Comparison with existing benchmarks}
\label{sec:comparison_with_existing_benchmarks}
\begin{table*}[t]
\small
\centering
\begin{adjustbox}{max width=2.0\columnwidth}

\begin{tabular}{ccccc}
\toprule
\textbf{Name}    & \textbf{Task}        & \textbf{Domain} & \textbf{\# Samples} & \textbf{Data Source}         \\ \midrule

Design2Code      & UI-to-Code    & Web pages &    484 & Real websites \\
Chart2Code       & Chart-to-Code  & Chart &  2,023 & - \\
Plot2Code        & Chart-to-Code                        & Chart           & 386                & Matplotlib gallery                  \\
ChartMimic       & Chart-to-Code                         & Chart           & 4,800              & ArXiv papers   \\
IWR-Bench        & Video-to-Code                         & Web pages          & 113                & Real websites                  \\
Interaction2Code & Images-to-Code                        & Web pages       & 374                & Github+C4                  \\
DashboardQA      & GUI Interactive QA                       & Dashboard       & 112                & Tableau Public                  \\
DashboardMimic   & GUI Interactive-to-Code                  & Dashboard       & 180                & Dash gallery+LLM aug. \\ \bottomrule
\end{tabular}

\end{adjustbox}
\caption{Comparison of DashboardMimic with existing  benchmarks. Design2Code focuses on generating code to reconstruct UI design. Plot2Code and ChartMimic focus on static chart-to-code, while IWRBench and Interaction2Code reconstruct code from human-annotated interaction traces; DashboardQA first proposes dashboard question answering. In contrast, DashboardMimic evaluates Dashboard2Code, where a GUI agent actively interacts with a dashboard at runtime and generates code that reproduces it in both behavior and visual appearance.}
\label{tab:benchmark_statistics}
\end{table*}
We compare our benchmark with existing related benchmarks in Table~\ref{tab:benchmark_statistics}.

\subsection{Benchmark Construction Details}
\label{sec:benchmark_construction_detail}
\subsubsection{Benchmark Filtering Principle}

\label{sec:benchmark_filtering_principle}
\begin{itemize}
  \item \textbf{P1 (Interactive \& dynamic).}
  The dashboard must be interactive; its full behavior is revealed and verified only through interaction.

  \item \textbf{P2 (No animations or pop-ups).}
  Exclude animations, modal dialogs, and other non-instant response behaviors.

  \item \textbf{P3 (Reverse-engineerable \& self-contained).}
  The dashboard must not rely on unreproducible external assets (e.g., audio/images) or data downloaded from external links, except for data that can be mocked and hard-coded. All relevant elements must be observable, and the code must be self-contained.

  \item \textbf{P4 (Unified framework).}
    We restrict implementations to a standard library set---\texttt{dash}, \texttt{plotly},
      \texttt{dash\_bootstrap\_components (dbc)}, \texttt{dash\_daq}, and \texttt{dash\_ag\_grid}---and avoid non-standard dependencies.

  \item \textbf{P5 (Realistic layout and visuals).}
  The dashboard should have a realistic layout and appearance without readability issues (e.g., occlusion/overlap).

  \item \textbf{P6 (Correct complexity labeling).}
  Each sample must match its assigned complexity subtype.
\end{itemize}

\subsubsection{Definitions of Different Interactive Logics}
\label{sec:interaction_logic_detail}
\emph{\textbf{Level 1: Atomic (One-to-One).}}
This level captures the minimal interaction unit: a single input component controls a single chart output through one callback function. It primarily tests basic understanding of callback wiring and interaction syntax.

\emph{\textbf{Level 2: Star \& Mesh Topology (Coupling \& Aggregation).}}
This level covers \emph{broadcasting} and \emph{aggregation} patterns. It includes \emph{one-to-many} logic (one input updates multiple charts), \emph{many-to-one} logic (multiple inputs jointly control one chart), and \emph{many-to-many} coupled mesh structures. The main challenge is to disentangle interactions among variables and correctly handle \emph{simultaneous updates}.

\emph{\textbf{Level 3: Inter-dependent (State \& Circularity).}}
This level features long-range dependency chains and explicit state management. It includes
\emph{chained callbacks}, where the output of one callback updates another input component and thereby triggers subsequent callbacks;
\emph{conditional visibility}, where the presence or visibility of a control is dynamically toggled based on other controls or intermediate outputs;
\emph{state dependency}, where an output depends not only on the currently triggered inputs but also on persistent state variables (e.g., stored selections) carried across interactions;
and \emph{circular callbacks}, where two or more components mutually update each other, forming feedback loops that require careful handling to avoid inconsistent states.
These behaviors cannot be understood by isolating individual inputs, and thus demand \emph{global state reasoning}.

\subsubsection{LLM-based Data Augmentation Details}
\label{sec:llm_aug_detail}
We show the prompt for LLM-Based Data augmentation in Figure~\ref{fig:data_gen_prompt}.

\subsubsection{Real-world Data License}
\label{sec:real_world_license}
The 58 real-world seed samples in our benchmark come from dash-example-index, which is under MIT-License.

\subsubsection{Mitigating Data Contamination}
\label{sec:decontamination}

Given that our real-world seed data are sourced from open-source repositories, there is a potential risk that large-scale models have been exposed to these samples during pre-training. To rigorously mitigate this risk and ensure the benchmark evaluates reasoning rather than memorization, we applied a strict \textbf{Rewrite-and-Refactor} pipeline to all collected seed dashboards.

As shown in our curation process, we employed a standardized LLM-based refactoring step guided by strict constraints to perturb the original code surface forms while preserving interaction logic:

\begin{itemize}
    \item \textbf{Semantic Anonymization:} We explicitly removed specific references to standard "toy datasets" (e.g., Iris, Titanic) that act as strong retrieval cues for LLMs. As per our curation protocol, all specific proper nouns and dataset names were replaced with generic, domain-realistic equivalents (e.g., renaming ``Iris Dataset'' to ``Botanical Dimensions Analysis'') to prevent the model from recalling the code via keyword association.
    \item \textbf{Data Slicing and Component Normalization:} To differentiate our samples from raw GitHub files, we enforced constraints on data density. This ensures that even if a model had seen the original implementation, the token sequence in our benchmark differs significantly, forcing the model to rely on the current visual and DOM context rather than memory.
\end{itemize}

\begin{table}[t]
\centering
\small
\begin{tabular}{lcc}
\toprule
Metric & Matched & Random \\
\midrule
5-gram Jaccard & 0.1724 & 0.0102 \\
3-gram Jaccard & 0.2509 & 0.0270 \\
MC Cosine      & 0.2069 & 0.0018 \\
\bottomrule
\end{tabular}
\caption{Similarity analysis between the original real-world code and the refactored code. ``Matched'' denotes similarity between each original sample and its corresponding refactored version, while ``Random'' denotes similarity between each original sample and a randomly sampled dashboard. We report lexical overlap using $n$-gram Jaccard similarity and semantic relatedness using mean-centered (MC) cosine similarity. The low Jaccard scores for matched pairs indicate substantial surface-form variation. The higher MC cosine similarity than the random baseline suggests that the underlying domain logic is largely preserved.}
\label{tab:data_leakage_similarity}
\end{table}
These perturbation strategies ensure that DashboardMimic serves as a test of generalization capability rather than a retrieval task.

Furthermore, to assess the risk of data contamination quantitatively, we conducted a rigorous similarity analysis between the original real-world code and the corresponding refactored code. We considered both lexical overlap and embedding-level semantic similarity by adopting 5-gram and 3-gram Jaccard similarity, together with mean-centered cosine similarity, which mitigates embedding anisotropy. We further compared the true matched pairs (Original vs. Refactored) against a random baseline constructed by pairing each original sample with a randomly sampled dashboard. As shown in Table~\ref{tab:data_leakage_similarity}, the matched pairs exhibit only limited surface-form overlap, with 5-gram and 3-gram Jaccard scores of 0.1724 and 0.2509, respectively, suggesting that our refactoring pipeline substantially alters the lexical realization of the code and thus reduces the likelihood of rote memorization. At the same time, the mean-centered cosine similarity remains markedly higher than that of the random baseline (0.2069 vs. 0.0018), indicating that the refactored code still preserves the core domain logic of the original dashboard programs.

\subsubsection{Benchmark Statistics}
\label{sec:benchmark_statistics}
\begin{table}[t]
    \small
    \centering
    \begin{adjustbox}{max width=1.0\columnwidth}
    \begin{tabular}{ccc}
    \toprule
    \textbf{Component} & \textbf{Count} & \textbf{Prevalence (\%)} \\
    \midrule
    Dropdown     & 112 & 42.2 \\
    RadioItems   &  51 & 30.6 \\
    Slider       &  70 & 33.3 \\
    RangeSlider  &  28 & 20.0 \\
    Checklist    &  32 & 20.1 \\
    \bottomrule
    \end{tabular}
    \end{adjustbox}
    \caption{Prevalence of common interactive components in the benchmark.}
    \label{tab:component_prevalence}
\end{table}

\begin{figure}[t]
  \centering
  \includegraphics[width=1.0\columnwidth]{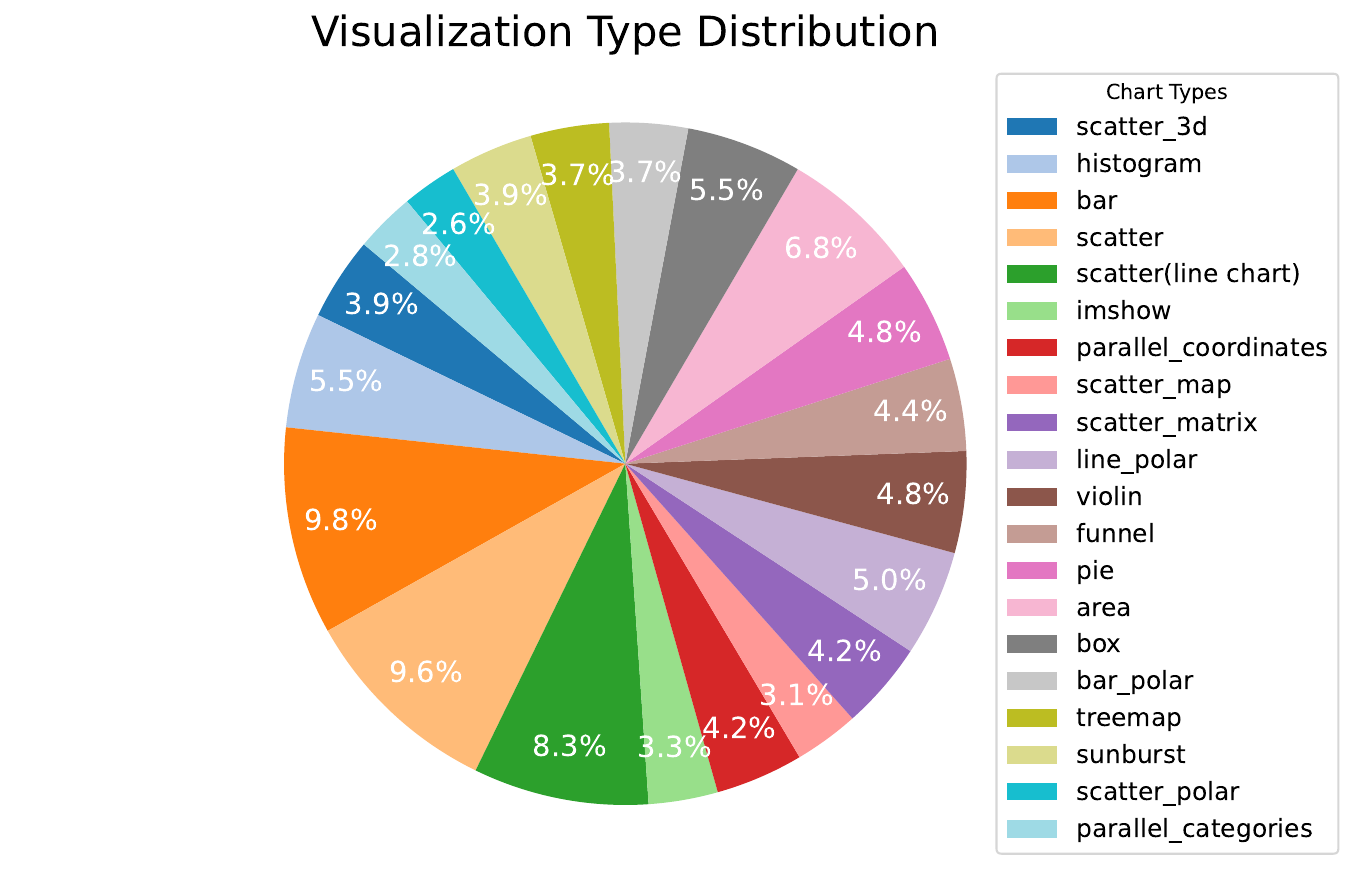}
  \caption {Visualization type distribution.}
  \label{fig:Visualization_Type_Distribution}
\end{figure}

\begin{figure}[t]
  \centering
  \includegraphics[width=0.6\columnwidth]{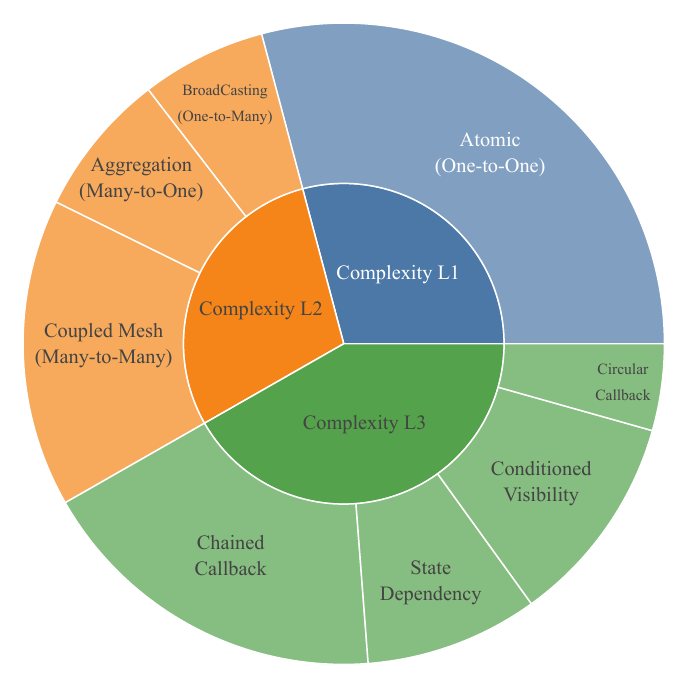}
  \caption {Visualization interaction logic distribution.}
  \label{fig:Visualization_Complexity_distribution}
\end{figure}
\emph{\textbf{UI component statistics.}}
As shown in Table~\ref{tab:component_prevalence}, the top-5 UI components include the most widely used dashboard controls—\texttt{Dropdown}, \texttt{RadioItems}, \texttt{Slider}, \texttt{RangeSlider}, and \texttt{Checklist}—each with substantial prevalence. This design reflects real-world dashboard interaction patterns and ensures broad coverage of common filtering and selection operations.

\emph{\textbf{Visualization type diversity.}}
As shown in the Figure~\ref{fig:Visualization_Type_Distribution}, the benchmark covers 20 common visualization types, with the categories represented in a relatively balanced manner.

\emph{\textbf{Interaction logic diversity.}}
As shown in the Figure~\ref{fig:Visualization_Complexity_distribution}, the benchmark covers 8 common dashboard interaction-logic types, grouped into 3 levels of complexity. Within each complexity level, the subtypes are distributed relatively well balanced.

\subsubsection{Discussion about other Dashboard Frameworks}
\label{sec:discuss_other_framework}
We also considered alternative dashboard frameworks, but found them less well aligned with Dashboard2Code. Proprietary BI systems such as Tableau and Looker do not provide a transparent, executable program target suitable for rigorous code-level evaluation. React-based stacks, including Tailwind CSS and Material UI, are highly expressive, but they introduce substantial presentation-layer variability through component composition and utility-class styling. By comparison, Plotly+Dash provides a more suitable abstraction for our setting, as its Python-based, callback-driven design exposes dashboard logic more directly and reduces interference from frontend presentation details.

\subsection{Evaluation Framework Details}
\label{sec:eval_framework_detail}
\subsubsection{Human Evaluation Processes and Guidelines}
\label{sec:human_eval_process_and_guideline}
To establish a rigorous ground truth for evaluation, we designed a streamlined front-end interface facilitating expert review. A panel of three graduate students majoring in computer science, each possessing extensive experience in data visualization, served as human evaluators. They assessed a diverse set of \textbf{90 dashboards} generated by three distinct models (GPT-5.1, Gemini 3 Pro, Claude Sonnet 4.5), randomly sampled to ensure representativeness.

Evaluators scored each dashboard based on \emph{Visual Fidelity} and \emph{Functional Consistency}. The final ground-truth score is derived from the average of these dimensions. To validate the reliability of our human annotations, we calculated the inter-annotator agreement. The evaluators demonstrated a strong consensus, achieving a \textbf{Pearson correlation coefficient of 0.6525}, confirming the consistency of the subjective scoring process.

We then analyzed the correlation between our automated metrics and these human judgments. As shown in Table~\ref{tab:metric_correlations}, all proposed metrics show statistically significant positive correlations ($p \ll 0.001$). Notably, our weighted combination metric achieves the highest alignment with human perception (Pearson $r=0.7807$, Spearman $r=0.7173$), demonstrating that our method serves as a robust proxy for human evaluation even under rigorous statistical testing.

\paragraph{Visual Fidelity Scoring Guidelines} The maximum score for this aspect is 5 and the detailed scoring guidelines are as follows:
\begin{itemize}[label=-]
    \item \textbf{Score 5: Pixel-Perfect }\\
    Almost identical to the original. Layout, component types, colors, font sizes, and spacing are nearly the same. Minor differences due to rendering engines are allowed.
    
    \item \textbf{Score 4: High Fidelity }\\
    The overall look is very similar. Layout and component placement are correct. There are minor style differences (e.g., slight color differences in charts, font size slightly off) but they don't affect the structure.
    
    \item \textbf{Score 3: Structural Match }\\
    The structure is correct, but the appearance differs. All charts and components are present, and the arrangement is correct. There are noticeable style issues (e.g., missing background color, misplaced legend, different point shapes in scatter plots).
    
    \item \textbf{Score 2: Component Error }\\
    Layout or core components are incorrect. The general structure is maintained, but there are clear defects:
    \begin{itemize}
        \item Incorrect chart type (e.g., bar chart changed to a line chart).
        \item Missing key components (e.g., a dropdown menu is missing).
        \item Layout misalignment (e.g., a two-column layout turned into a vertical arrangement).
    \end{itemize}
    
    \item \textbf{Score 1: Severe Distortion }\\
    Significant breakdown. Only minimal text or elements are retained, the interface is chaotic, elements overlap, or the interface contains error messages or scrambled text.
    
    \item \textbf{Score 0: Irrelevant }\\
    Completely irrelevant or blank. The generated interface is either blank or contains entirely incorrect content (e.g., drawing a picture results in a text box).
\end{itemize}

\paragraph{Functional Consistency Scoring Guidelines} The maximum score for this aspect is 5 and the detailed scoring guidelines are as follows:

\begin{itemize}[label=-]
    \item \textbf{Score 5: Behavior Match }\\
    Logic and feedback are fully synchronized. The changes in the generated interface match the ground truth exactly.
    
    \item \textbf{Score 4: Logic Correct }\\
    Core functionality is correct, but visual feedback has minor flaws.
    \begin{itemize}
        \item the data filtering logic is correct (e.g., trends update properly), but there are minor issues with visual feedback.
    \end{itemize}
    
    \item \textbf{Score 3: Partial Success }\\
    Partial interaction failure. The main chart updates, but secondary information does not:
    \begin{itemize}
        \item For example, clicking a point on the map updates the bar chart on the right, but the “Current Selected City: X” title does not change.
    \end{itemize}
    
    \item \textbf{Score 2: Wrong Logic }\\
    The interaction responds, but the result is incorrect:
    \begin{itemize}
        \item e.g. the task is “Highlight,” but it performs a “Filter” action, causing other data to disappear.
    \end{itemize}
    
    \item \textbf{Score 1: No Response }\\
    The interaction fails. After performing an action, there is no change in the interface (e.g., the screenshot remains static), indicating that the callback was not triggered or there is a bug in the code.
    
    \item \textbf{Score 0: Crash/Broken }\\
    The interaction causes an error. After performing an action, the chart disappears, becomes blank, or an error message such as a Python Traceback appears.
\end{itemize}

\begin{table}[t]
\normalsize
\centering
\begin{adjustbox}{max width=1.0\columnwidth}
\begin{tabular}{lccc}
\toprule
\textbf{Metric} & \textbf{Pearson} & \textbf{Spearman} & \textbf{p-value} \\
\midrule
Semantic & 0.6165 & 0.5824 & $<$ 0.0001 \\
Figure Sim & 0.6274 & 0.5719 & $<$ 0.0001 \\
LLM-Visual-Eval & 0.6476 & 0.6387 & $<$ 0.0001 \\
LLM-Behavior-Eval & 0.6671 & 0.6192 & $<$ 0.0001 \\
\bottomrule
\end{tabular}
\end{adjustbox}
\caption{Correlation between different metrics and human evaluation scores.}
\label{tab:metric_correlations}
\end{table}
\begin{table*}[t]
\centering
\begin{adjustbox}{max width=2.0\columnwidth}
\begin{tabular}{llccc}
\toprule
\textbf{Generator} & \textbf{Evaluator} & \textbf{LLM Sem.} & \textbf{LLM Vis.} & \textbf{LLM Behav.} \\
\midrule
Gemini 3 Pro & Gemini 3 Flash & 81.9 & 70.3 & 77.1 \\
Gemini 3 Pro & GPT-5.1      & 70.3 & 72.2 & 76.8 \\
\addlinespace
GPT-5.1      & Gemini 3 Flash & 61.4 & 62.0 & 50.9 \\
GPT-5.1      & GPT-5.1      & 53.8 & 62.1 & 52.1 \\
\bottomrule
\end{tabular}
\end{adjustbox}
\caption{Robustness analysis of LLM-as-Judge with different evaluators. All models are provided with DOM information during generation. The first row in each group represents the standard evaluation setting, while the second uses GPT-5.1 as a cross-evaluator.}
\label{tab:model_performance}
\end{table*}

\subsubsection{Correlation between Human Evaluation and Automated Evaluation}
\label{sec:human_correlation}

We show the correlations between human evaluation and each individual automatic evaluation metric in Table~\ref{tab:metric_correlations}. 
To obtain the final metric, we follow the approach of IWR-Bench, using an optimization algorithm to derive weights that maximize the correlation between the final metric and human evaluations. As shown in Sec~\ref{sec:final_eval_metric}, we find that the optimal weights are generally uniform. The resulting Pearson correlation is 0.7807, and the Spearman correlation is 0.7173.

In practice, in behavior-critical settings (interaction correctness/reliability) , users can up-weight task behavior metric/code-semantic metric while in visual-fidelity-critical settings up-weight visual similarity metrics (fig-sim/LLM-Visual-sim). We recommend \emph{\textbf{always reporting sub-metrics}} to reflect the performance in different metrics.

\subsubsection{Robustness of LLM-as-judge}
\label{sec:robustness_llm_as_judge}
To eliminate the potential bias caused by using a single LLM as a judge and the possible inflation of scores due to homogeneous models, we also use the GPT-5.1 model to evaluate the generation results of Gemini3-Pro as well as those of GPT-5.1 itself on all the samples of our benchmark. The experimental results are shown in Table~\ref{tab:model_performance}. It can be observed that the scores for LLM-Visual-Eval and LLM-Behavior-Eval are nearly identical, and the scoring trends for LLM-Semantic-Eval are also very similar for both models. This demonstrates the robustness of our LLM-As-Judge approach.

To mitigate the potential self-preference bias inherent in single-model evaluations, we performed a cross-validation analysis using GPT-5.1 as an external judge. We evaluated the outputs of both Gemini 3 Pro and GPT-5.1, where all responses were generated with the assistance of DOM information. As shown in Table~\ref{tab:model_performance}, the evaluation metrics exhibit high consistency regardless of the judge model used. Specifically, the scores for LLM-Vis. and LLM-Behav. show minimal variance between the two evaluators. Although LLM-Sem. scores show slight fluctuations, the relative performance ranking between models remains unchanged. These results strongly validate the robustness and reliability of our LLM-as-Judge evaluation framework.

\subsubsection{Evaluation Task Annotation Details}
\label{sec:eval_task_anno_detail}
\textbf{Annotation Principles.} To ensure the tasks are both comprehensive and efficient, annotators followed three core principles: 
(1) \textbf{Full Coverage}, ensuring every interactive component and its underlying callback functions are triggered at least once; 
(2) \textbf{Visual Significance}, which incorporates the \textit{equivalence partitioning} method from black-box testing. For continuous controls (e.g., Sliders), annotators selected representative values that yield distinct visual state transitions rather than minor shifts, effectively grouping functionally equivalent inputs; and 
(3) \textbf{Independence}, where each task is designed to be atomic, starting from the initial state to avoid error propagation between test cases. This process resulted in a total of 450 high-quality interaction tasks.

\textbf{Quality Assurance.} The quality of the instructions was verified through a two-stage process. 
First, we conducted a ``Golden-Run'' where an automated agent executed all tasks on all the ground-truth dashboards. A 100\% success rate was achieved, confirming the precision of the natural language instructions. 
Second, we randomly sampled 100 dashboards generated by different models and ran the same workflow. When failures or mismatched outcomes occurred, we confirmed they were attributable to issues in the generated dashboards rather than errors in the annotation task specifications or evaluation procedure. 
Third, we cross-verified the instructions against the internal logic of the Python code to ensure that no hidden states or conditional callbacks were overlooked.

\begin{figure*}[t]
    \centering
    \begin{subfigure}[b]{0.31\textwidth}
        \centering
        \includegraphics[width=\textwidth]{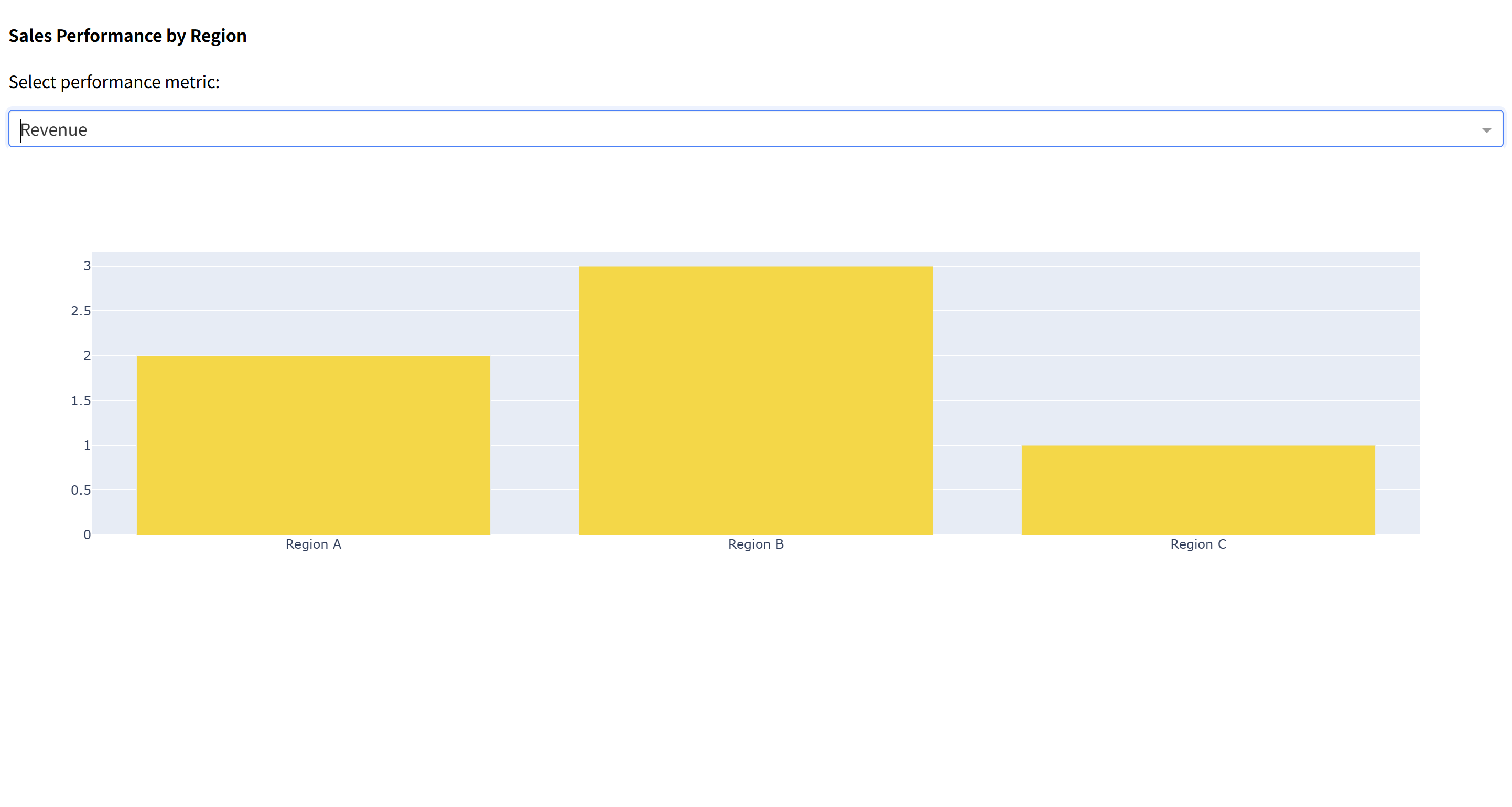}
        \caption{Initial state}
        \label{fig:task_case_2a}
    \end{subfigure}
    \hfill
    \begin{subfigure}[b]{0.31\textwidth}
        \centering
        \includegraphics[width=\textwidth]{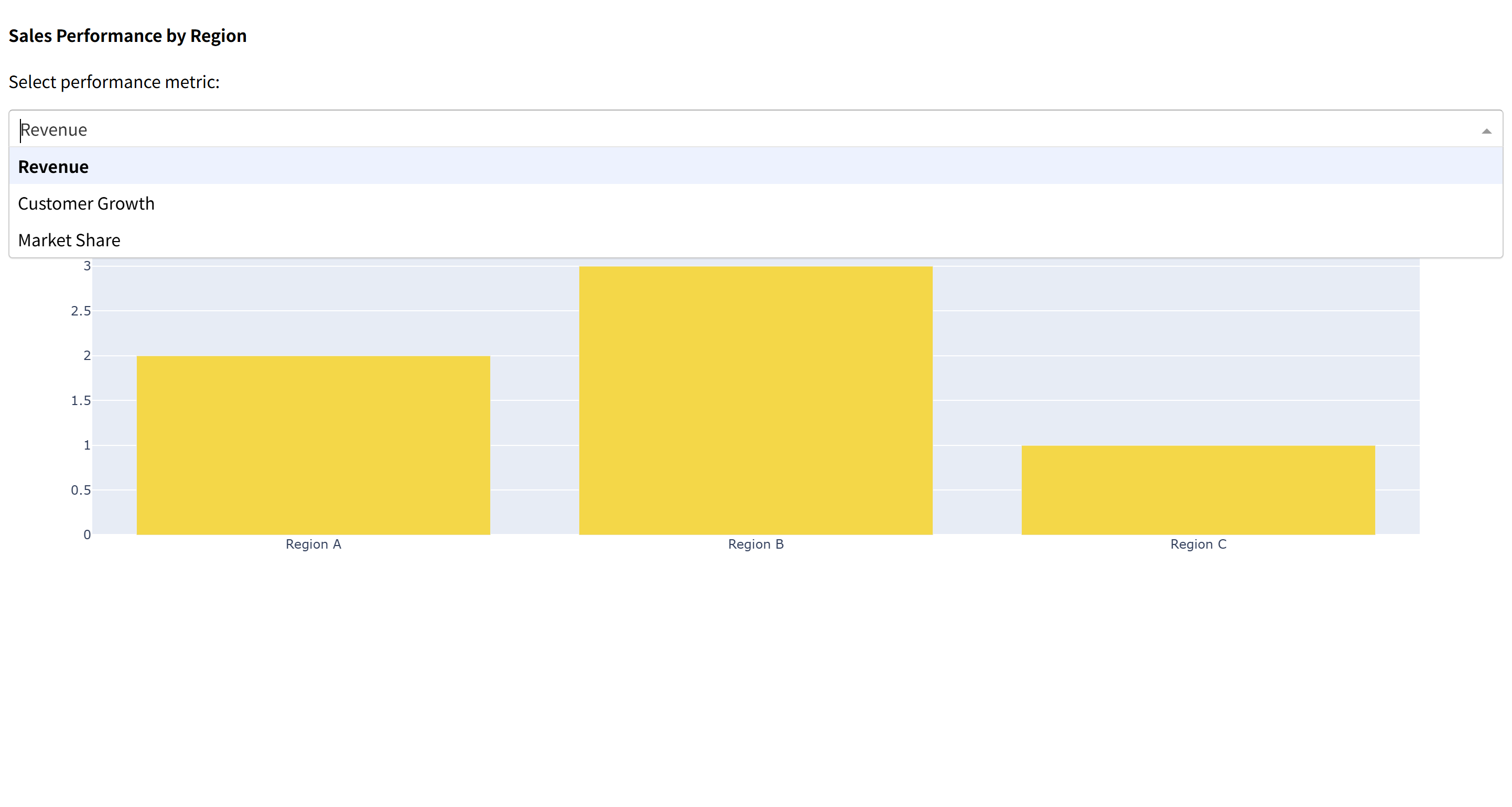}
        \caption{Open dropdown}
        \label{fig:task_case_2b}
    \end{subfigure}
    \hfill
    \begin{subfigure}[b]{0.31\textwidth}
        \centering
        \includegraphics[width=\textwidth]{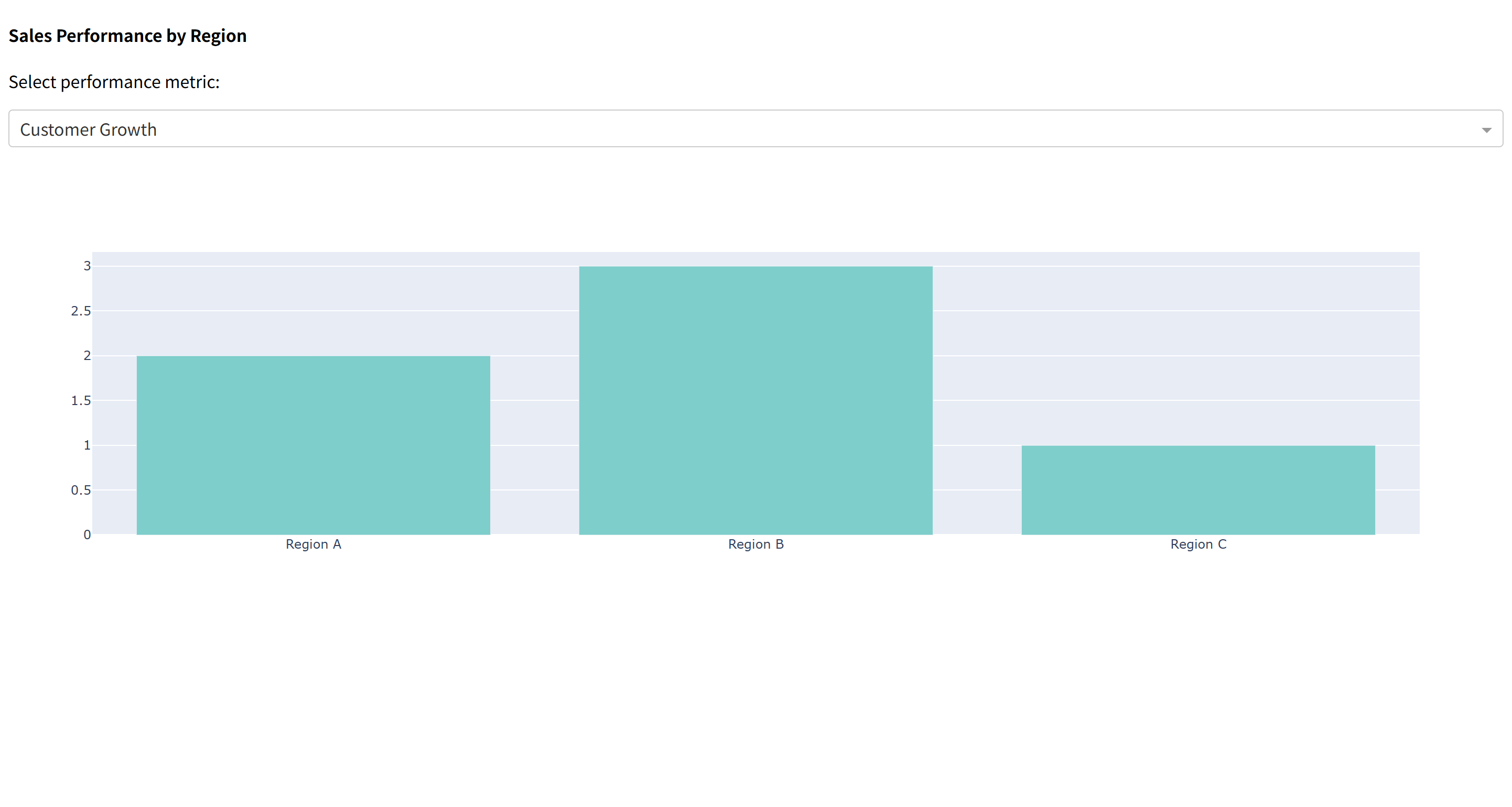}
        \caption{Select `Customer Growth`}
        \label{fig:task_case_2c}
    \end{subfigure}
    
    \caption{Example of a deterministic test trajectory for the Sales Performance dashboard. The task consists of: (1) opening the performance-metric-dropdown to display available options, and (2) selecting Customer Growth, resulting in updated bar chart colors and values.}
    \label{fig:task_case_1}
\end{figure*}

\begin{figure*}[t]
    \centering
    \begin{subfigure}[t]{0.31\textwidth}
        \centering
        \includegraphics[width=\textwidth]{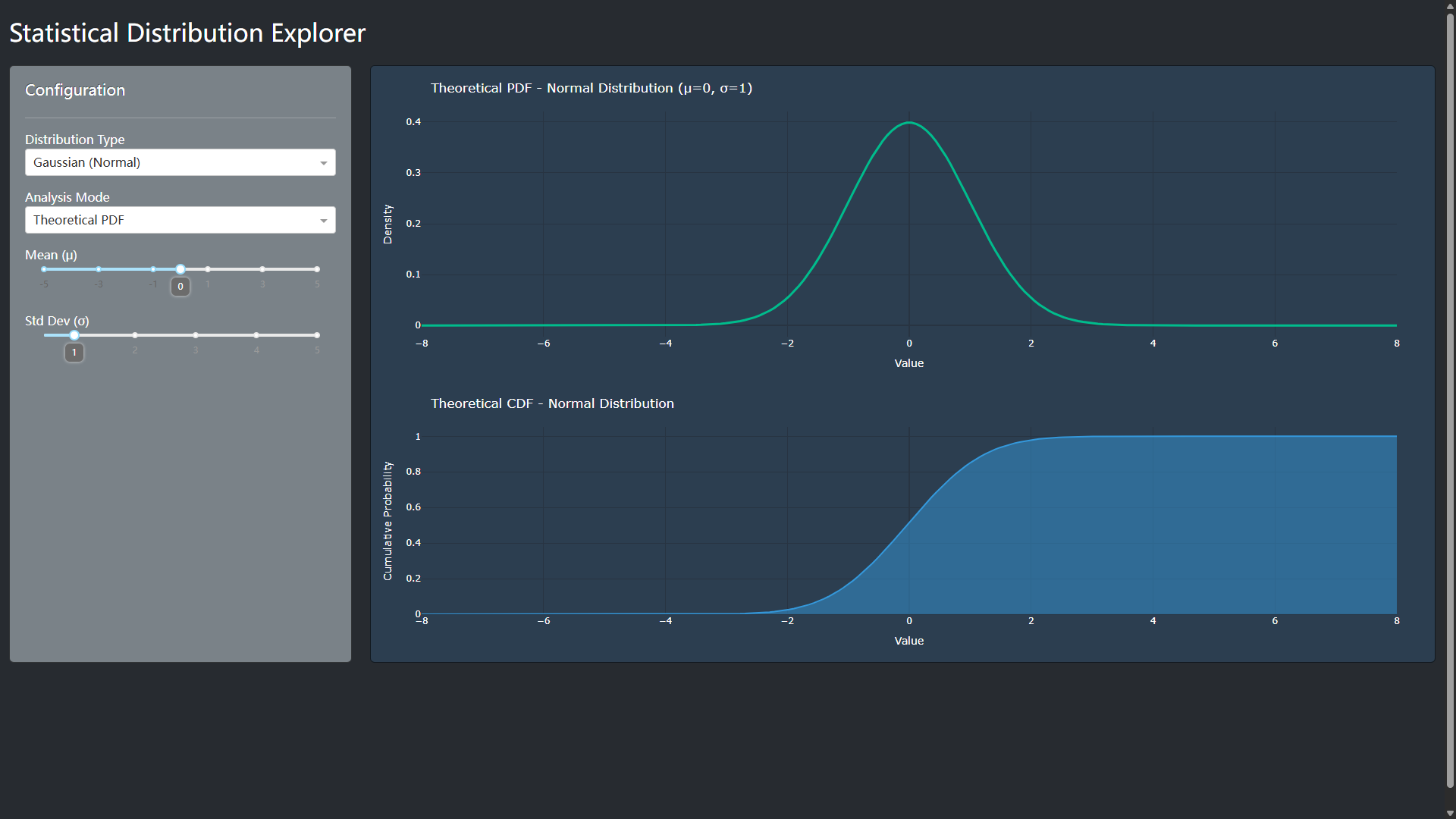}
        \caption{Initial state}
        \label{fig:task_case_2a}
    \end{subfigure}
    \hfill
    \begin{subfigure}[t]{0.31\textwidth}
        \centering
        \includegraphics[width=\textwidth]{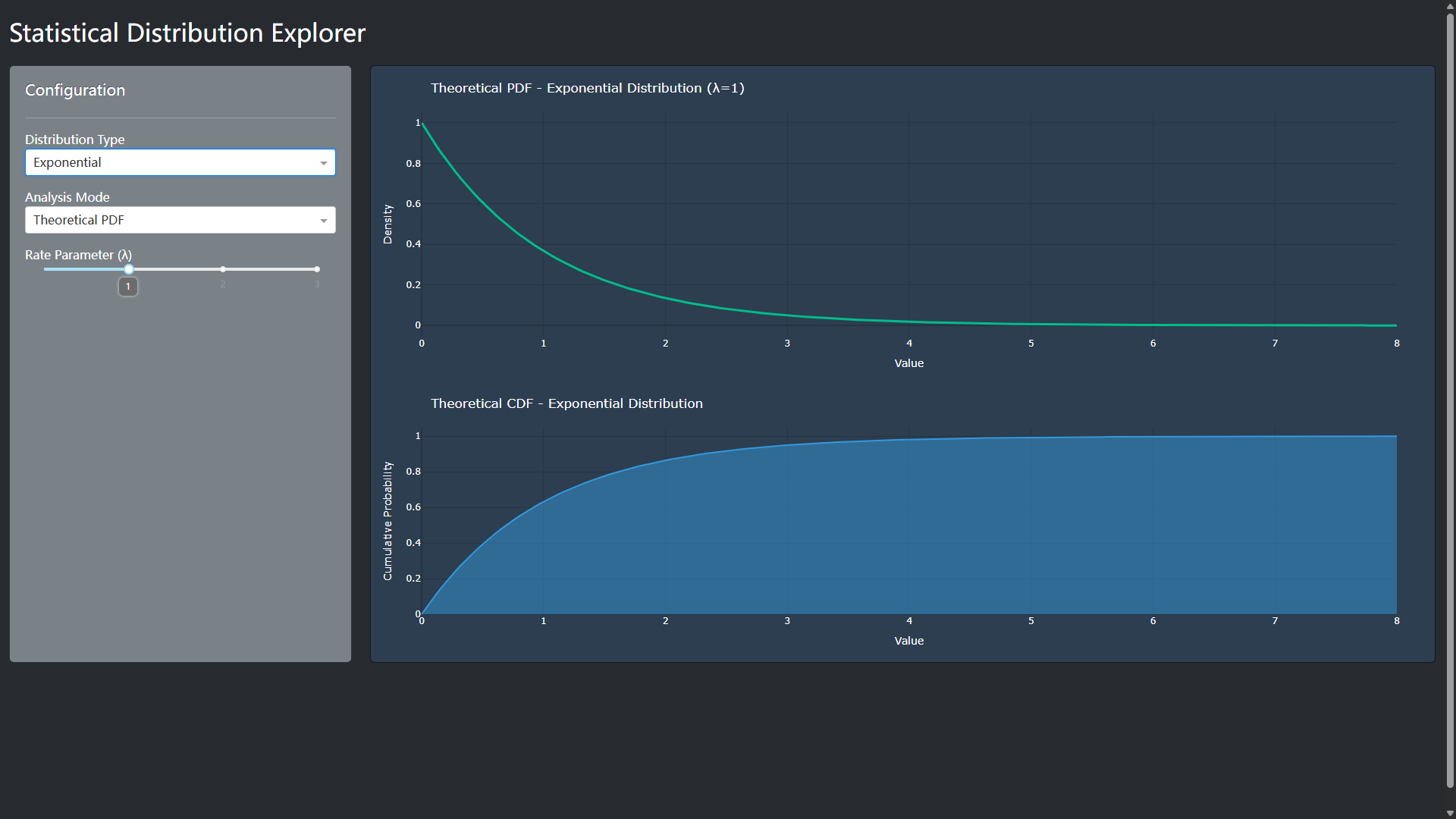}
        \caption{Select `Exponential` in `dist-dropdown`}
        \label{fig:task_case_2b}
    \end{subfigure}
    \hfill
    \begin{subfigure}[t]{0.31\textwidth}
        \centering
        \includegraphics[width=\textwidth]{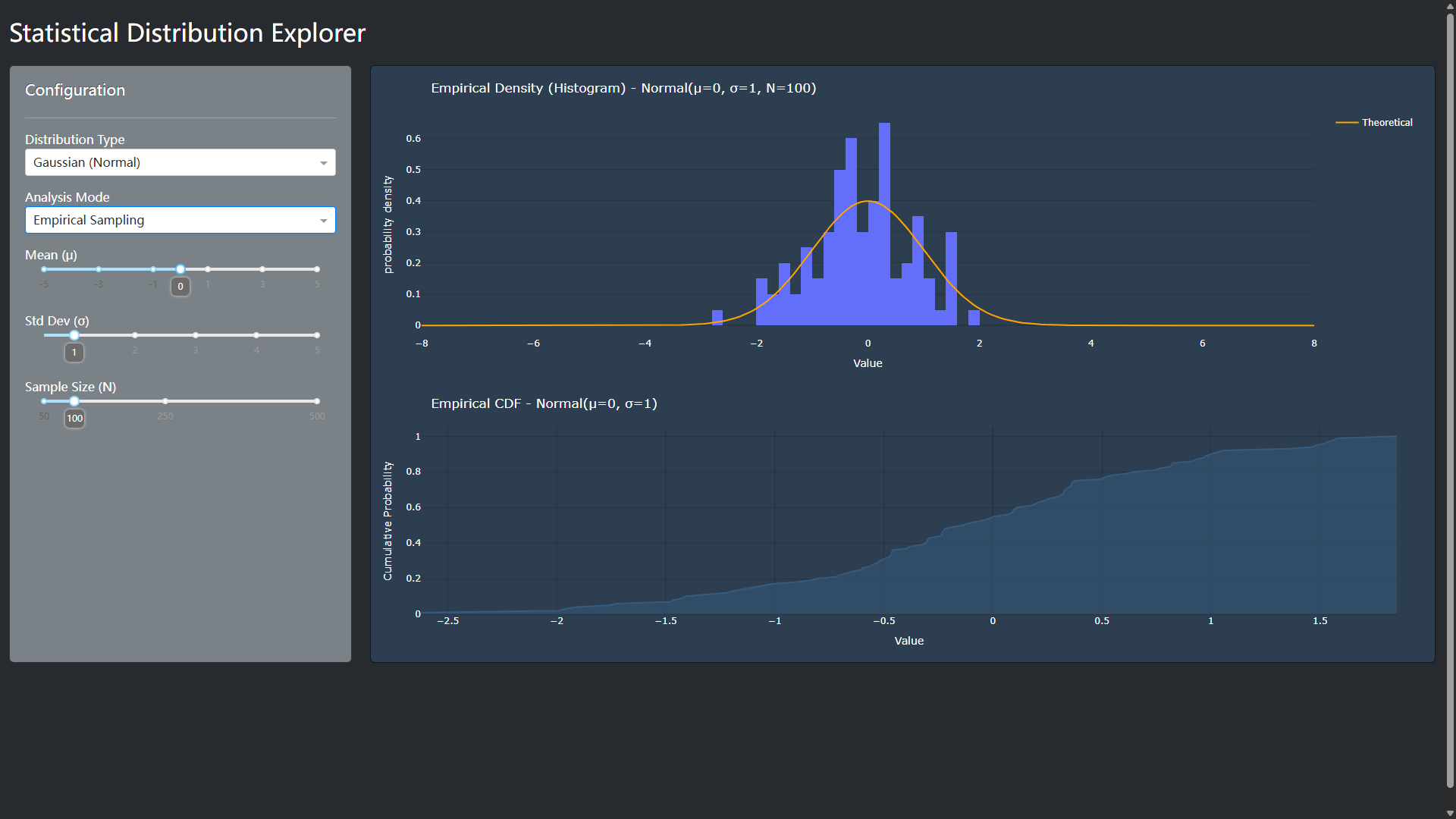}
        \caption{Select `Empirical Sampling` in `mode-dropdown`}
        \label{fig:task_case_2c}
    \end{subfigure}

    \vspace{8pt}

    \begin{subfigure}[t]{0.31\textwidth}
        \centering
        \includegraphics[width=\textwidth]{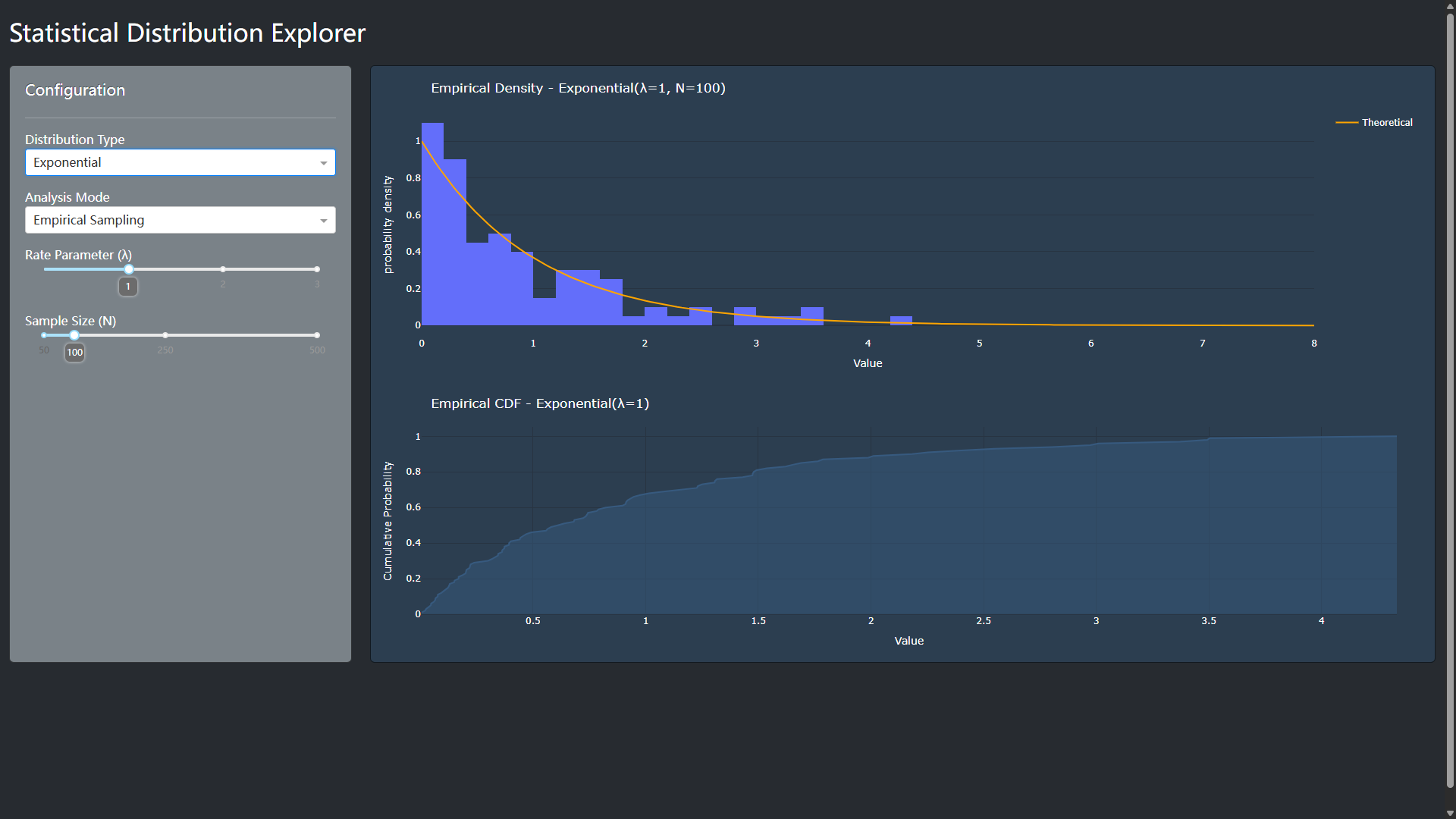}
        \caption{Select `Empirical Sampling` then `Exponential`}
        \label{fig:task_case_2d}
    \end{subfigure}
    \hfill
    \begin{subfigure}[t]{0.31\textwidth}
        \centering
        \includegraphics[width=\textwidth]{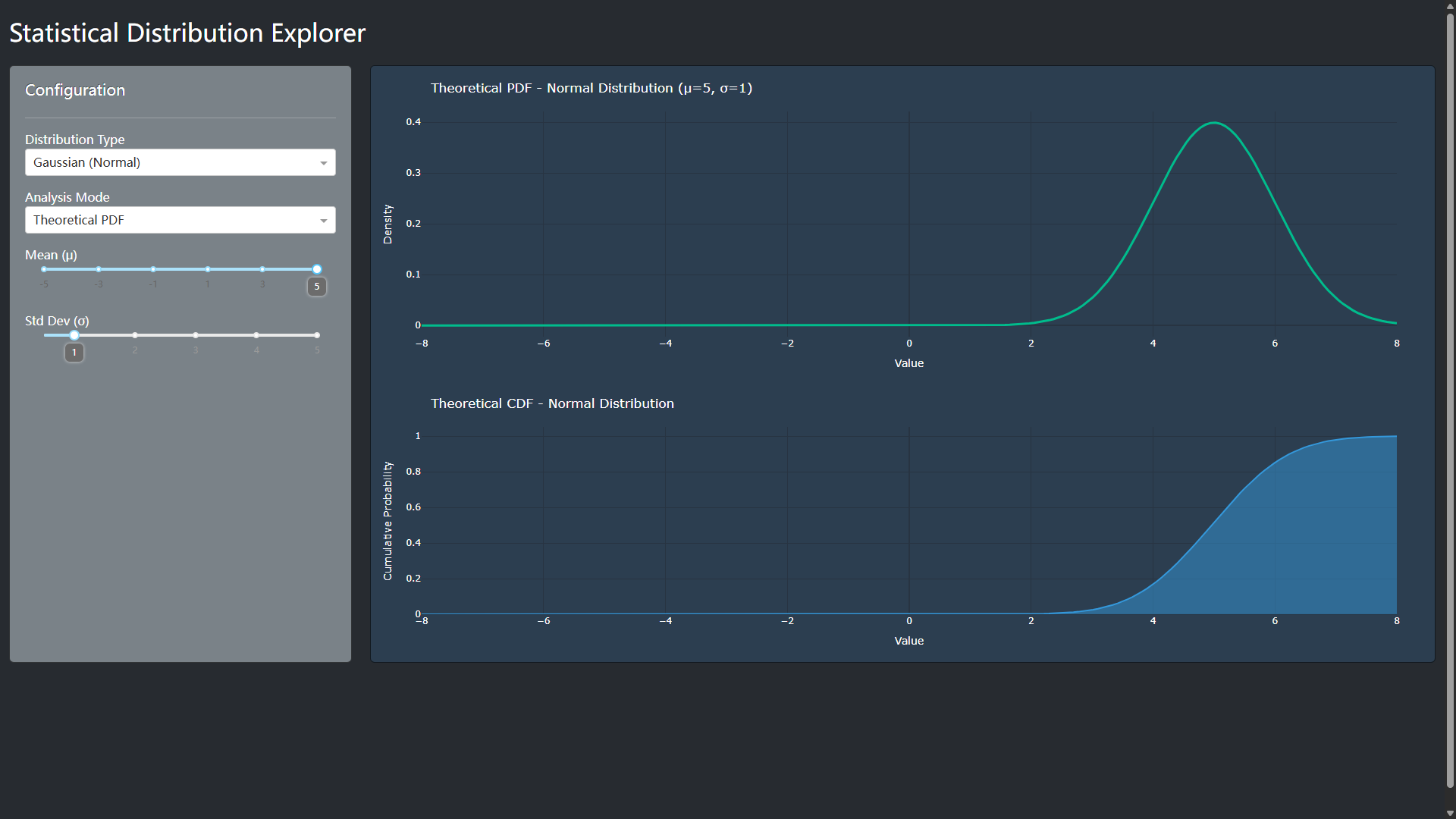}
        \caption{Drag `Mean ($\mu$)` to value 5}
        \label{fig:task_case_2e}
    \end{subfigure}
    \hfill
    \begin{subfigure}[t]{0.31\textwidth}
        \centering
        \includegraphics[width=\textwidth]{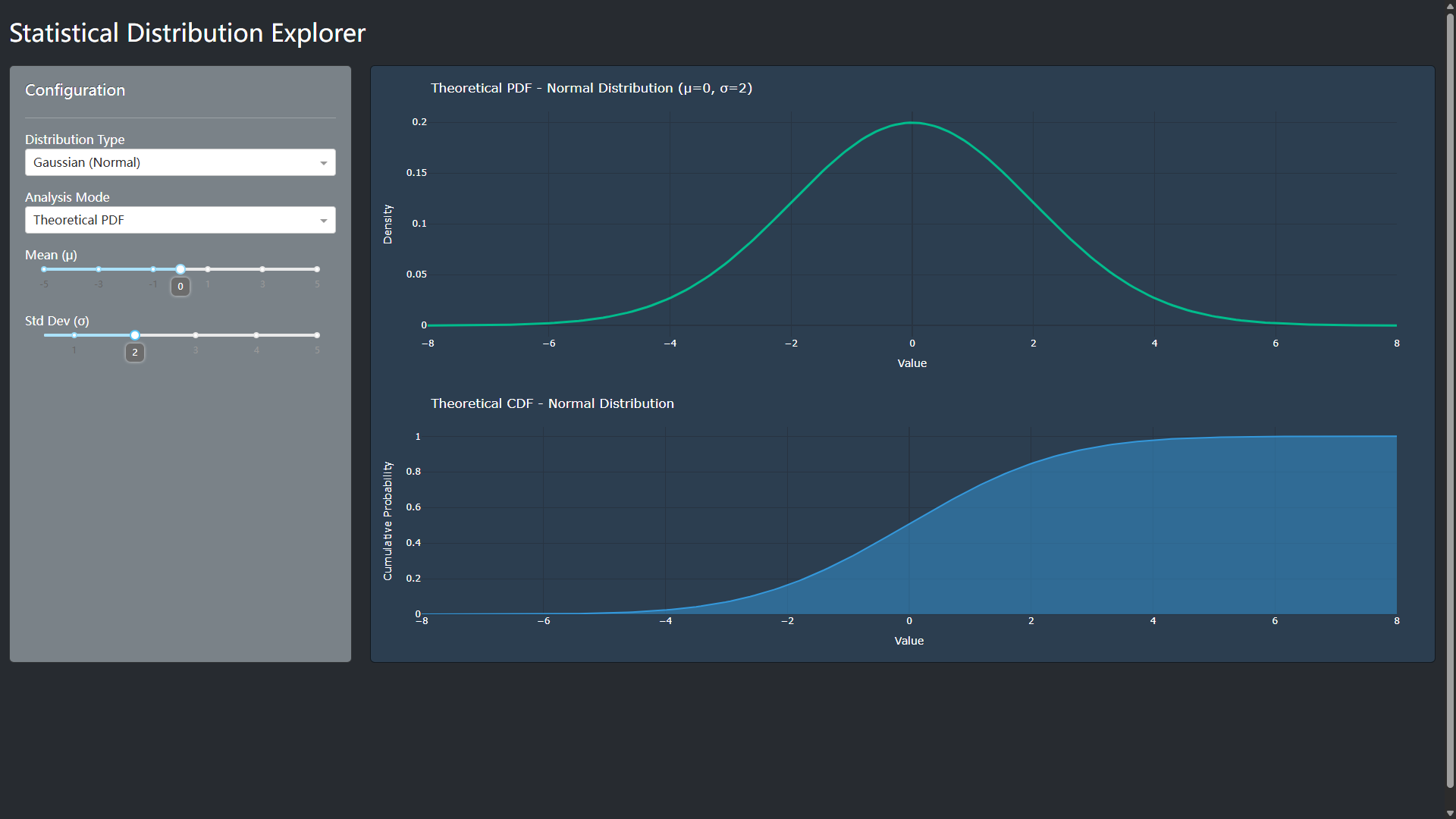}
        \caption{Drag `Std Dev ($\sigma$)` to value 2}
        \label{fig:task_case_2f}
    \end{subfigure}

    \caption{Experimental trajectory for the Statistical Distribution Explorer. The sequence demonstrates (a) initial default settings, (b-d) switching between theoretical and empirical analysis for Gaussian and Exponential distributions, and (e-f) dynamic adjustment of normal distribution parameters via sliders.}
    \label{fig:task_case_2}
\end{figure*}

We show the evaluation task annotation guidelines in Figure~\ref{fig:task_guidance} and show evaluation task examples in Figure~\ref{fig:task_case_1} and Figure~\ref{fig:task_case_2}.

\subsubsection{Figure Similarity Calculation Details}
\label{sec:fig_sim_detail}
\textbf{Style.} For each trace type, we collect the set of colors used by its traces. Following the color evaluation protocol in ChartMimic, we first match traces of the same type and then compute the L2 similarity between their color values.

\textbf{Data.} For each trace type, we sort data points, align dimensions via interpolation, and compute the L2 similarity between the aligned data arrays.

\textbf{Type and Text.} We extract trace types/names and textual objects directly from the JSON, and compute F1 scores for both.
We aggregate the four sub-scores into a single metric:

\begin{equation}
\begin{aligned}
\mathrm{Figure Similarity} \;=\;& 0.3\,\mathrm{Data} + 0.2\,\mathrm{Style} \\
&+ 0.2\,\mathrm{Type} + 0.3\,\mathrm{Text}.
\end{aligned}
\end{equation}

\subsubsection{Task Executor Implementation Details}
To facilitate dynamic interaction evaluation, we developed a robust Automated Task Executor that orchestrates headless browsers to perform natural language instructions using a multi-modal agent. This agent leverages a comprehensive observation space---comprising the accessibility tree, action history, and a distinctive dual-frame visual input (contrasting the previous and current screenshots) to verify action effects and automatically detect execution stalemates. We formalize all spatial interactions into a normalized $[0, 1000]$ coordinate system; empirically, we find that this strategy slightly enhances the spatial reasoning and task execution rates of the Gemini 3 Flash model. The framework operates in a high-concurrency environment, enforcing strict state isolation by resetting the dashboard between tasks to ensure reproducible evaluation.

\subsubsection{Related Prompts}
\label{sec:related_prompt_eval}
We show the prompt for LLM-Semantic-Eval in Figure~\ref{fig:semantic_eval_prompt}, prompt for LLM-Visual-Eval in Figure~\ref{fig:fidelity_eval_prompt}, prompt for LLM-Task-Behavior-Eval in Figure~\ref{fig:behavior_eval_prompt}, prompt for task executor evaluation agent in Figure~\ref{fig:executor_prompt}.

\subsection{Experimental Details}
\label{sec:exp_details}

\subsubsection{Interaction Environment Details}
\label{sec:interaction_environment_detail}
To evaluate the agent's capability to explore and interact with dynamic dashboards, we engineered a lightweight, high-fidelity interaction environment. This environment is built upon Selenium WebDriver integrated with Google Chrome, ensuring that the agent perceives the dashboard exactly as a human user would.

\paragraph{Viewport \& Coordinate System.}
We enforce a fixed viewport resolution of $1920 \times 1080$ pixels. This standardization ensures consistent visual feedback across different models and runs, eliminating artifacts caused by responsive layout shifts. The coordinate system is mapped 1:1 to pixels, where $(0, 0)$ represents the top-left corner.

\paragraph{Observation Space.}
At every time step $t$, the environment provides two modalities to the agent. The inclusion of structured text data is configurable based on the experimental setting:
\begin{itemize}
    \item \textbf{Visual Feedback:} A high-resolution screenshot ($S_t$) capturing the current state of the dashboard, including tooltips, hover states, and updated charts.
    \item \textbf{Structured Feedback (Optional):} A simplified, role-based accessibility list derived from our DOM extraction strategy (described later), which provides precise bounding box coordinates ($[x, y, w, h]$) for interactive elements.
\end{itemize}

\begin{figure*}[t]
    \centering
    \begin{tcolorbox}[
        colback=white,
        colframe=gray!50!black,
        title=\textbf{Comparison: Visual View vs. Pruned DOM},
        fonttitle=\bfseries
    ]
    \small
    \textbf{1. Visual View (Screenshot Crop)} \\
    \begin{center}
       \includegraphics[width=0.95\linewidth]{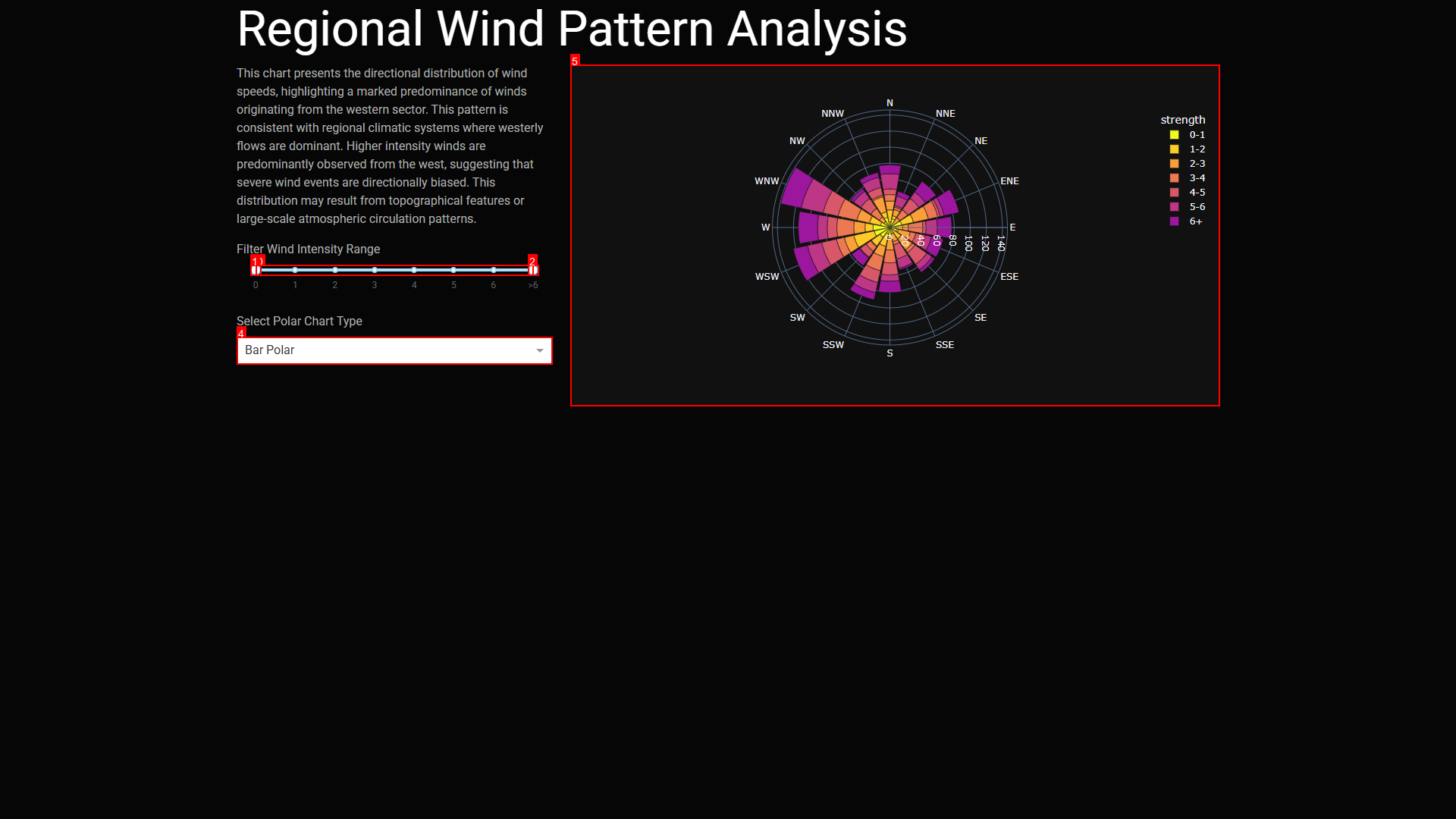} 
    \end{center}
    
    \vspace{5pt}
    \hrule
    \vspace{5pt}

    \textbf{2. Pruned DOM Representation (Input to Agent)}
    \begin{verbatim}
[
  {
    "box": [337, 349, 366, 14],
    "id": 0,
    "role": "slider_container"
  },
  {
    "box": [330, 349, 14, 14],
    "id": 1,
    "role": "slider_handle"
  },
  {
    "box": [696, 349, 14, 14],
    "id": 2,
    "role": "slider_handle"
  },
  {
    "box": [312, 444, 416, 36],
    "id": 3,
    "role": "dropdown_container"
  },
  {
    "box": [312, 444, 416, 36],
    "id": 4,
    "role": "dropdown_input"
  },
  {
    "box": [752, 85, 856, 450],
    "id": 5,
    "role": "chart_container"
  }
]

    \end{verbatim}
    \end{tcolorbox}
    \caption{Visualizing the DOM Pruning Strategy. While the visual view (Top) shows complex data points, the Pruned DOM (Bottom) abstracts the chart into a single container while preserving precise coordinates for the interactive dropdown and button.}
    \label{fig:dom_pruning_viz}
\end{figure*}

\paragraph{DOM Extraction \& Abstraction Strategy}
We show the example DOM tree in Figure~\ref{fig:dom_pruning_viz}. A raw HTML DOM tree for a data-intensive dashboard is often prohibitively large and noisy. Plotly charts typically render thousands of SVG nodes (e.g., one \texttt{<path>} per data point), which can easily overflow the context window of Multi-modal LLMs (MLLMs).
Instead of using the raw DOM, we implemented a custom JavaScript-based Heuristic Extraction Strategy that filters and flattens the DOM into a concise list of actionable elements. We process the DOM via the following rules:

\begin{itemize}
\item{Noise Reduction.} We aggressively prune internal structures of complex widgets. For data tables (e.g., Dash DataTable, AgGrid), we ignore individual cells and headers, retaining only the macro container. Similarly, internal SVG components of charts (e.g., \texttt{modebar}, \texttt{rc-slider-mark}) and purely decorative wrappers are discarded.

\item{Visibility \& Interaction Filtering.} We strictly filter out invisible elements based on CSS properties (\texttt{display}, \texttt{visibility}) and dimensions ($width/height \le 0$). Crucially, our extractor handles special cases for data visualization libraries (e.g., Plotly), where interactive layers often have \texttt{opacity: 0} but \texttt{pointer-events: auto}.

\item{Role Identification.}
Instead of generic HTML tags, we map elements to semantic roles relevant to dashboard interaction. Based on class name heuristics, elements are categorized into roles such as \texttt{chart\_container}, \texttt{data\_table}, \texttt{slider\_container}, \texttt{dropdown\_option}, or \texttt{axis\_filter\_track}.

\item{Output Format.}
The final output is not a tree, but a flat JSON list containing the element's id, role, tag, and its bounding box $[x, y, w, h]$. This format reduces token usage by approximately 95\% while preserving essential spatial grounding information.

\end{itemize}





\begin{figure*}[t]
  \centering
  \begin{subfigure}{\textwidth}
    \centering
    \includegraphics[width=0.95\linewidth]{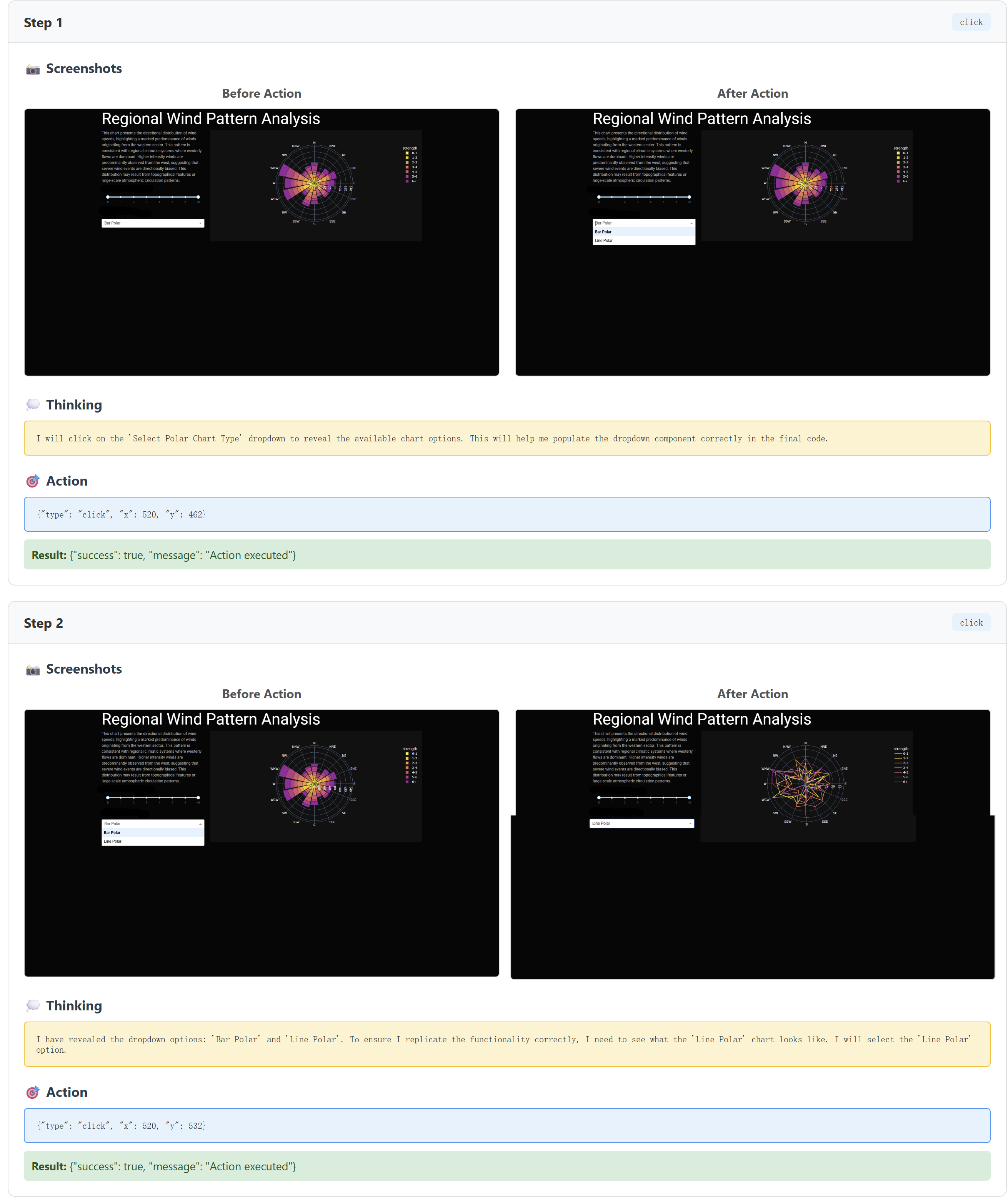}
    \caption{The agent begins by identifying interactive UI components within the "Regional Wind Pattern Analysis" dashboard. In Step 1, it performs a targeted click on the 'Select Polar Chart Type' dropdown to reveal the available data visualization modes, identifying two options: 'Bar Polar' and 'Line Polar'. In Step 2, the agent selects the 'Line Polar' option to trigger a state change, observing how the visual encoding shifts from radial bars to interconnected data points. This sequence allows the model to map the relationship between the dropdown controller and the rendering logic of the polar coordinate system.}
  \end{subfigure}
  \caption{Generation cases (Part I)} 
  \label{fig:gen_case_part1}
\end{figure*}

\begin{figure*}[t]
  \ContinuedFloat 
  \centering
  \begin{subfigure}{\textwidth}
    \centering
    \includegraphics[width=0.95\linewidth]{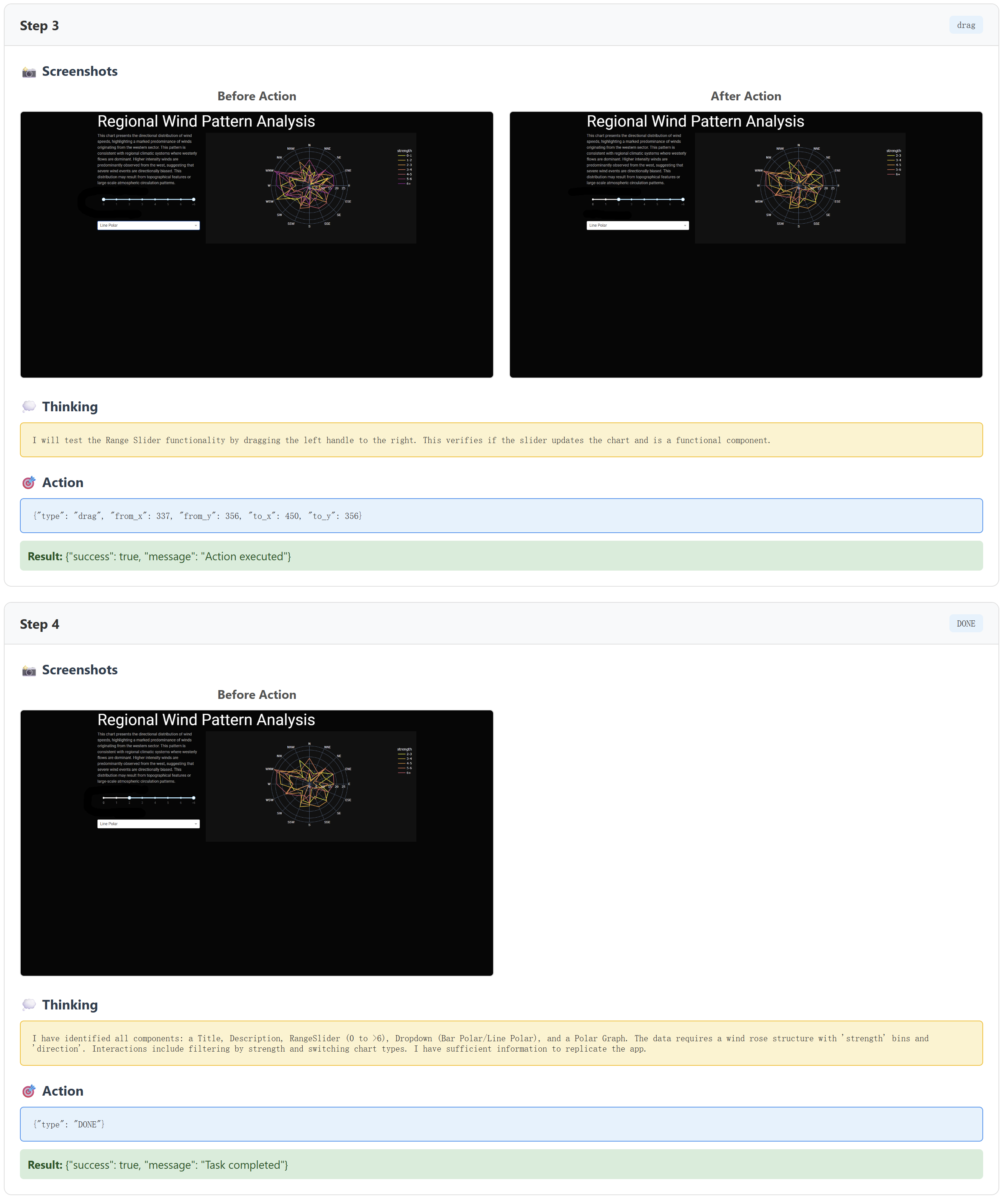}
    \caption{The agent shifts focus to data filtering mechanisms and final requirement gathering. In Step 3, it interacts with the 'Filter Wind Intensity Range' slider by dragging the left handle, observing real-time updates to the chart to confirm that the range slider dynamically filters the wind strength datasets. Finally, in Step 4, the agent synthesizes its findings, identifying the complete application stack: a title/description header, a range-based slider (0 to >6), a dual-mode dropdown, and a wind rose polar graph. Having mapped the data structure (strength bins and directional vectors) and the interaction logic, the agent concludes the exploration phase, ready for code replication.}
  \end{subfigure}
  \caption{Generation cases (Part II)}
  \label{fig:gen_case_part2}
\end{figure*}

\subsubsection{Evaluation Setup Details}
\label{sec:eval_setup_detail}
We use an MM-React-style framework for evaluation. At each step, the multimodal model receives the current dashboard screenshot, optional DOM tree, historical screenshots, and actions as input. The model outputs its reasoning and selects an action from its action space. For all evaluations, we use default inference parameters for the models. We further use a resolution compression strategy to reduce overhead. 

To reduce overhead, we use a resolution of \(1920 \times 1080\) for the current screenshot, and low-resolution (compressed by a factor of 4) screenshots for historical steps without retaining prior reasoning. We found that this compression does not affect the model's final performance, as shown in Table~\ref{tab:ablation_results}. 

Possible actions include: \textbf{Success:} Outputs \texttt{Done} and generates the final code.
\textbf{Failure:} Outputs \texttt{Fail} and abandons the task if the action is unparseable.
\textbf{Interaction limit:} The agent must stop when reaching the pre-defined interaction limit (25 in our evaluation setup). This scenario was not observed in our experiments.
\textbf{Interaction:} Performs an action such as clicking or dragging.

If an interaction is performed, the dashboard returns a new screenshot and optional DOM tree, which are appended to the model's input for the next step.

\paragraph{Agent Payload Details}
To facilitate long-horizon reasoning within context limits, we employ a configurable structured prompt payload. The payload consists of three distinct blocks:

\begin{enumerate}
    \item \textbf{System Instruction:} A static definition of the agent's role, action space (Click, Scroll, Drag, etc.), and strict JSON output format.
    
    \item \textbf{Conversation History (Context Compression):} A chronological log of previous turns. To optimize memory usage and enhance reasoning reliability, we implement an optional Context Compression mechanism. When enabled, screenshots in the user's history messages are resized to a lower resolution ($480 \times 270$) before being stored.
    
    \textit{Empirical Benefit:} Our experiments indicate that this strategy serves a dual purpose: it reduces token consumption and, more importantly, channels the model's attention toward the current observation. By reducing the visual saliency of past states, the agent is less prone to hallucinations derived from obsolete visual details, thereby improving interaction accuracy.
    
    \item \textbf{Current Observation:} The active input for the current step $t$. This includes:
    \begin{itemize}
        \item The High-Resolution screenshot ($1920 \times 1080$) to ensure fine-grained visual details are legible for the immediate task.
        \item (Optional) The extracted DOM list is injected as text into the user prompt, explicitly informing the agent of the exact coordinates of valid interactive zones.
    \end{itemize}
\end{enumerate}
\paragraph{Prompts and Examples} We show the prompt for evaluation in Figure~\ref{fig:eval_prompt}, the example for evaluation generation process in Figure~\ref{fig:gen_case_part1} (Part I) and Figure~\ref{fig:gen_case_part2} (Part II).

\subsection{More Experiments}
\label{sec:more_exp}
\begin{table*}[t]
\normalsize
\centering
\begin{adjustbox}{max width=2.0\columnwidth}
\begin{tabular}{lccccc}
\toprule
\textbf{Model} & \textbf{Code Exec.} & \textbf{Task Exec.} & \textbf{Comp. Cov.} & \textbf{LLM Sem.} & \textbf{LLM Behav.} \\
\midrule
Gemini 3 Pro   & 97.78 & 94.17 & 91.60 & 81.89 & 77.10 \\
Gemini 3 Pro (w/o Thought) & 90.00 & 85.30 & 85.22 & 72.22 & 65.80 \\
Gemini 3 Pro (w/o Compression) & 93.89 & 90.77 & 88.12 & 76.78 & 69.90 \\
Gemini 3 Pro (w/o Interaction)   & 95.56 & 82.82 & 87.43 & 73.67 & 62.00 \\

\bottomrule
\end{tabular}
\end{adjustbox}
\caption{Ablation experiment results on Gemini 3 Pro with DOM under different conditions.}
\label{tab:ablation_results}
\end{table*}

\subsubsection{Ablation Study}
\label{sec:ablation_study}
In this section we conduct ablation studies of our method. Results are shown in Table~\ref{tab:ablation_results}.
\paragraph{Effect of Native Image Resolution (w/o Compression).}  
We modified the evaluation setup to use the native resolution for historical images, instead of applying the low-resolution compression. We observed a slight decrease in performance across various metrics when using Native image resolution. This suggests that our compression strategy is appropriate and does not significantly impact the model’s performance on this task.

\paragraph{Effect of Action Thinking Mode (w/o Thought).}  
We removed the "thinking" step before the model outputs an action at each step of the evaluation, which led to a noticeable decline in performance. This demonstrates the importance of thoughtful reasoning at each step of exploration for better task performance.

\paragraph{Effect of Using Only Initial Screenshot for Reconstruction (w/o Interaction).}  
To confirm that our benchmark indeed requires model interaction to fully reconstruct a dashboard, we tested the Gemini model’s ability to generate code with only the initial screenshot of the dashboard. We found a significant drop in performance across various metrics, indicating that interaction is necessary for accurate dashboard reconstruction. However, this performance decline is not catastrophic. We discovered that in some cases, the text information in certain dataset samples provided enough clues about the dashboard’s interaction logic. We did not remove or modify these samples, as we believe that inferring interaction logic based on text is also a model capability. The impact of text on model performance is further discussed in Sec~\ref{sec:text_robustness}.

\subsection{Text Anonymization Experiment Details}
\label{sec:text_anoy}
We show the prompt for anonymize the text in the dash app in Figure~\ref{fig:Anonymize_prompt}. Note that the LLM is only allowed to modify the textual element in app.layout and keep all the others the same. After the process, we manually recheck each modified code to ensure that only the text in app.layout has changed. 

We show an anonymized example in Figure~\ref{fig:text_anomize_figure}. 
We observed that Gemini3-Pro exhibits identical exploration paths on both anonymized and non-anonymized dashboards (as shown in (b)). However, as depicted in (c), the model is only able to accurately reconstruct the interaction logic on the non-anonymized dashboard. This suggests that the model may rely heavily on textual cues, rather than truly understanding and replicating the underlying interaction logic.

\begin{figure*}[t]
  \centering
  \includegraphics[width=2.0\columnwidth]{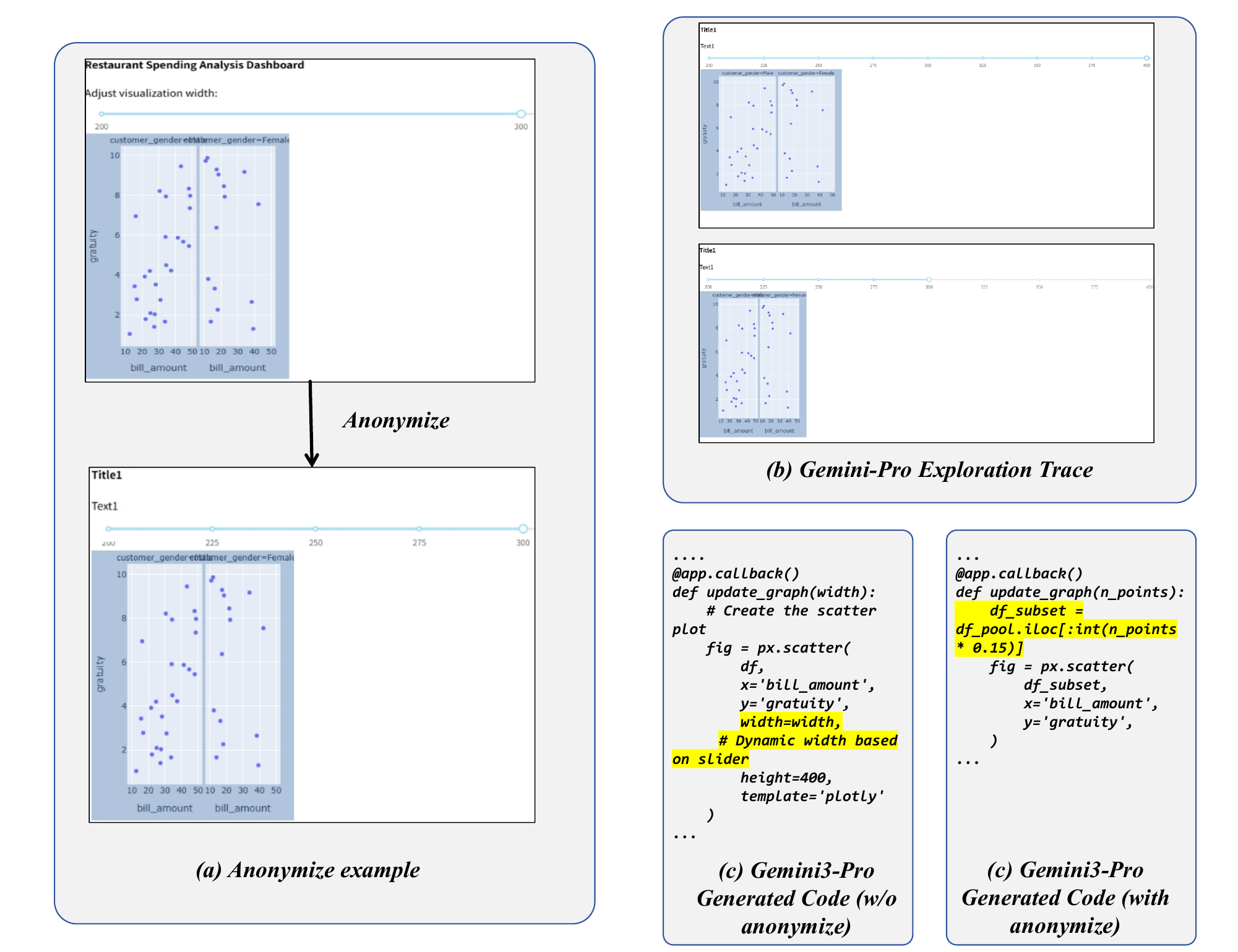}
  \caption {Text anonymize example.}
  \label {fig:text_anomize_figure}
\end{figure*}

\begin{figure*}[t]
    \centering
    \begin{promptbox}[title=System Instruction for Interactive Dashboard Reconstruction Agent]
    \small
    \textbf{\# Role} \\
    You are an expert Python developer reverse-engineering Plotly Dash apps. \\
    Your goal: Explore a live dashboard screenshot (1920x1080) and generate the exact Python code to recreate it.

    \vspace{5pt}
    \textbf{\# Environment \& Rules} \\
    - \textbf{Viewport:} 1920x1080 (0,0 is Top-Left). \\
    - \textbf{Coordinate System:} 
      - x-axis: Horizontal position (0 at left, 1919 at right). 
      - y-axis: Vertical position (0 at top, 1079 at bottom). \\
    - \textbf{Max Steps:} 25. \\
    - \textbf{Single Action:} Output EXACTLY ONE action per turn. \\
    - \textbf{Process:} Explore components to reveal data/states before generating code. \\
    - \textbf{Output:} A single JSON object. No conversational text.

    \vspace{5pt}
    \textbf{\# Action Space} \\
    You must output \textbf{EXACTLY ONE} action per turn from the list below: \\
    1. \texttt{click}: Trigger a primary mouse click at a specific coordinate. Params: \texttt{\{"x": int, "y": int\}} \\
    2. \texttt{double\_click}: Trigger a double-click event. Params: \texttt{\{"x": int, "y": int\}} \\
    3. \texttt{move\_mouse\_to}: Move mouse to a specific coordinate. Params: \texttt{\{"x": int, "y": int\}} \\
    4. \texttt{scroll}: Perform a vertical scroll action. If x/y are provided, moves mouse there first. Params: \texttt{\{"amount": int, "x": int (optional), "y": int (optional)\}} (-up, +down) \\
    5. \texttt{drag}: Perform a click-and-drag operation from start to end coordinates. Params: \texttt{\{"from\_x": int, "from\_y": int, "to\_x": int, "to\_y": int\}} \\
    6. \texttt{replace\_text}: Click at a coordinate, select all existing text (Ctrl+A), and type new text. Params: \texttt{\{"x": int, "y": int, "text": str\}} \\
    7. \texttt{mark}: visually mark a coordinate for reference in the next step. Params: \texttt{\{"x": int, "y": int\}} \\
    8. \texttt{DONE}: Exploration finished. Params: None. (Requires \texttt{code} field) \\
    9. \texttt{FAIL}: Cannot proceed. Params: None.

    \vspace{5pt}
    \textbf{\# Response Format (Strict JSON)} \\
    Respond ONLY with a single JSON object. Do not include any explanations or conversational text.

    \vspace{5pt}
    \textbf{State 1: Exploration} \\
    \texttt{\{} \\
    \quad \texttt{"thought": "Brief reasoning. What element am I testing?",} \\
    \quad \texttt{"action": \{ "type": "click", "x": 500, "y": 300 \},} \\
    \quad \texttt{"code": null} \\
        \texttt{\}} \\
    \textbf{State 2: Generation (When action is DONE)} \\
    \texttt{\{} \\
    \quad \texttt{"thought": "I have identified all components. Generating code.",} \\
    \quad   \texttt{"action": \{ "type": "DONE" \},} \\
    \quad   \texttt{"code": "import plotly...from dash import...Full runnable code here"} \\
    \texttt{\}} \\
    \vspace{5pt}
    \textbf{\# Technical Stack \& Constraints (STRICT)} \\
    Replicate the dashboard exactly using ONLY these libraries: \\
    1. \textbf{Layout \& UI:} \texttt{dash\_bootstrap\_components} as \texttt{dbc}, \texttt{dash\_daq} as \texttt{daq}, \texttt{dash\_ag\_grid} as \texttt{dag}. \\
       - \textit{Forbidden:} Do NOT use \texttt{dash\_mantine\_components}. \\
       - \textbf{Constraint:} ALL figures MUST be wrapped in \texttt{dcc.Graph} components. \\
    2. \textbf{Core Dash:} \texttt{dash} (html, dcc, Input, Output, State, ctx, MATCH, ALL). \\
    3. \textbf{Plotly:} \texttt{plotly.graph\_objects} as \texttt{go}, \texttt{plotly.express} as \texttt{px}. \\
    4. \textbf{Others:} \texttt{pandas}, \texttt{numpy}, \texttt{json}, \texttt{math}, \texttt{datetime}, \texttt{random}. \\
    5. \textbf{Data Strategy:} \\
     - You do NOT have access to the backend data. \\
     - You MUST generate \textbf{synthetic data} using \texttt{pandas}/\texttt{numpy} that visually matches the charts. \\
    6. \textbf{Initial State Fidelity:} Ensure input components and Graph views are initialized with the EXACT values shown in the original screenshot.

    \vspace{5pt}
    \textbf{\#\#\# GOAL \& STRATEGY} \\
    1. \textbf{Explore Thoroughly:} You should interact with components (e.g. Open the dropdown) to reveal ANY \textbf{hidden states} before coding. \\
    2. \textbf{High-Fidelity Replication:} Your generated code must strictly match the screenshot. The generated code MUST implement the interactive logic (Callbacks) and Data Trend Fidelity.
    \end{promptbox}
    \caption{System Instruction for Interactive Dashboard Reconstruction Agent.}
    \label{fig:eval_prompt}
\end{figure*}

\begin{figure*}[h]
    \centering
    \begin{promptbox}[title=Task Annotation Guidance]
    \small
    \textbf{\# Interaction Selection Principles} \\
    1. \textbf{Selection Components:} For \texttt{Dropdowns}, \texttt{RadioItems}, or \texttt{Checkboxes}, select significant non-default values that trigger visible changes. \\
    2. \textbf{Continuous Controls:} For \texttt{Sliders} or \texttt{Numeric Inputs}, do not create multiple tasks for different values. Set a single value that covers the majority of the data distribution. \\
    3. \textbf{Prioritize "Trigger" Components:} Focus on interactions with \texttt{Buttons}, \texttt{Tabs}, and master control panels over minor aesthetic filters.

    \vspace{8pt}
    \textbf{\# Task Annotation Strategy} \\
    4. \textbf{Minimalism:} Annotate the minimum number of tasks (typically 1 to 5) required to verify all unique interactive logic paths. Avoid redundant testing of similar functional units. \\
    5. \textbf{Imperative Phrasing:} Use the imperative mood for instructions (e.g., "Click...", "Select..."). Do not include explanations of purpose or expected outcomes. \\
    6. \textbf{Independence:} Ensure all tasks are atomized and independent. There must be no sequential or temporal dependencies between tasks. \\
    7. \textbf{Initial State Assumption:} Assume each task begins from the application's default initial state. Do not generate tasks that reset a component to its default value.

    \vspace{8pt}
    \textbf{\# Constraint Checklist for Reproducibility} \\
    8. \textbf{Meaningful Impact:} Verify if every task is essential for a specific callback or state change. \\
    9. \textbf{Value Diversity:} Remove redundant tasks with different data values but identical logic. \\
    10. \textbf{Explicitness:} Ensure descriptions are reproducible (e.g., Use "Select the second row" instead of "Pick a row"; use "Set range to [0.5, 1.0]" instead of "Set a narrow range"). \\
    11. \textbf{Efficiency:} Ensure the task can be completed within a maximum of 5 interaction steps.
    \end{promptbox}
    \caption{Task Annotation Guidance for the Interactive Evaluation Framework.}
    \label{fig:task_guidance}
\end{figure*}

\begin{figure*}[h]
    \centering
    \begin{promptbox}[title=Visual Fidelity Evaluation Prompt]
    \small
    You are an expert UI judge evaluating the fidelity of a Generated Dashboard Screenshot (GEN) against a Ground Truth Dashboard screenshot (GT). \\
    Your goal is to assess how well the generated screenshot specifically recreated the visual appearance of the ground truth screenshot.

    \vspace{5pt}
    Image 1: Ground Truth (GT) \\
    Image 2: Generated (GEN)

    \vspace{5pt}
    Please evaluate the GEN image based on the following specific criteria. \\
    \textbf{Ignore minor differences in window size or exact axis tick values, as rendering engines may vary.}

    \vspace{5pt}
    \textbf{\#\#\# Scoring Criteria:}
    \begin{enumerate}[leftmargin=15pt, nosep]
        \item \textbf{Layout \& Components (0-20)}: Are all figures and interactive components present? Does the arrangement match the GT image?
        \item \textbf{Chart Types (0-20)}: Are all chart types correct (e.g., Bar vs Line vs Scatter)?
        \item \textbf{Text Content (0-10)}: Do the Main Titles, Axis Titles, Legend Texts and Annotations match?
        \item \textbf{Data \& Grouping (0-20)}: Do the data trends look identical? Is the number of bars/lines/groups the same?
        \item \textbf{Style \& Aesthetics (0-20)}: Does the GEN match the GT in terms of colors, marker types, legends, grids, and backgrounds?
        \item \textbf{Clarity (0-10)}: Is the layout clean? Are there any overlapping elements or broken CSS?
    \end{enumerate}

    \vspace{5pt}
    \textbf{\#\#\# Output Format:} \\
    Return a JSON object with specific scores and specific comments for each dimension.
    
    \vspace{5pt}
    \texttt{\{} \\
    \quad \texttt{"layout\_score": int,} \\
    \quad \texttt{"layout\_comment": "string",} \\
    \quad \texttt{"chart\_type\_score": int,} \\
    \quad \texttt{"chart\_type\_comment": "string",} \\
    \quad \texttt{"text\_content\_score": int,} \\
    \quad \texttt{"text\_content\_comment": "string",} \\
    \quad \texttt{"data\_fidelity\_score": int,} \\
    \quad \texttt{"data\_fidelity\_comment": "string",} \\
    \quad \texttt{"style\_score": int,} \\
    \quad \texttt{"style\_comment": "string",} \\
    \quad \texttt{"clarity\_score": int,} \\
    \quad \texttt{"clarity\_comment": "string"} \\
    \texttt{\}}
    \end{promptbox}
    \caption{Visual Fidelity Evaluation Prompt for generated dashboard assessment.}
    \label{fig:fidelity_eval_prompt}
\end{figure*}

\begin{figure*}[h]
    \centering
    \begin{promptbox}[title=Dynamic Behavior Evaluation Prompt]
    \small
    Evaluate if the Generated Dashboard (GEN) behaves exactly like the Ground Truth (GT) for the user task: ``\{task\_description\}''.

    \vspace{5pt}
    Images provided in order:
    \begin{enumerate}[leftmargin=15pt, nosep]
        \item GT Start (Before Task)
        \item GT End (After Task)
        \item GEN Start (Before Task)
        \item GEN End (After Task)
    \end{enumerate}

    \vspace{5pt}
    \textbf{Your task}: Compare the CHANGES from Image 1$\rightarrow$2 (GT Delta) with the CHANGES from Image 3$\rightarrow$4 (GEN Delta).

    \vspace{5pt}
    \textbf{Focus ONLY on behavior consistency:}
    \begin{itemize}[leftmargin=15pt, nosep]
        \item Did the same data values change in both?
        \item Did the same visual elements get highlighted/updated?
        \item Did the same controls respond to the interaction?
        \item Did the dashboard update in the expected way?
    \end{itemize}

    \vspace{5pt}
    \textbf{Do NOT evaluate:}
    \begin{itemize}[leftmargin=15pt, nosep]
        \item Overall visual quality (handled separately)
        \item Layout integrity (handled separately)
        \item Style matching (handled separately)
    \end{itemize}

    \vspace{5pt}
    Rate ONLY the dynamic behavior consistency on 0-10:
    \begin{itemize}[leftmargin=15pt, nosep]
        \item \textbf{10}: Perfect behavior match, GEN responds exactly like GT
        \item \textbf{7-9}: Very similar behavior with minor differences
        \item \textbf{4-6}: Correct general direction but noticeable differences
        \item \textbf{1-3}: Behavior differs significantly
        \item \textbf{0}: No behavior change or completely wrong response
    \end{itemize}

    \vspace{5pt}
    \textbf{Output JSON format only:} \\
    \texttt{\{} \\
    \quad \texttt{"dynamic\_behavior\_consistency\_score": int (0-10),} \\
    \quad \texttt{"reasoning": "brief explanation focusing on behavior changes"} \\
    \texttt{\}}
    \end{promptbox}
    \caption{Dynamic Behavior Evaluation Prompt for interaction assessment.}
    \label{fig:behavior_eval_prompt}
\end{figure*}

\begin{figure*}[h]
    \centering
    \begin{promptbox}[title=System Instruction for LLM-Semantic-Eval]
    \small
    You are a Senior Python Code Reviewer specializing in Plotly Dash applications. \\
    Your task is to compare "Generated Code" against "Ground Truth Code" to evaluate their \textbf{Functional Equivalence}.

    \vspace{5pt}
    \textbf{\#\#\# Goal} \\
    Determine if the Generated Code implements the same interactive logic and data visualization behavior as the Ground Truth. \\
    \textbf{Do not obsess over variable names or comment styles.} Focus on the semantic logic of \texttt{app.layout} and \texttt{@app.callback}.

    \vspace{5pt}
    \textbf{\#\#\# Ground Truth Code:} \\
    \texttt{\{code\_gt\}}

    \vspace{5pt}
    \textbf{\#\#\# Generated Code:} \\
    \texttt{\{code\_gen\}}

    \vspace{5pt}
    \textbf{\#\#\# Evaluation Criteria} \\
    Analyze the code based on two dimensions: \\
    1. \textbf{Layout Structure}: Are the component hierarchies (Divs, Graphs, Buttons) semantically identical? \\
    2. \textbf{Callback Logic}: Do the callbacks listen to the same Inputs, update the same Outputs, and perform equivalent data transformations?

    \vspace{5pt}
    \textbf{\#\#\# Scoring Rubric} \\
    \begin{itemize}[leftmargin=15pt, nosep]
        \item \textbf{"Functional Equivalent" (Score: 100)}: Logically identical. Produces exact same UI and handles interactions correctly.
        \item \textbf{"Minor Discrepancy" (Score: 80)}: Mostly correct. Minor issues like default values or style parameters.
        \item \textbf{"Moderate Discrepancy" (Score: 60)}: Core logic present, but noticeable gaps (e.g., secondary filter ignored).
        \item \textbf{"Significant Defect" (Score: 40)}: UI exists, but interactive logic (Callbacks) has major flaws or errors.
        \item \textbf{"Critical Defect" (Score: 20)}: Fails significantly. Missing key components; callbacks are hallucinated.
        \item \textbf{"Mismatch" (Score: 0)}: Syntactically broken, irrelevant, or empty.
    \end{itemize}

    \vspace{5pt}
    \textbf{\#\#\# Output Format} \\
    You must respond with a SINGLE JSON object: \\
    \texttt{\{} \\
    \quad \texttt{"reasoning": "Step-by-step analysis of Layout and Callbacks...",} \\
    \quad \texttt{"category": "Functional Equivalent",} \\
    \quad \texttt{"score": 100} \\
    \texttt{\}}
    \end{promptbox}
    \caption{LLM-Semantic-Eval Prompt for functional consistency check.}
    \label{fig:semantic_eval_prompt}
\end{figure*}

\begin{figure*}[h]
    \centering
    \begin{promptbox}[title=System Instruction for Task Executor Agent]
    \small
    You are an expert UI Automation Agent for benchmarking Plotly Dash applications. \\
    Your goal is to execute a specific User Task on the current screen.

    \vspace{5pt}
    \textbf{\# Input format} \\
    1. \textbf{User Task}: The natural language instruction. \\
    2. \textbf{Action History}: A list of actions you have executed in the current session so far. \\
    3. \textbf{Screenshots}: Previous Screen and Current Screen. \\
    4. \textbf{Accessibility Tree}: DOM structure. \\
       \textbf{IMPORTANT}: The bounding box within is formalized as \textbf{[ymin, xmin, ymax, xmax]} and \textbf{normalized to 0-1000}.

    \vspace{5pt}
    \textbf{\# Action Space (Coordinates are normalized 0-1000)} \\
    1. \texttt{"click": \{"type": "click", "x": int, "y": int\}} - Trigger a primary mouse click. \\
    2. \texttt{"double\_click": \{"type": "double\_click", "x": int, "y": int\}} - Trigger a double-click event. \\
    3. \texttt{"move\_mouse\_to": \{"type": "move\_mouse\_to", "x": int, "y": int\}} - Move mouse to coordinate. \\
    4. \texttt{"scroll": \{"type": "scroll", "amount": int, "x": int, "y": int\}} - Perform vertical scroll. \\
    5. \texttt{"drag": \{"type": "drag", "from\_x": int, "from\_y": int, "to\_x": int, "to\_y": int\}} - Drag action. \\
    6. \texttt{"replace\_text": \{"type": "replace\_text", "x": int, "y": int, "text": str\}} - Select all and type. \\
    7. \texttt{"DONE": \{"type": "DONE"\}} - Use ONLY when task is fully completed visually. \\
    8. \texttt{"FAIL": \{"type": "FAIL", "reason": "str"\}} - The task is impossible to complete. \\
    9. \texttt{"RESET": \{"type": "RESET", "reason": "str"\}} - Request a page reload.

    \vspace{5pt}
    \textbf{\# Response Format} \\
    Respond ONLY with a single JSON object. Do not include any explanations. \\
    \textbf{Example 1: Click (Center of screen)} \\
    \texttt{\{ "type": "click", "x": 500, "y": 500 \}}

    \vspace{5pt}
    \textbf{\# Rules} \\
    - Output strict JSON only. \\
    - \textbf{Verification}: Compare 'Previous Screen' and 'Current Screen'. \\
    - \textbf{Progress Tracking}: Check \textbf{Action History} to avoid repeating steps. \\
    - \textbf{Zero Guessing}: ONLY generate an action if the target is visible or in the Tree. \\
    - \textbf{Coordinate System}: All input and output coordinates must be 0-1000. \\
    - \textbf{Stalemate Detection}: If actions don't change the screen, consider returning "FAIL". \\
    - \textbf{Confident Termination}: Return "FAIL" if the dashboard is broken; do not try to fix it.
    \end{promptbox}
    \caption{System Instruction for Task Executor Agent.}
    \label{fig:executor_prompt}
\end{figure*}

\begin{figure*}[h]
    \centering
    \begin{promptbox}[title=System Instruction for Data Generator Agent]
    \small
    \textbf{system\_prompt} = f""" \\
    You are a Python Dash expert generating a benchmark dataset. \\
    Your goal is to generate a fully functional, self-contained Dash app.

    \vspace{5pt}
    \textbf{**Constraints:**} \\
    1. \textbf{Libraries:} \texttt{dash}, \texttt{plotly}, \texttt{pandas}. Optional: \texttt{dash\_bootstrap\_components}, \texttt{dash\_daq}, \texttt{dash\_ag\_grid}. \\
    2. \textbf{Data:} Generate dummy data inside the code using pandas. Do not load external files. \\
    3. \textbf{Execution:} The code must be runnable as a single script. \\
    4. \textbf{Style:} \{theme\_instruction\} \\
    5. \textbf{Viewport:} The entire dashboard layout should fit within a single view on a 1920x1080 screen.

    \vspace{5pt}
    \textbf{**Task Specifications:**} \\
    - \textbf{Domain:} \{selected\_domain\} \\
    - \textbf{Complexity Level:} \{level\_name\}

    \vspace{5pt}
    \textbf{**Available Candidates:**} \\
    - \textbf{Candidate Components:} \{', '.join(selected\_components)\} \\
    - \textbf{Candidate Charts:} \{', '.join(selected\_charts)\}

    \vspace{5pt}
    \textbf{**Callback Logic Requirement (CRITICAL):**} \\
    \{callback\_instruction\}

    \vspace{5pt}
    \textbf{**Creative \& Realism Guidelines:**} \\
    1. \textbf{Selection Strategy:} Pick AT MOST 1-2 components/charts that make sense for the story. \textbf{Less is More (Critical!!!)}. \\
    2. \textbf{Micro-Narrative:} Invent a specific user story (e.g., "Analyzing Q3 Churn Rate"). \\
    3. \textbf{Data Strategy:} 
       - \textbf{DROPDOWN LIMIT:} Max 8 options. 
       - Use trends/clusters instead of random noise (fixed seed 42).
       - \textbf{Avoid External Data:} No \texttt{px.data}, \texttt{read\_csv}, or URLs.
       - Control data points (\textasciitilde 10 per figure) to avoid clutter.

    \end{promptbox}
    \caption{Data Generator Prompt with Embedded Testing Guidelines.}
    \label{fig:data_gen_prompt}
\end{figure*}

\begin{figure*}[h]
    \centering
    \begin{promptbox}[title=Prompt for Anonymize the text in dash app layout.]
    \small

You are given a piece of Dash app code. Please modify the code as per the following requirements and output the complete code:

A. For all the UI components on `app.layout`, anonymize the labels (the content actually displayed in the UI) and corresponding titles. The anonymized text should be generic and have no business-specific meaning, such as `Dropdown1`, `Title1`, etc. Ensure that the components do not reveal any specific interaction logic.

B. For all the text components on `app.layout` (including titles, descriptions, hints, etc.), also anonymize them and change them to generic names like `Title1`, `Text1`, etc., while maintaining UI readability.

C. Additionally, please meet the following extra requirements:

\begin{itemize}
\item * For dropdown menus, anonymize their title text, and render the options in the dashboard as `Option1`, `Option2`, `Option3`, and so on.

\item  * For sliders or range sliders, anonymize their title text, and avoid displaying the scale content explicitly in the rendered dashboard.

\item  * For radio items or checklists, anonymize their title text, and also anonymize their option content.
\end{itemize}

Notes:

1. Do not modify the code logic, variable names, callback functions, chart contents, or data processing parts.

2. Only modify the displayed text in the layout to anonymized labels.

3. The anonymized text should be meaningful (not empty), but should be as generic as possible so that it’s hard to infer any specific functionality or logic from the text.

Here is the code:
\{python code\}

    \end{promptbox}
    \caption{Prompt for Anonymize the text in dash app layout.}
    \label{fig:Anonymize_prompt}
\end{figure*}

\subsection{Benchmark Cases}
\label{sec:benchmark_cases}
\begin{figure*}[t]
    \centering
    \begin{subfigure}[b]{0.48\textwidth}
        \centering
        \includegraphics[width=\textwidth]{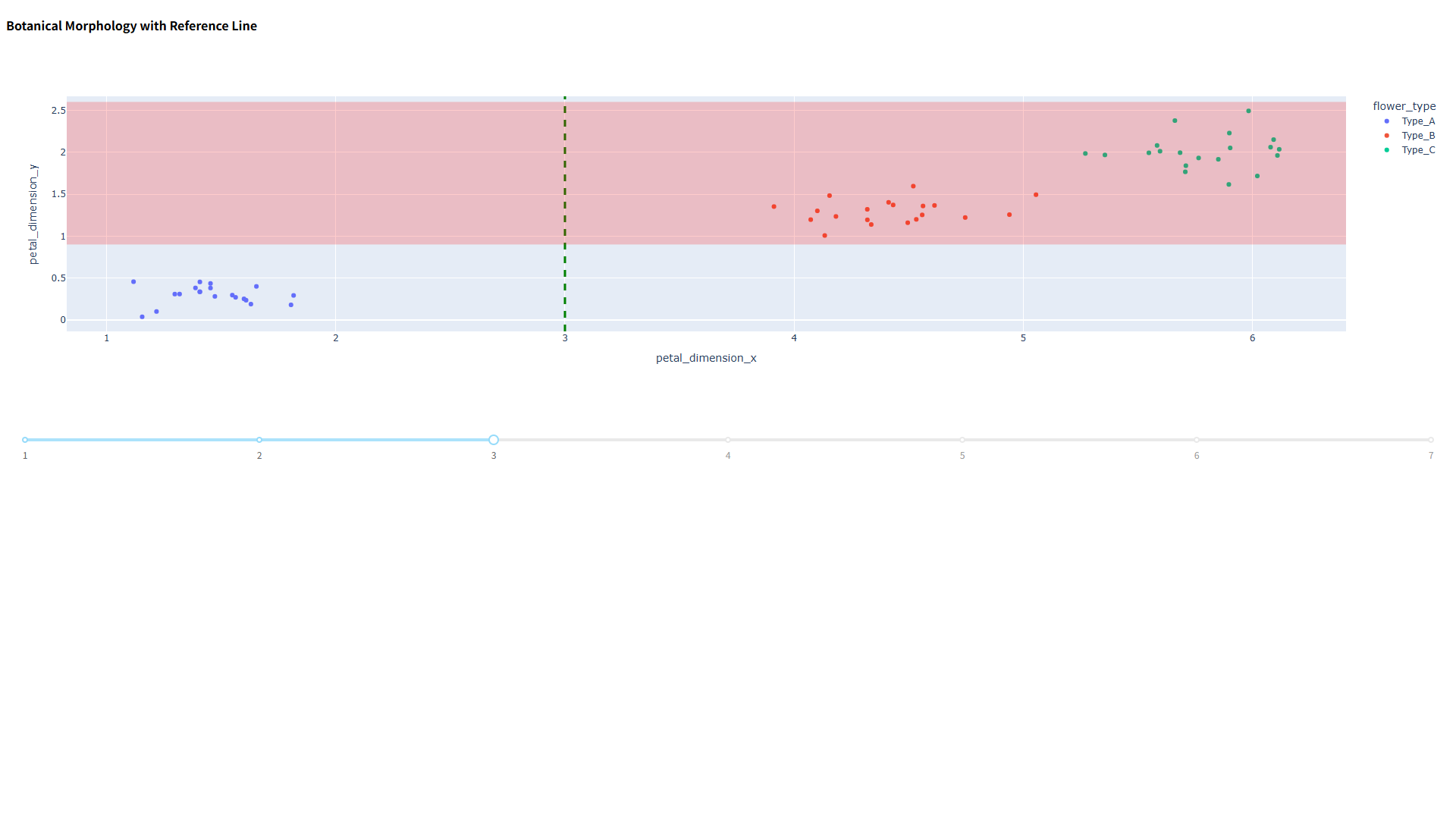}
        \caption{Position at $x=3$}
        \label{fig:task_step1}
        
    \end{subfigure}
    \hfill
    \begin{subfigure}[b]{0.48\textwidth}
        \centering
        \includegraphics[width=\textwidth]{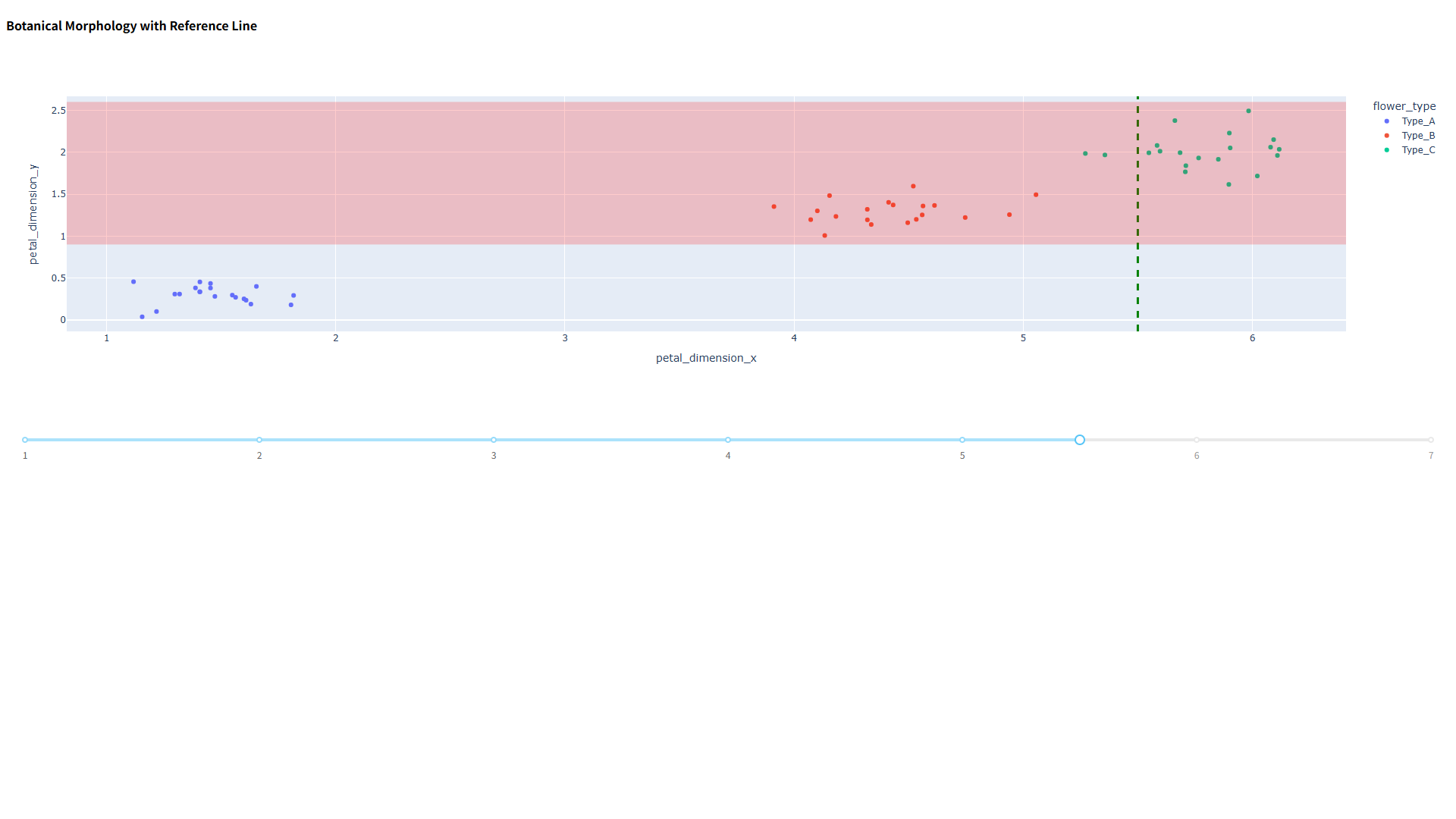}
        \caption{Position at $x=5.5$}
        \label{fig:task_step2}
    \end{subfigure}
    
    \caption{Level 1 (Atomic) Complexity Example. The application demonstrates a 1-to-1 interaction where (a) the vertical reference line is positioned at $x=3$, and (b) is updated to $x=5.5$ via a single slider input and a single callback function.}
    \label{fig:interaction_example}
\end{figure*}

\begin{figure*}[t]
    \centering
    \begin{subfigure}[b]{0.48\textwidth}
        \centering
        \includegraphics[width=\textwidth]{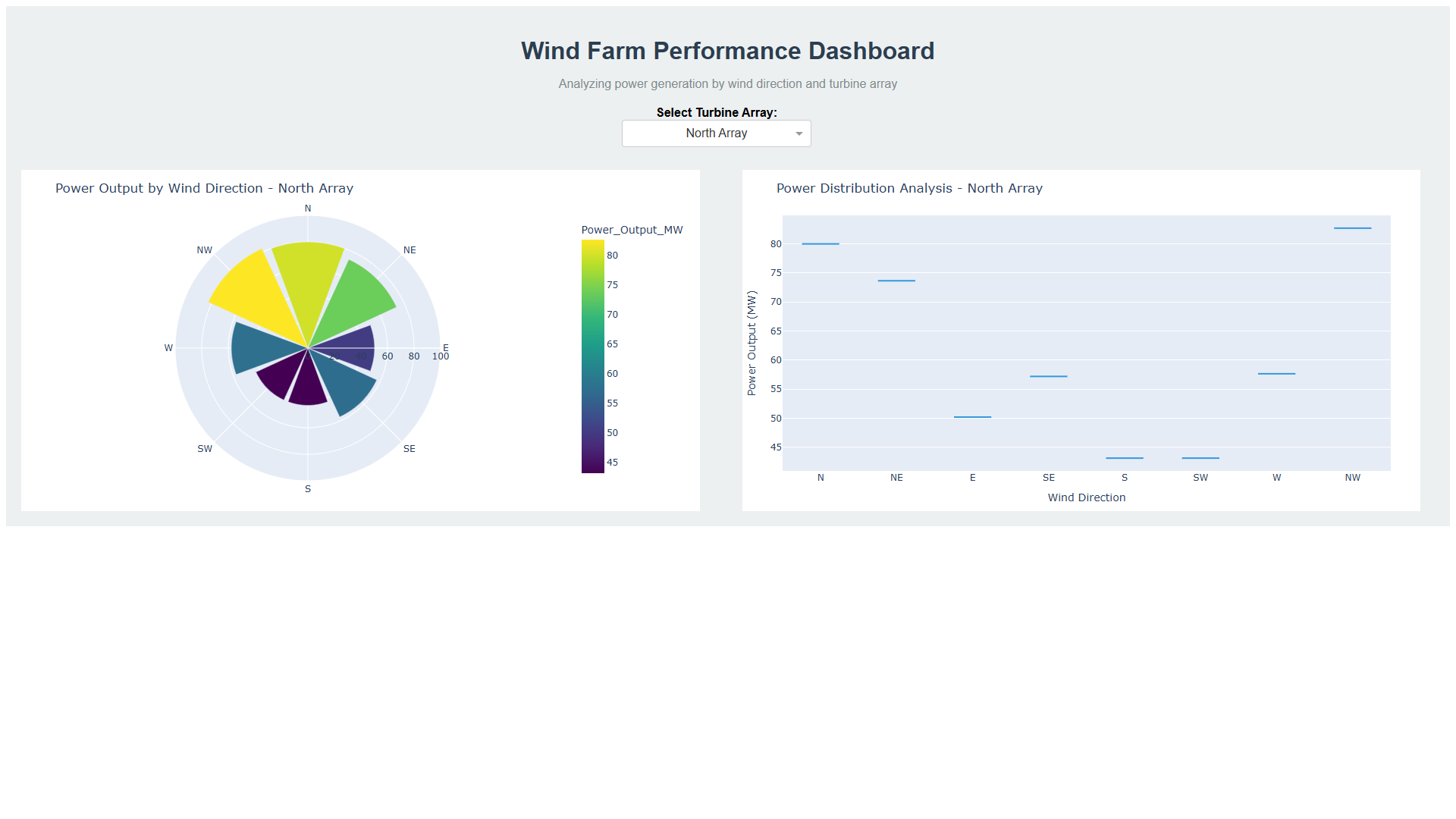}
        \caption{Initial state (North Array)}
        \label{fig:wind_task_step1}
    \end{subfigure}
    \hfill
    \begin{subfigure}[b]{0.48\textwidth}
        \centering
        \includegraphics[width=\textwidth]{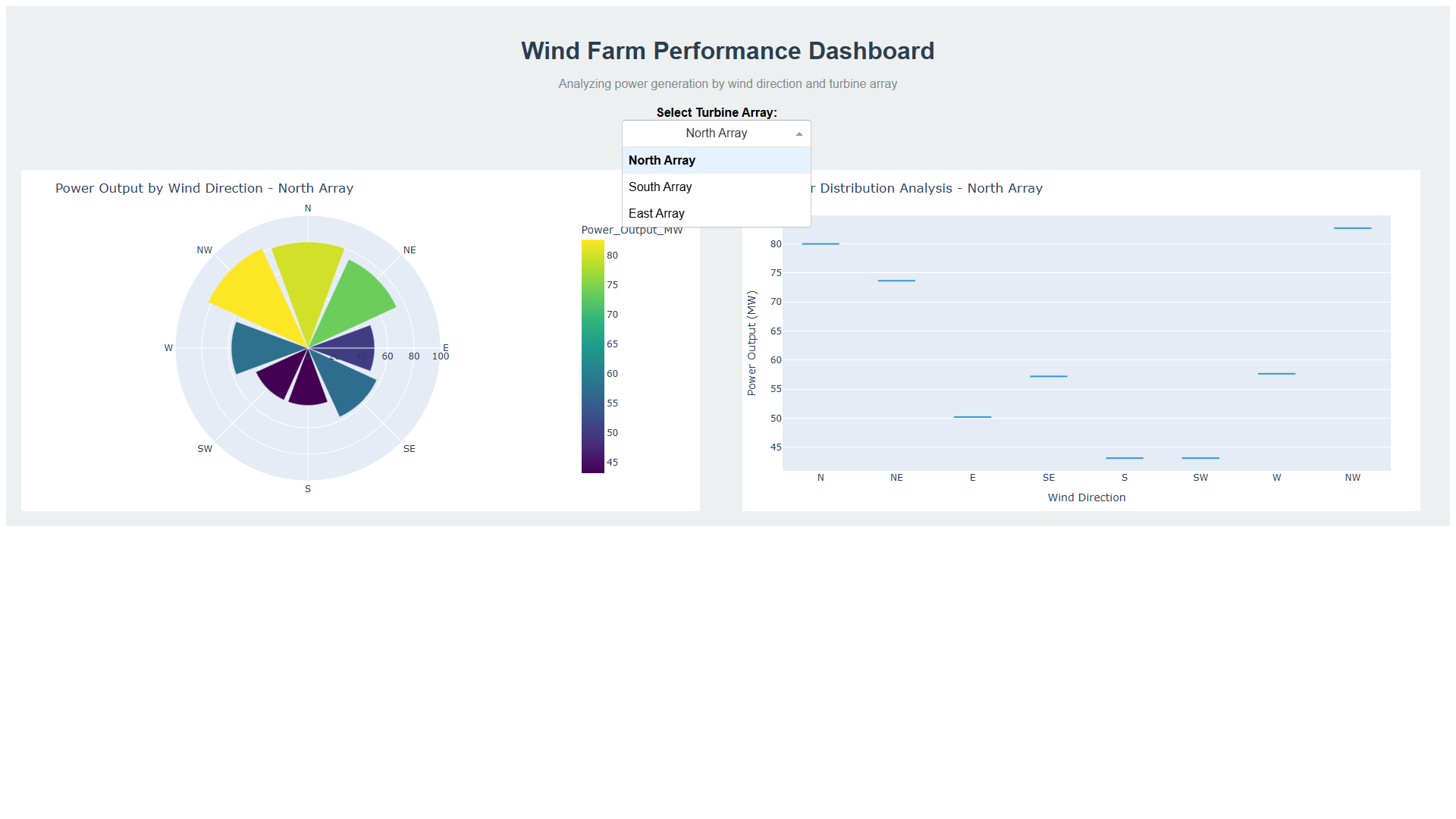}
        \caption{Broadcasting interaction}
        \label{fig:wind_task_step2}
    \end{subfigure}
    
    \caption{Level 2 (Broadcasting) Complexity Example. The application demonstrates a 1-to-Many interaction pattern where a single "Global Controller" (the \texttt{dcc.Dropdown} for Turbine Array selection) simultaneously updates two distinct visual outputs: a polar bar chart and a directional scatter plot. This tests the agent's ability to manage multiple output callbacks triggered by a single input change.}
    \label{fig:wind_farm_dashboard}
\end{figure*}

\begin{figure*}[t]
    \centering
    \begin{subfigure}[b]{0.32\textwidth}
        \centering
        \includegraphics[width=\textwidth]{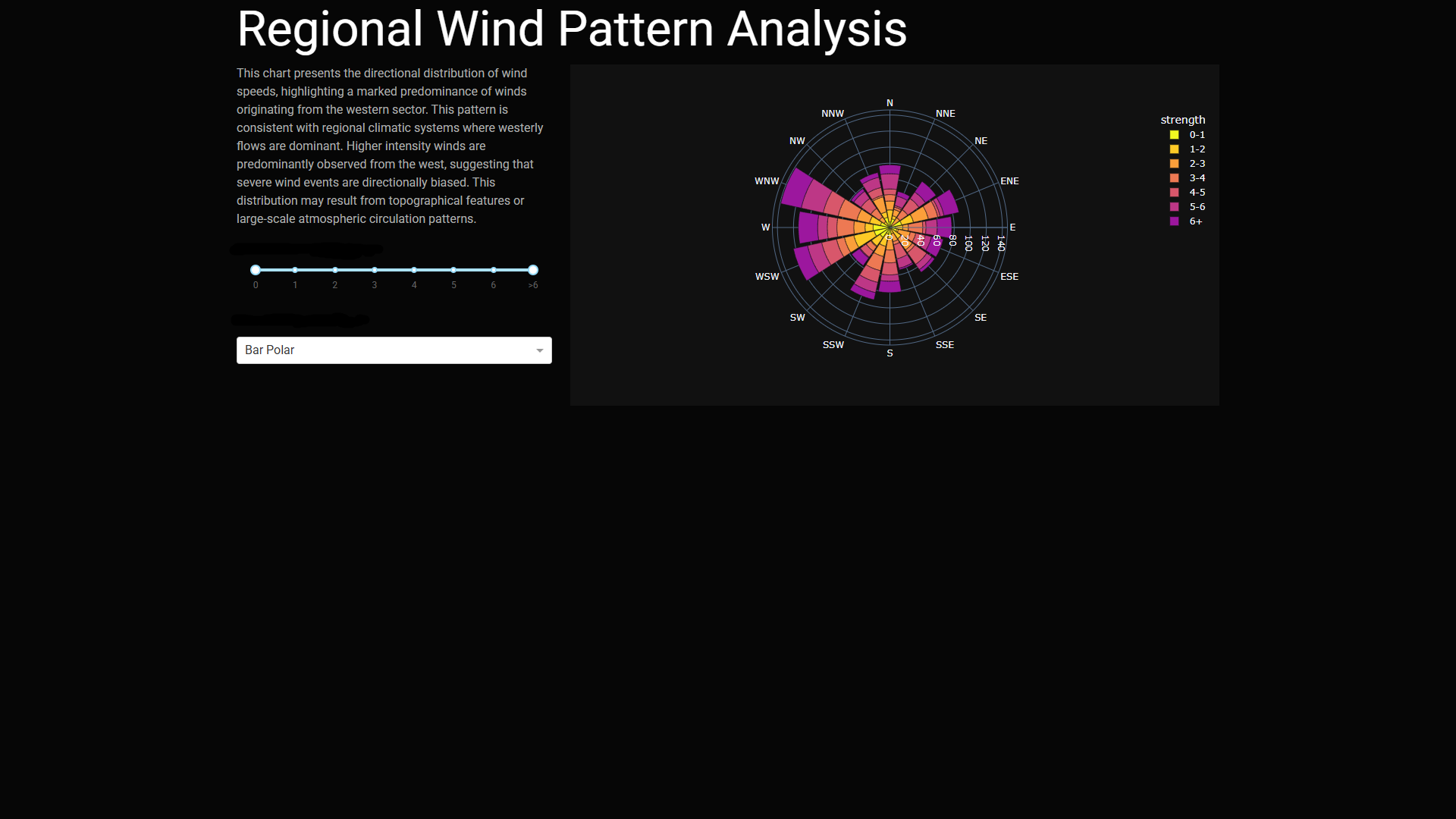}
        \caption{Initial state (Bar Polar)}
        \label{fig:agg_step1}
    \end{subfigure}
    \hfill
    \begin{subfigure}[b]{0.32\textwidth}
        \centering
        \includegraphics[width=\textwidth]{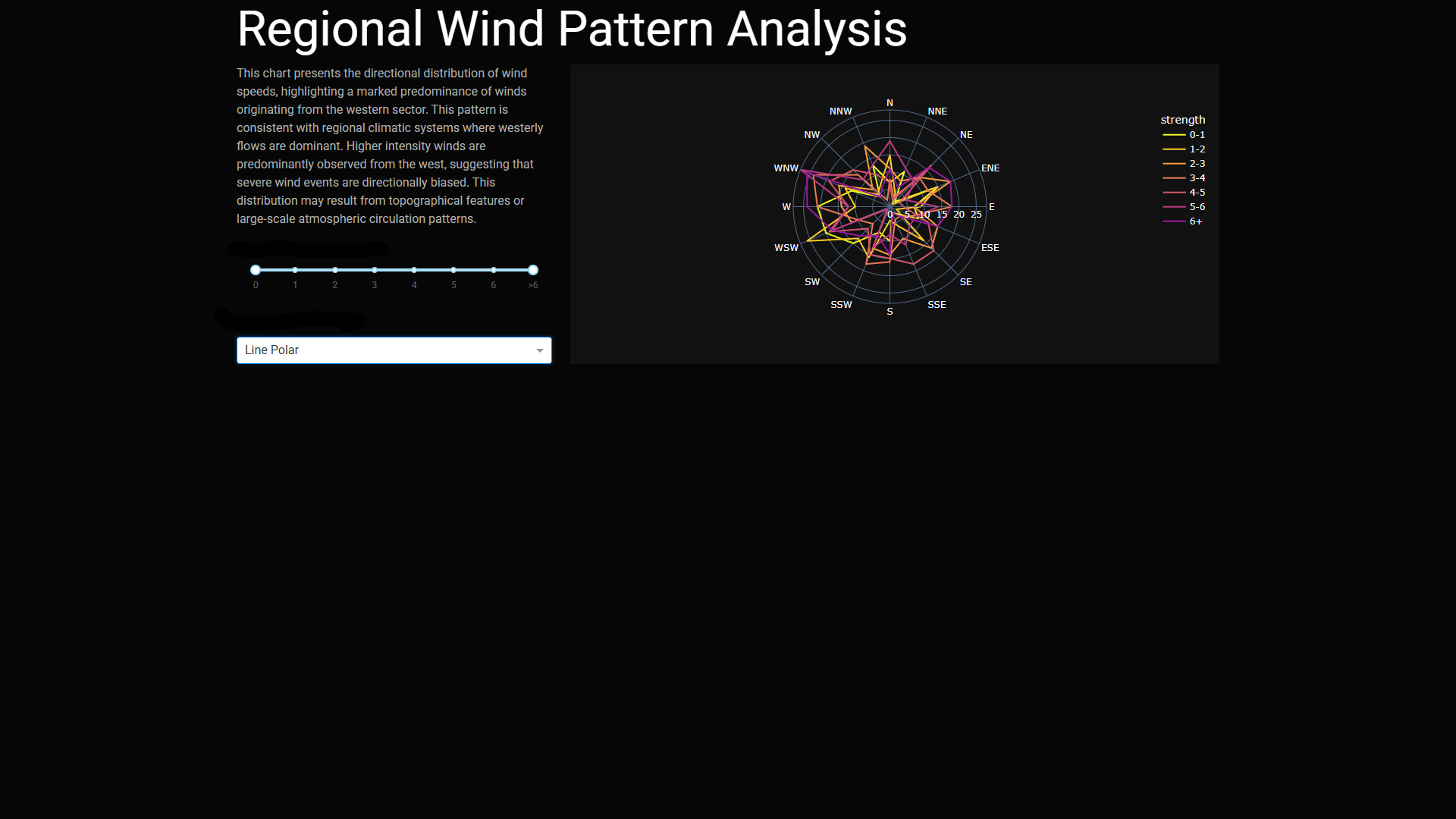}
        \caption{Chart type interaction}
        \label{fig:agg_step2}
    \end{subfigure}
    \hfill
    \begin{subfigure}[b]{0.32\textwidth}
        \centering
        \includegraphics[width=\textwidth]{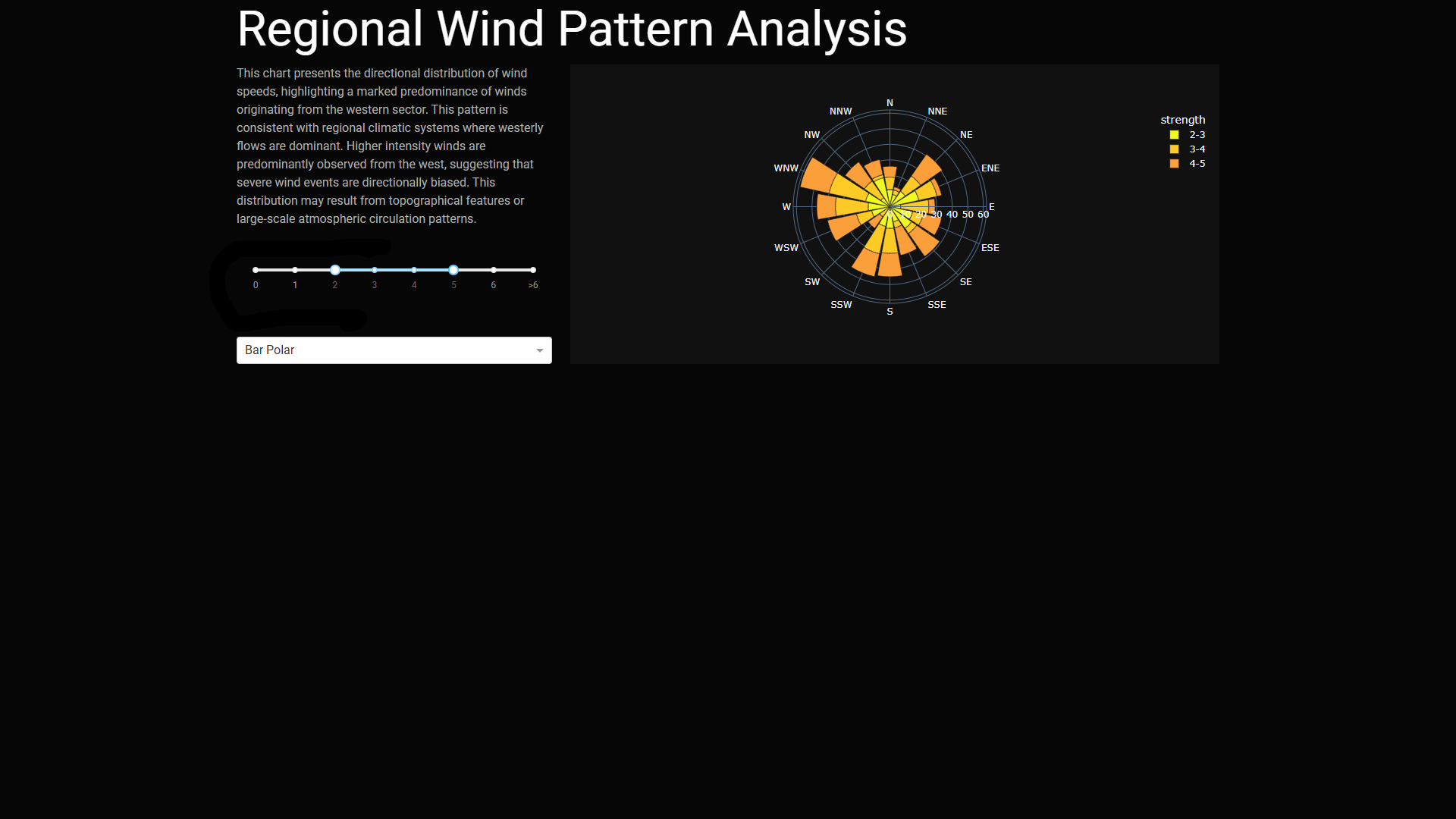}
        \caption{Combined range filter}
        \label{fig:agg_step3}
    \end{subfigure}
    
    \caption{Level 2 (Aggregation) Complexity Example. The application demonstrates a Many-to-1 interaction pattern where multiple independent inputs—a Range Slider for wind intensity and a Dropdown for chart type—jointly update a single visual output. This configuration tests the agent's ability to handle multi-input synchronization and conditional rendering within a single callback function.}
    \label{fig:wind_pattern_aggregation}
\end{figure*}

\begin{figure*}[t]
    \centering
    \begin{subfigure}[b]{0.32\textwidth}
        \centering
        \includegraphics[width=\textwidth]{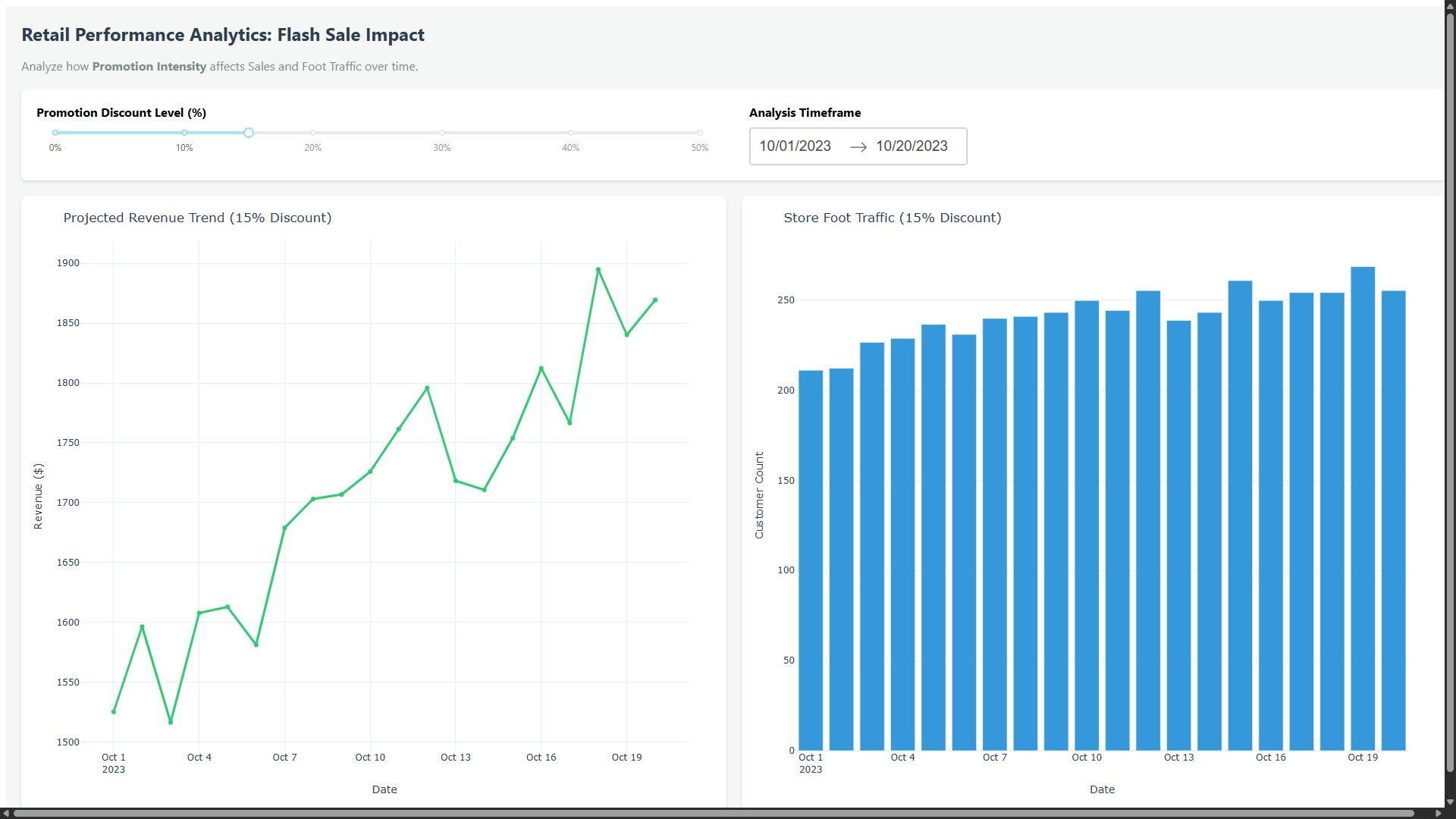}
        \caption{Baseline (15\% Discount)}
        \label{fig:mesh_step1}
    \end{subfigure}
    \hfill
    \begin{subfigure}[b]{0.32\textwidth}
        \centering
        \includegraphics[width=\textwidth]{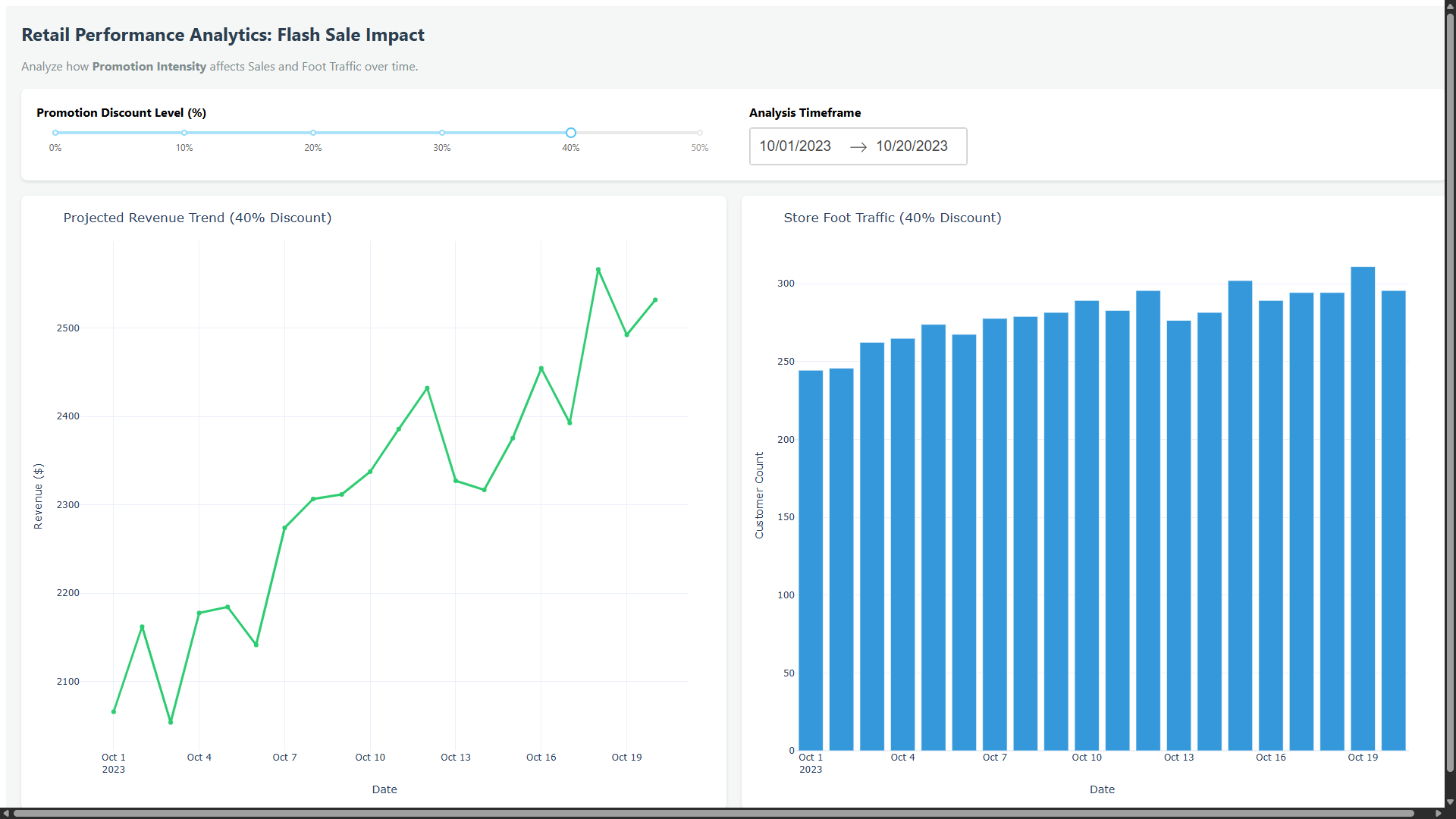}
        \caption{Scaling Input A (40\% Discount)}
        \label{fig:mesh_step2}
    \end{subfigure}
    \hfill
    \begin{subfigure}[b]{0.32\textwidth}
        \centering
        \includegraphics[width=\textwidth]{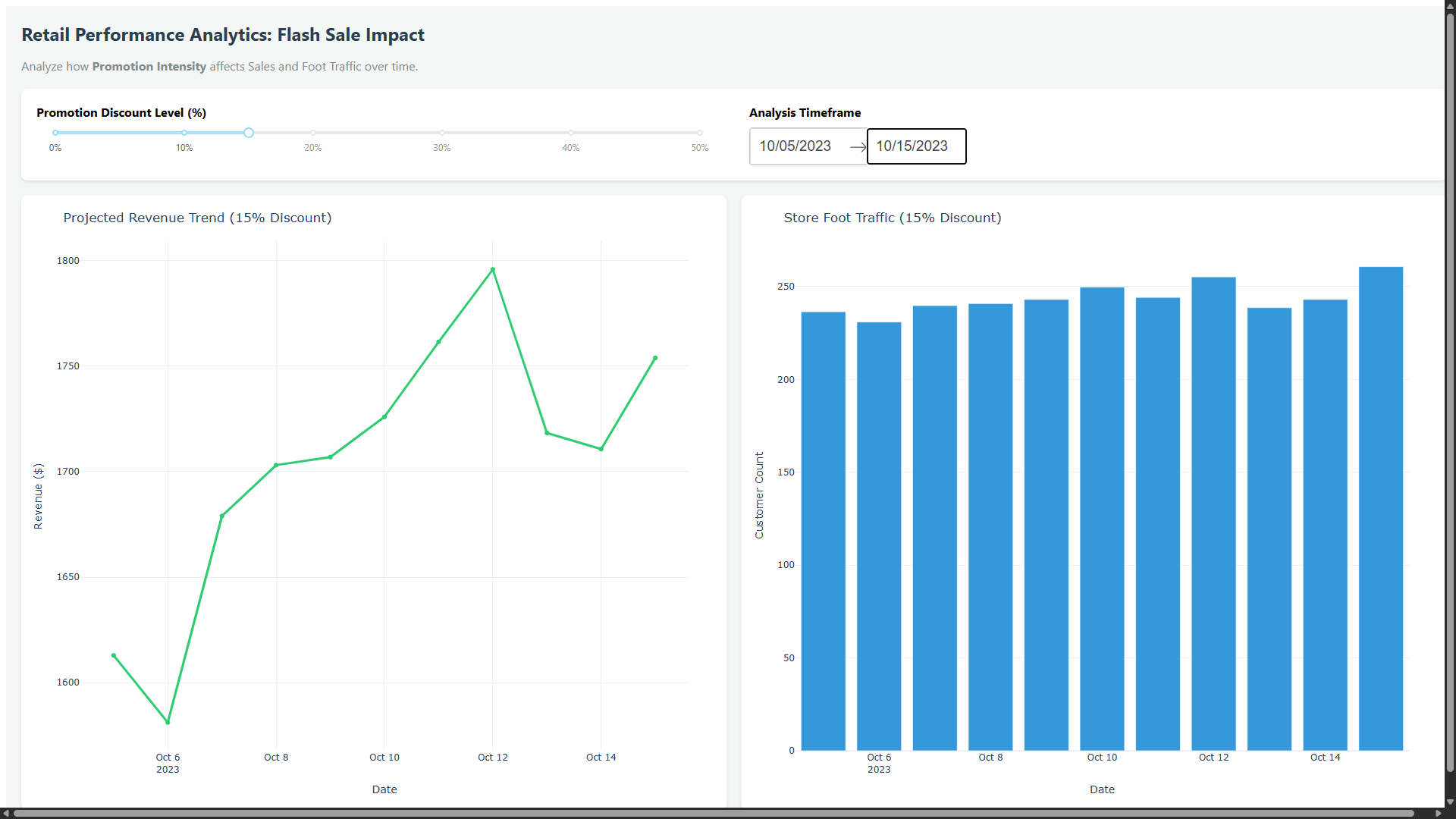}
        \caption{Filtering Input B (Timeframe)}
        \label{fig:mesh_step3}
    \end{subfigure}
    
    \caption{Level 2 (Coupled Mesh) Complexity Example. This dashboard demonstrates a Many-to-Many interaction pattern where at least two independent inputs—Promotion Discount Level and Analysis Timeframe—jointly update at least two distinct visual outputs (Revenue Trend and Foot Traffic). This setup evaluates the agent's proficiency in managing complex state dependencies where multiple triggers affect the global state of multiple components simultaneously.}
    \label{fig:retail_performance_mesh}
\end{figure*}

\begin{figure*}[t]
    \centering
    \begin{subfigure}[b]{0.48\textwidth}
        \centering
        \includegraphics[width=\textwidth]{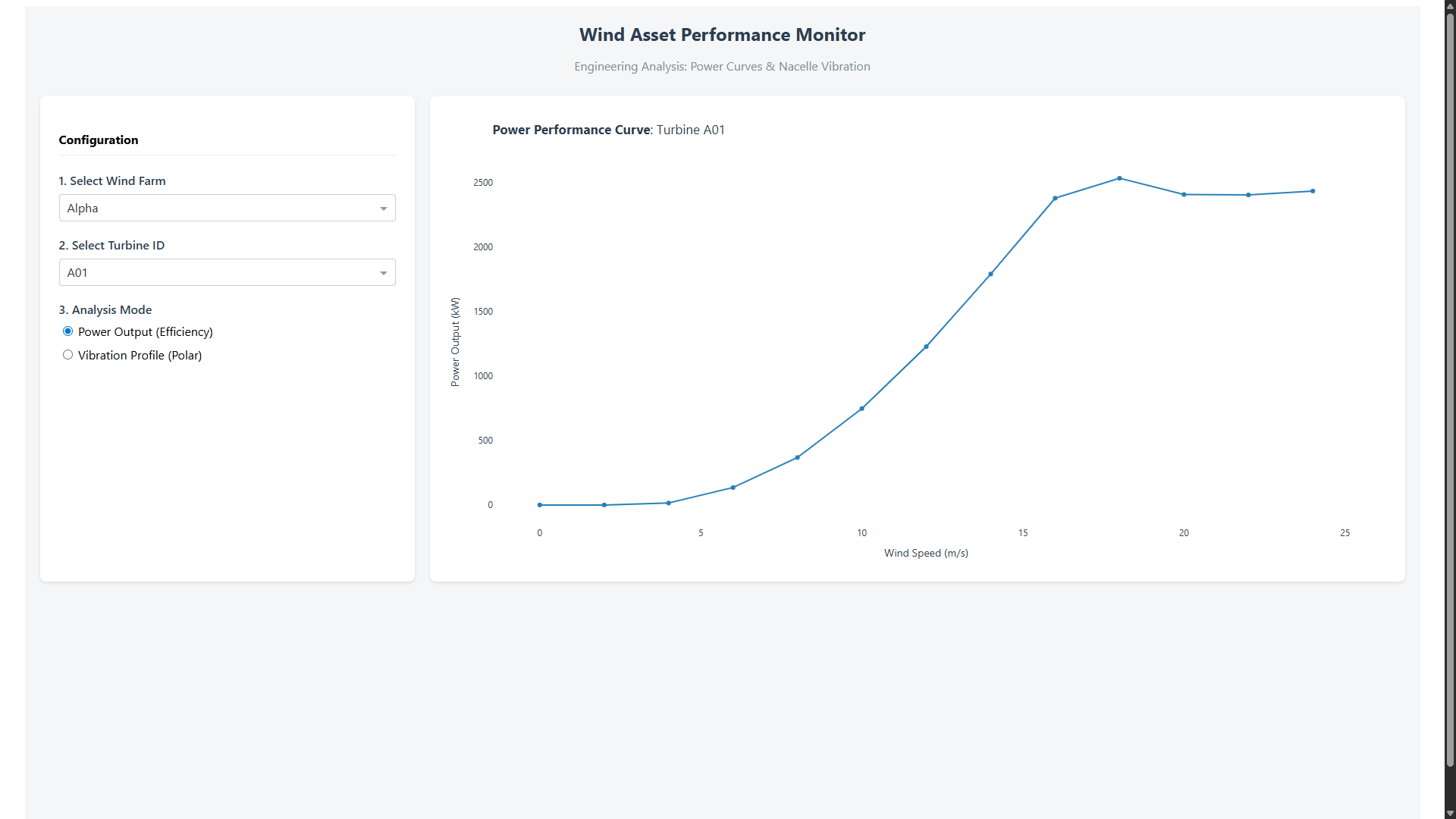}
        \caption{Initial state: Farm 'Alpha' selected}
        \label{fig:chain_step1}
    \end{subfigure}
    \hfill
    \begin{subfigure}[b]{0.48\textwidth}
        \centering
        \includegraphics[width=\textwidth]{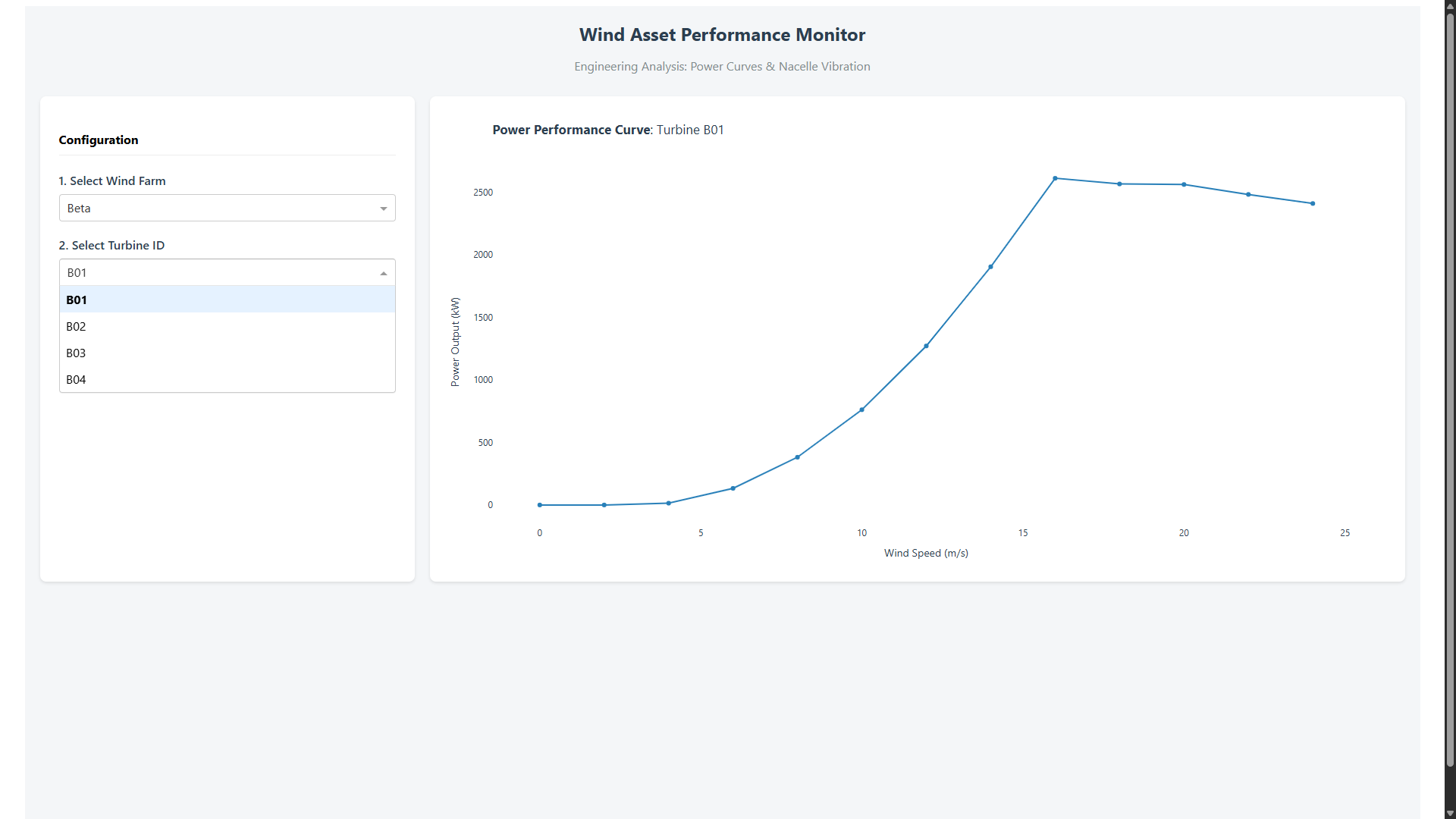}
        \caption{Chained update: Farm 'Beta' selected}
        \label{fig:chain_step2}
    \end{subfigure}
    
    \caption{Level 3 (Chained Callbacks) Complexity Example. The application demonstrates inter-dependent input components where the options in the 'Turbine ID' dropdown are dynamically filtered based on the 'Wind Farm' selection. Figure (b) specifically shows the active interaction where selecting a new farm triggers a callback to update child component values, illustrating a multi-step dependency chain that the agent must resolve to ensure data consistency across the UI.}
    \label{fig:wind_monitor_chained}
\end{figure*}

\begin{figure*}[t]
    \centering
    \begin{subfigure}[b]{0.48\textwidth}
        \centering
        \includegraphics[width=\textwidth]{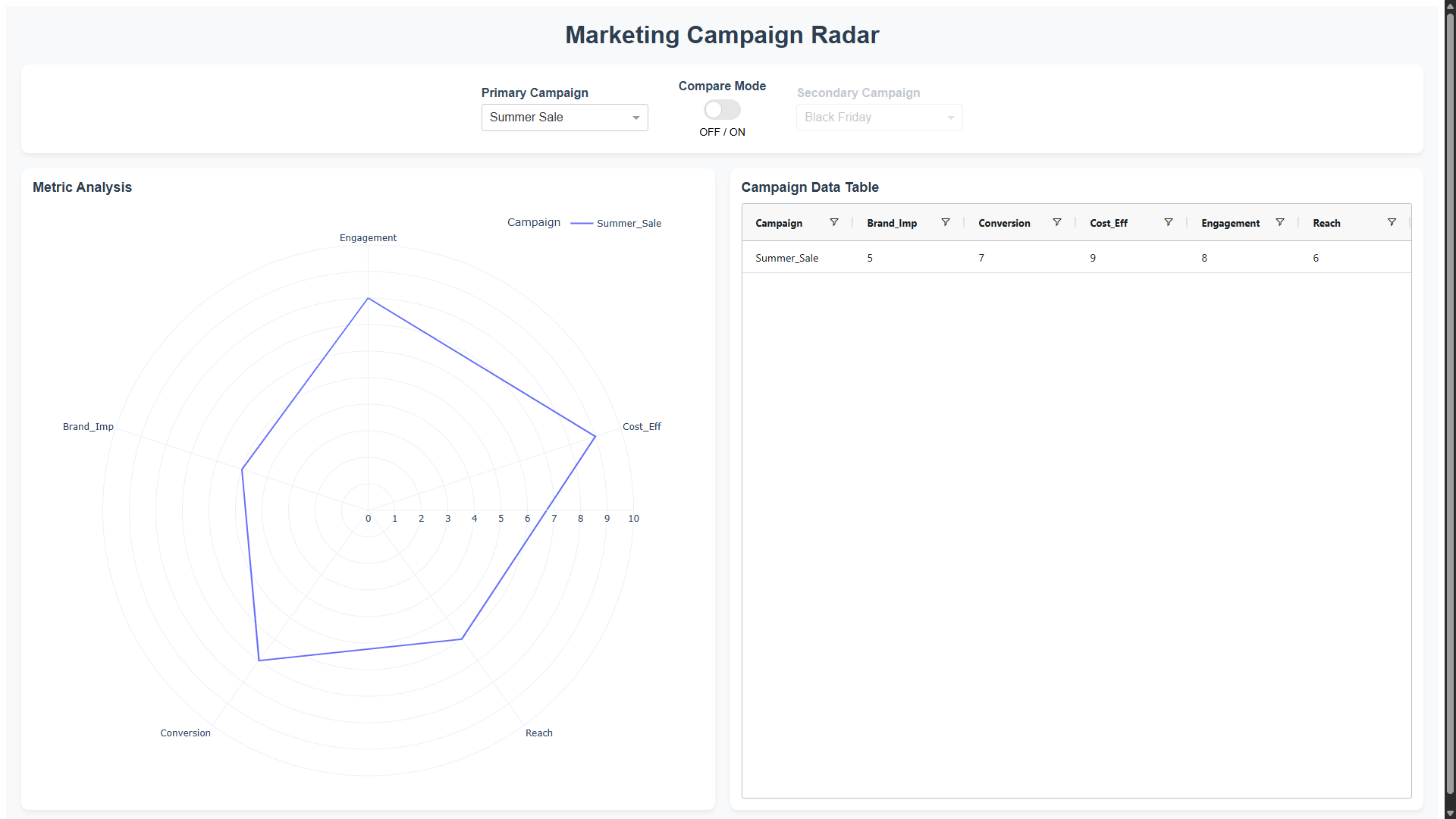}
        \caption{Single Mode (Compare Off)}
        \label{fig:cond_step1}
    \end{subfigure}
    \hfill
    \begin{subfigure}[b]{0.48\textwidth}
        \centering
        \includegraphics[width=\textwidth]{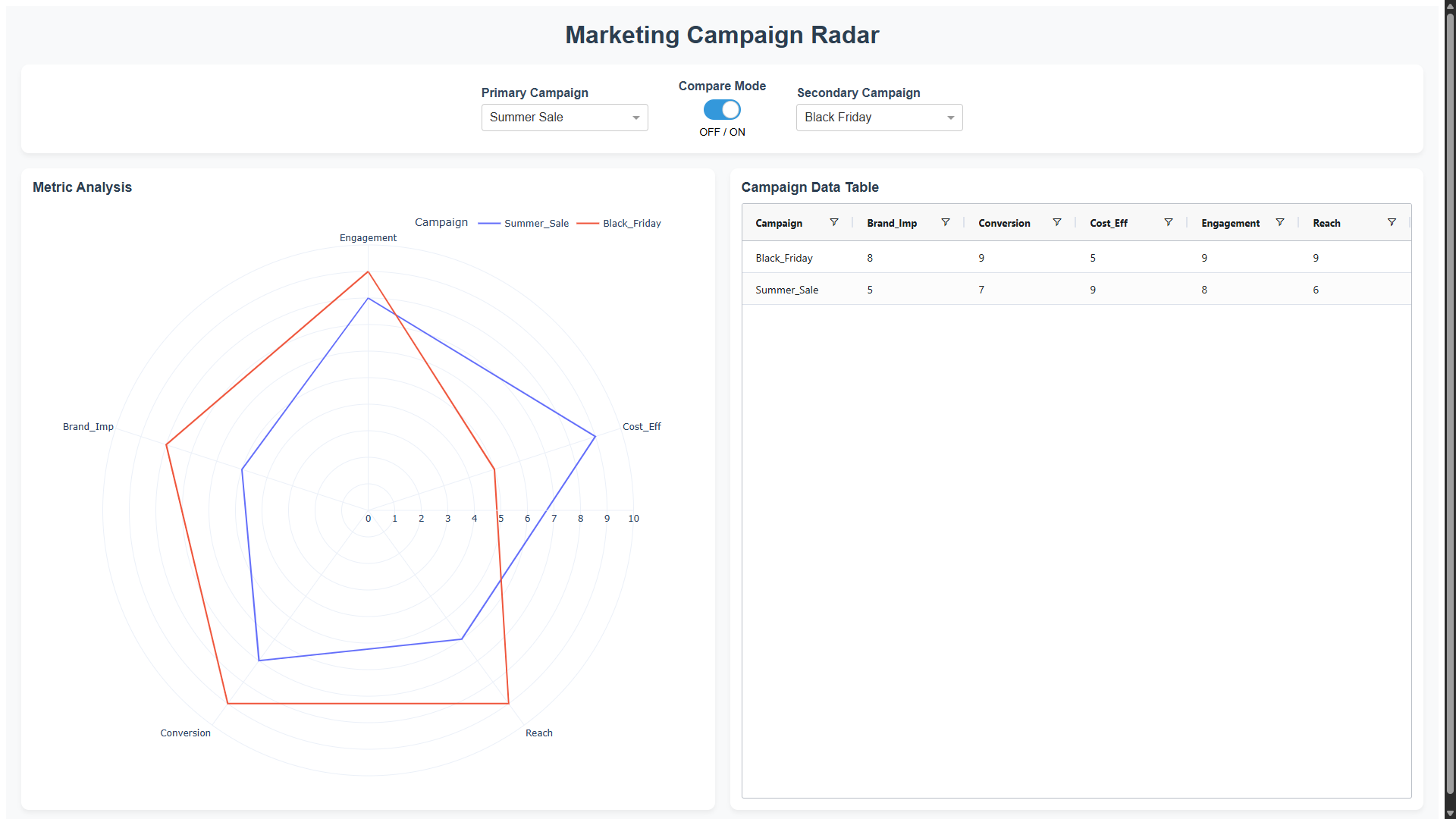}
        \caption{Comparison Mode (Compare On)}
        \label{fig:cond_step2}
    \end{subfigure}
    
    \caption{Level 3 (Conditional Visibility \& State) Complexity Example. The application implements a toggle-based conditional visibility pattern: when 'Compare Mode' is deactivated (a), the secondary input is disabled and the output displays a single metric set. Activating the toggle (b) dynamically enables the 'Secondary Campaign' dropdown and triggers a multi-trace update on both the radar chart and the data table. This requires the agent to handle component 'disabled' states and conditional data merging logic.}
    \label{fig:marketing_radar_conditional}
\end{figure*}

\begin{figure*}[t]
    \centering
    \begin{subfigure}[b]{0.48\textwidth}
        \centering
        \includegraphics[width=\textwidth]{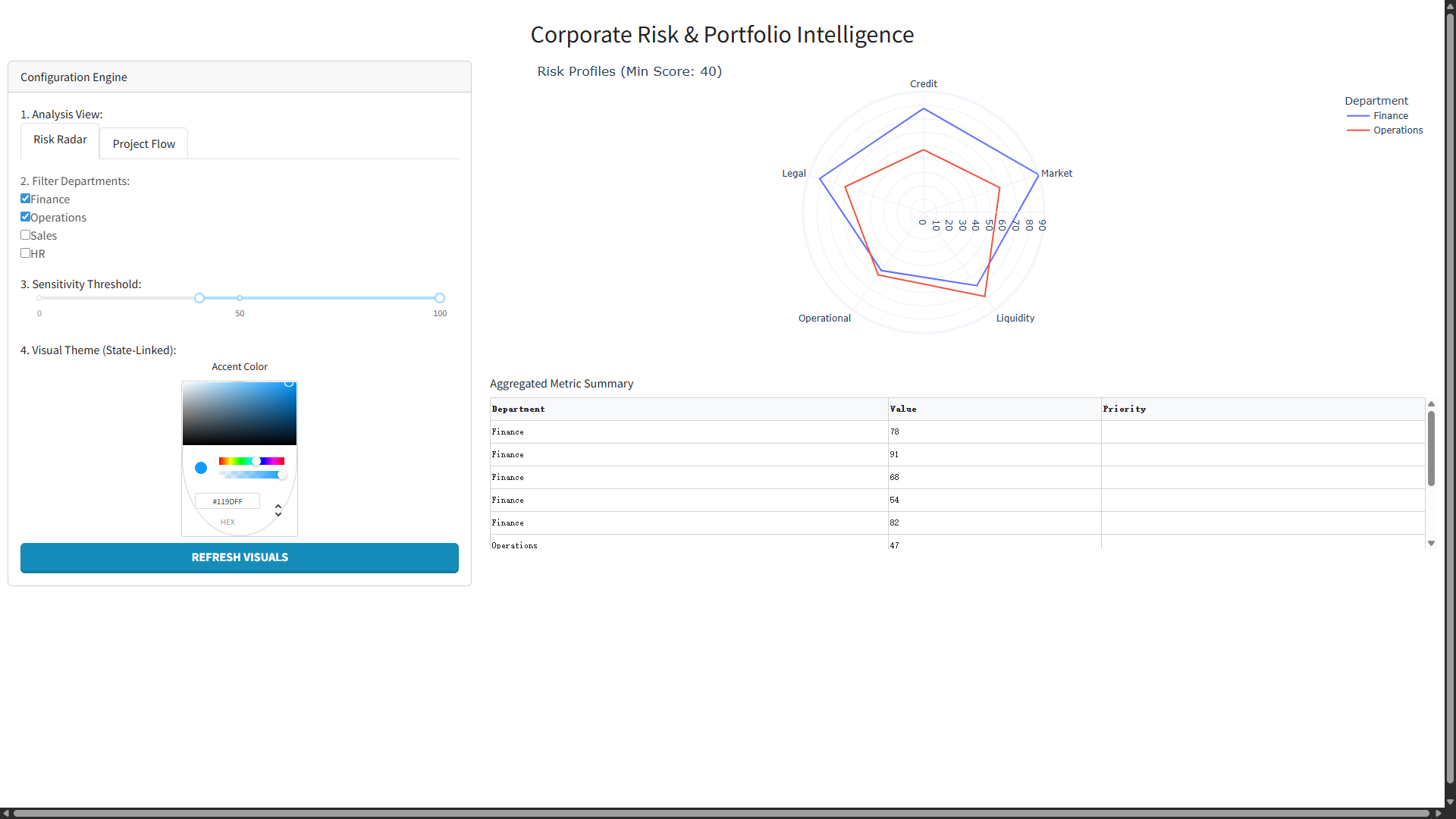}
        \caption{Risk Radar View with specific state}
        \label{fig:state_step1}
    \end{subfigure}
    \hfill
    \begin{subfigure}[b]{0.48\textwidth}
        \centering
        \includegraphics[width=\textwidth]{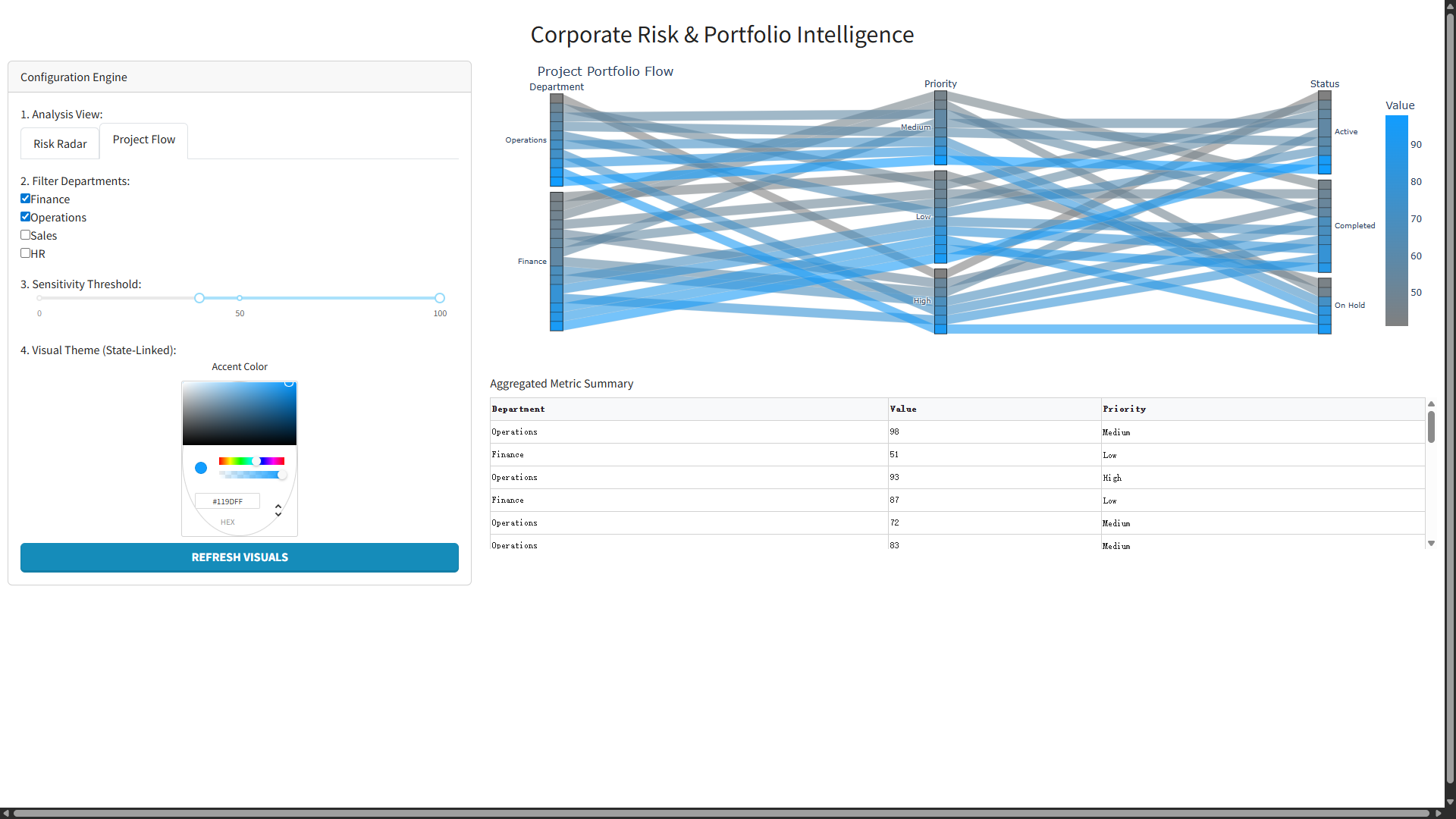}
        \caption{Project Flow View with shared state}
        \label{fig:state_step2}
    \end{subfigure}
    
    \caption{Level 3 (State Dependency) Complexity Example. The dashboard implements a state-managed logic chain where visualizations are decoupled from immediate inputs. Changes to filters and themes are captured as "States" and only applied to the UI upon the "REFRESH VISUALS" trigger. This requires the agent to manage a global state across conditional chart schemas (Radar and Sankey) rather than responding to isolated input changes.}
    \label{fig:corporate_risk_intelligence}
\end{figure*}

\begin{figure*}[t]
    \centering
    \begin{subfigure}[b]{0.32\textwidth}
        \centering
        \includegraphics[width=\textwidth]{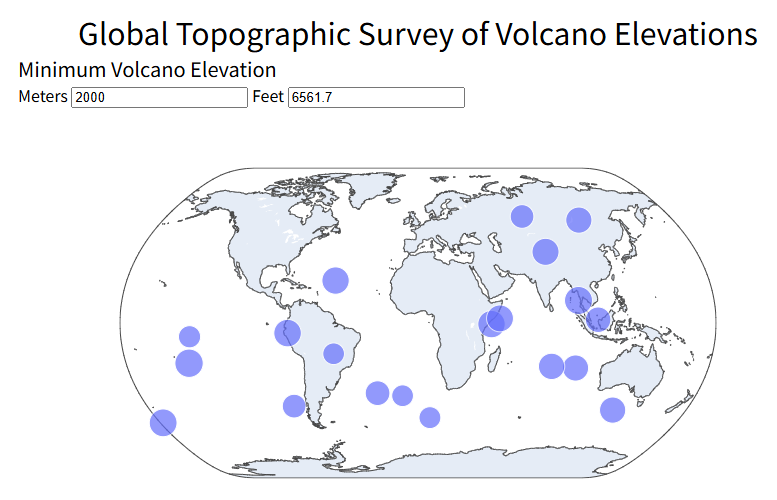}
        \caption{Baseline: 2000m / 6561.7ft}
        \label{fig:circular_step1}
    \end{subfigure}
    \hfill
    \begin{subfigure}[b]{0.32\textwidth}
        \centering
        \includegraphics[width=\textwidth]{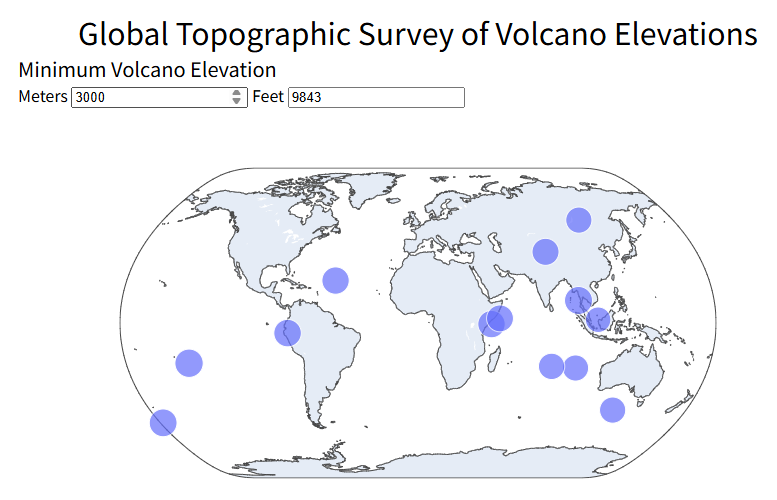}
        \caption{Update Meters $\rightarrow$ Feet}
        \label{fig:circular_step2}
    \end{subfigure}
    \hfill
    \begin{subfigure}[b]{0.32\textwidth}
        \centering
        \includegraphics[width=\textwidth]{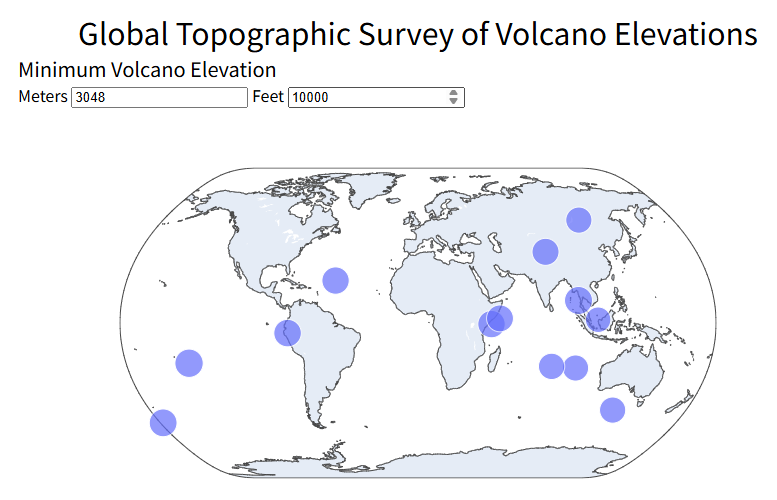}
        \caption{Update Feet $\rightarrow$ Meters}
        \label{fig:circular_step3}
    \end{subfigure}
    
    \caption{Level 3 (Circular Callback) Complexity Example. This survey application implements synchronized dual-unit inputs for volcano elevation filtering. The logic features a circular dependency where an update to the 'Meters' input (b) automatically recalculates and populates the 'Feet' field, and vice versa (c). This evaluates the agent's ability to implement bi-directional state synchronization without triggering infinite loops, while simultaneously filtering the geographic map output.}
    \label{fig:volcano_circular_sync}
\end{figure*}

Figures~\ref{fig:interaction_example} through~\ref{fig:volcano_circular_sync} illustrate the spectrum of interaction complexity implemented in our benchmark, progressing from simple atomic updates to complex conditional and circular dependencies.

\subsection{Error Cases of SOTA model}
\label{sec:error_case_sota}

\begin{figure*}[t]
    \centering
    \begin{subfigure}[b]{0.48\textwidth}
        \centering
        \includegraphics[width=\textwidth]{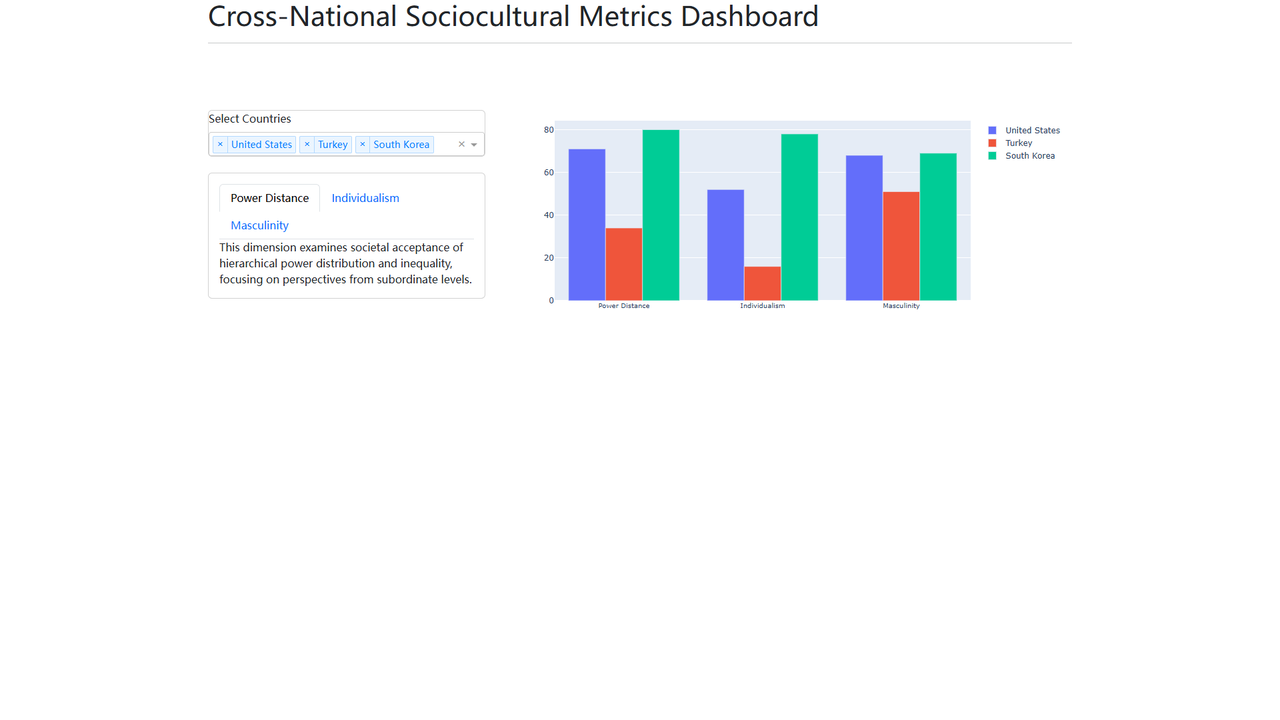}
        \caption{Ground Truth}
        \label{fig:error_gt}
    \end{subfigure}
    \hfill
    \begin{subfigure}[b]{0.48\textwidth}
        \centering
        \includegraphics[width=\textwidth]{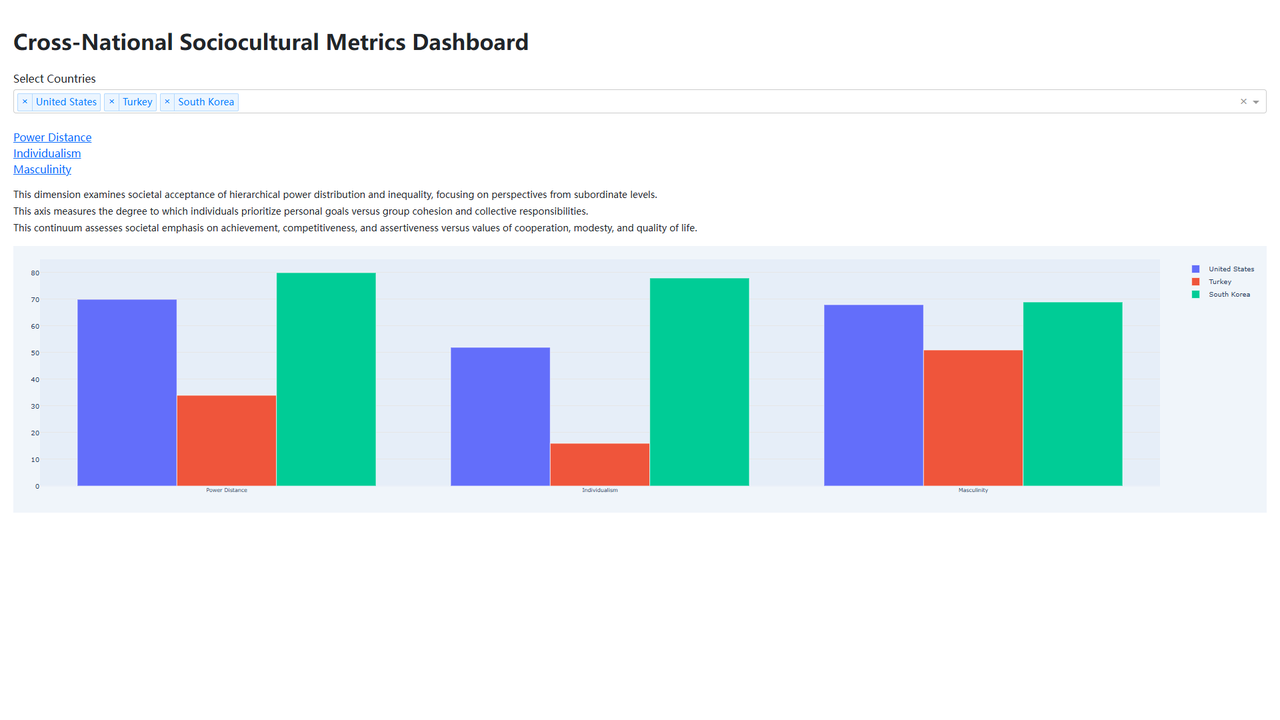}
        \caption{Model Generation}
        \label{fig:error_model}
    \end{subfigure}
    
    \caption{Example of Type II Error: Presentation Mismatch. While the generated dashboard (\subref{fig:error_model}) maintains functional parity with the ground truth (\subref{fig:error_gt}), it exhibits significant aesthetic deviations in layout (horizontal vs. vertical component stacking) and styling, illustrating a failure in visual fidelity despite logical correctness.}
    \label{fig:presentation_mismatch_error}
\end{figure*}

\subsubsection{Presentation dimension - Presentation Mismatch (Styling \& Layout Deviations)}
As illustrated in Figure~\ref{fig:presentation_mismatch_error}, our benchmark captures nuanced failures in the reverse-engineering process. A representative case is the Type II error (Presentation Mismatch), where the model successfully reconstructs the dashboard's interactive logic but fails to adhere to its aesthetic specifications. Specifically, comparing the ground truth (Figure~\ref{fig:error_gt}) with the model's output (Figure~\ref{fig:error_model}), we observe a shift from a compact sidebar layout to a vertically stacked arrangement, alongside deviations in CSS styling, which significantly impacts visual similarity without compromising core functional utility.

\begin{figure*}[t]
    \centering
    \begin{subfigure}[b]{0.48\textwidth}
        \centering
        \includegraphics[width=\textwidth]{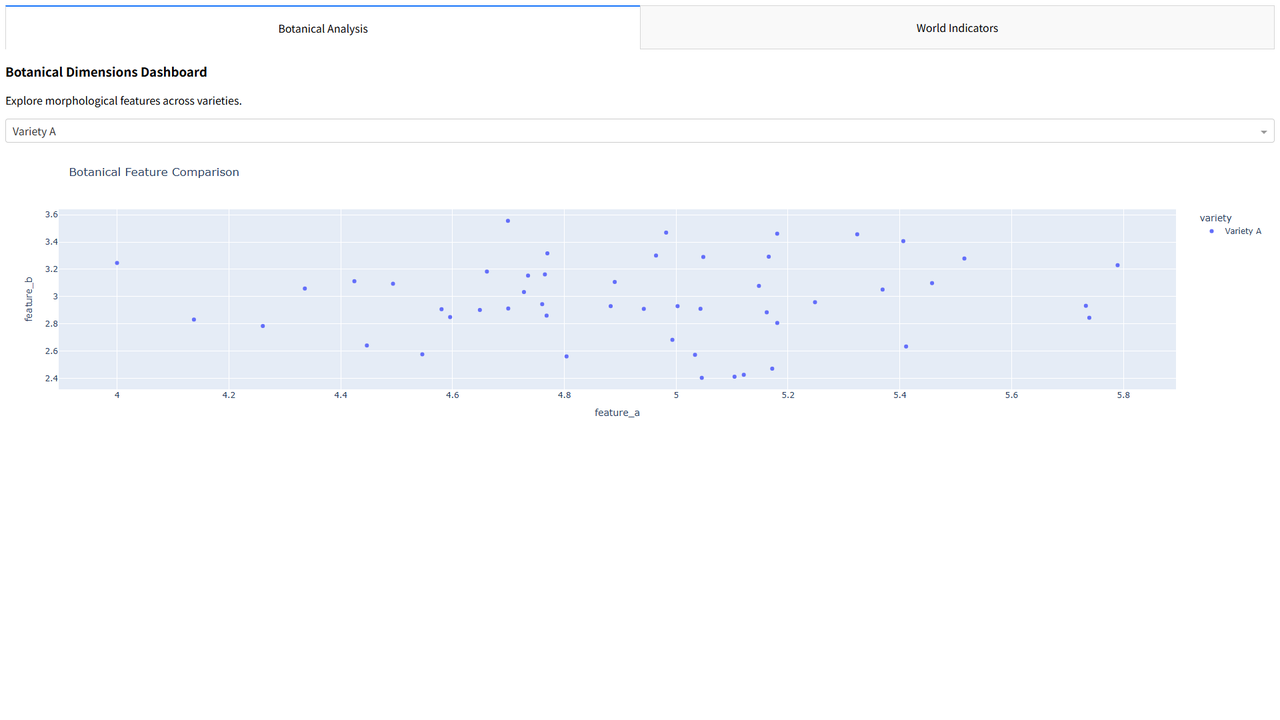}
        \caption{Ground Truth: State A}
        \label{fig:error3_gt_a}
    \end{subfigure}
    \hfill
    \begin{subfigure}[b]{0.48\textwidth}
        \centering
        \includegraphics[width=\textwidth]{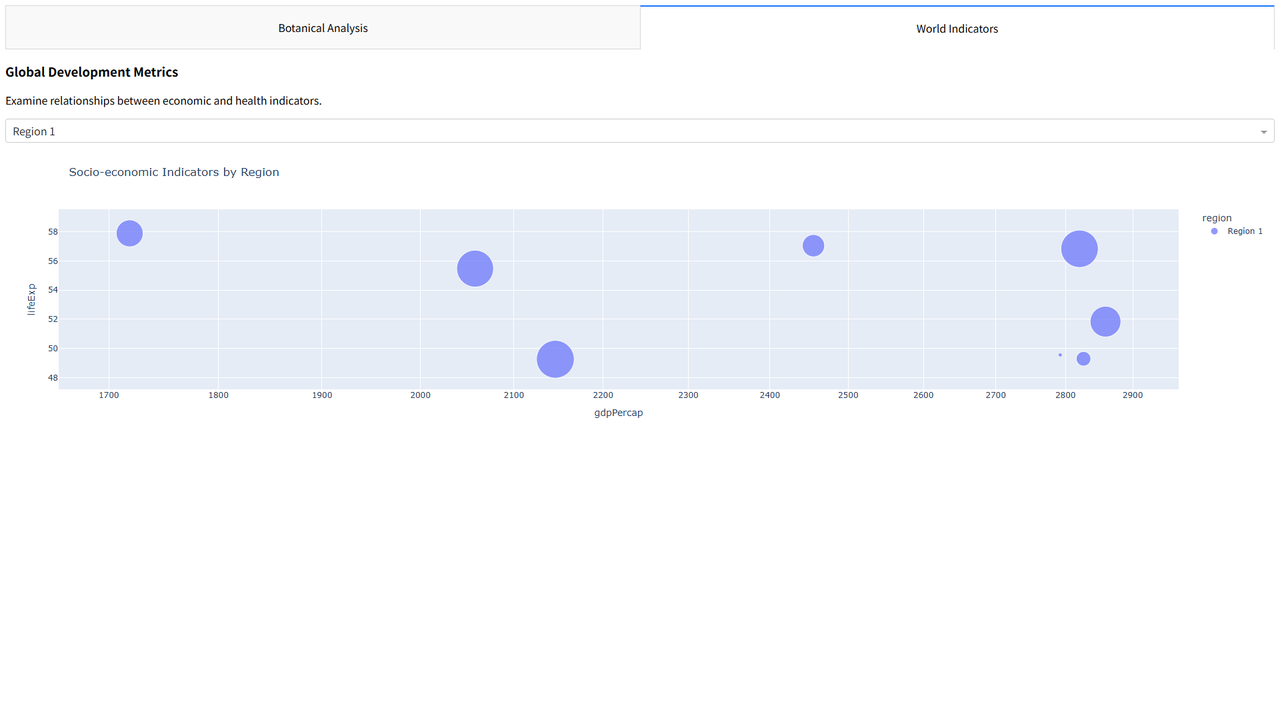}
        \caption{Ground Truth: State B}
        \label{fig:error3_gt_b}
    \end{subfigure}

    \vspace{2mm} 

    \begin{subfigure}[b]{0.48\textwidth}
        \centering
        \includegraphics[width=\textwidth]{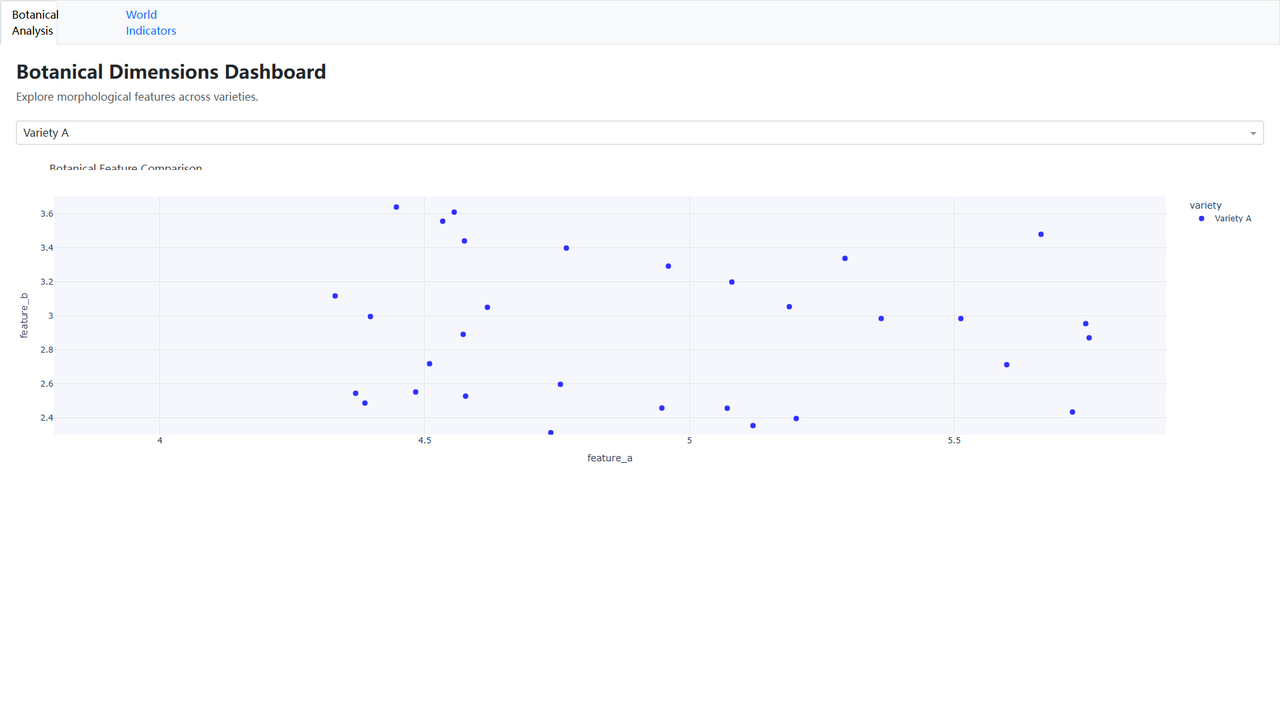}
        \caption{Generated: State A}
        \label{fig:error3_gen_a}
    \end{subfigure}
    \hfill
    \begin{subfigure}[b]{0.48\textwidth}
        \centering
        \includegraphics[width=\textwidth]{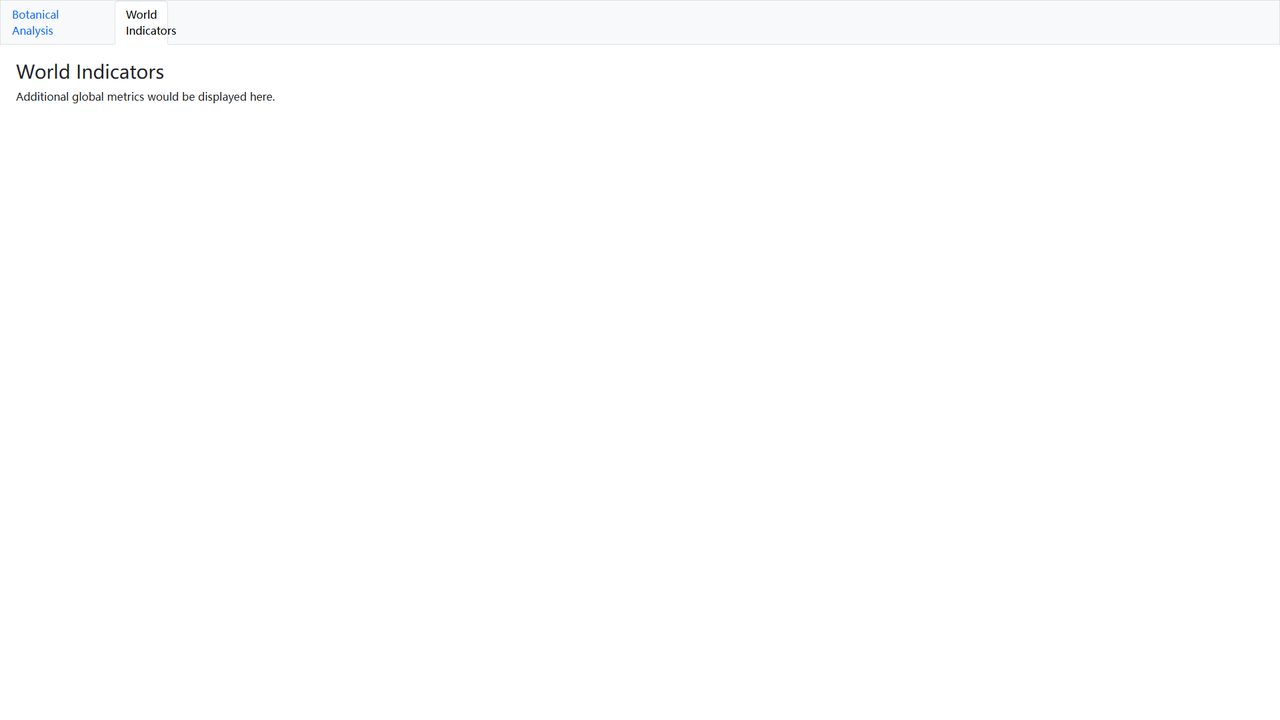}
        \caption{Generated: State B (Content Discovery Failure)}
        \label{fig:error3_gen_b}
    \end{subfigure}

    \caption{Example of Type III Error: Missing Interactive Components (Content Discovery Failure). Here, the model correctly identifies the presence of the Tab component and even replicates the switching mechanism (the header updates in \subref{fig:error3_gen_b}). However, it fails to populate the content for the \emph{World Indicators} state. Instead of the complex bubble chart seen in the Ground Truth (\subref{fig:error3_gt_b}), the model generates a generic placeholder. This indicates a failure in the exploration phase: the agent likely failed to traverse and scrape the DOM state associated with the second tab, resulting in an empty shell.}
    \label{fig:interaction_blindness_error}
\end{figure*}
\subsubsection{Logical dimension - Missing Interactive Components}
As demonstrated in Figure~\ref{fig:interaction_blindness_error}, Type III errors represent a failure in State Exploration and Content Discovery.

In this scenario, unlike Type II (where the visual style is wrong) or Type V (where the logic is hallucinated), the model successfully reconstructs the interactive skeleton—the tabs are functional and the view switches. However, the agent fails to replicate the target content associated with the secondary state.

As seen in (\subref{fig:error3_gen_b}), the complex visualization required for "World Indicators" is replaced by a hallucinated placeholder ("Additional global metrics..."). This suggests that during the reverse-engineering process, the agent failed to successfully traverse or scrape the specific DOM state of the target dashboard's second tab. Consequently, while the UI component exists, the critical data visualization component is missing entirely.

\begin{figure*}[t]
    \centering
    \begin{subfigure}[b]{0.48\textwidth}
        \centering
        \includegraphics[width=\textwidth]{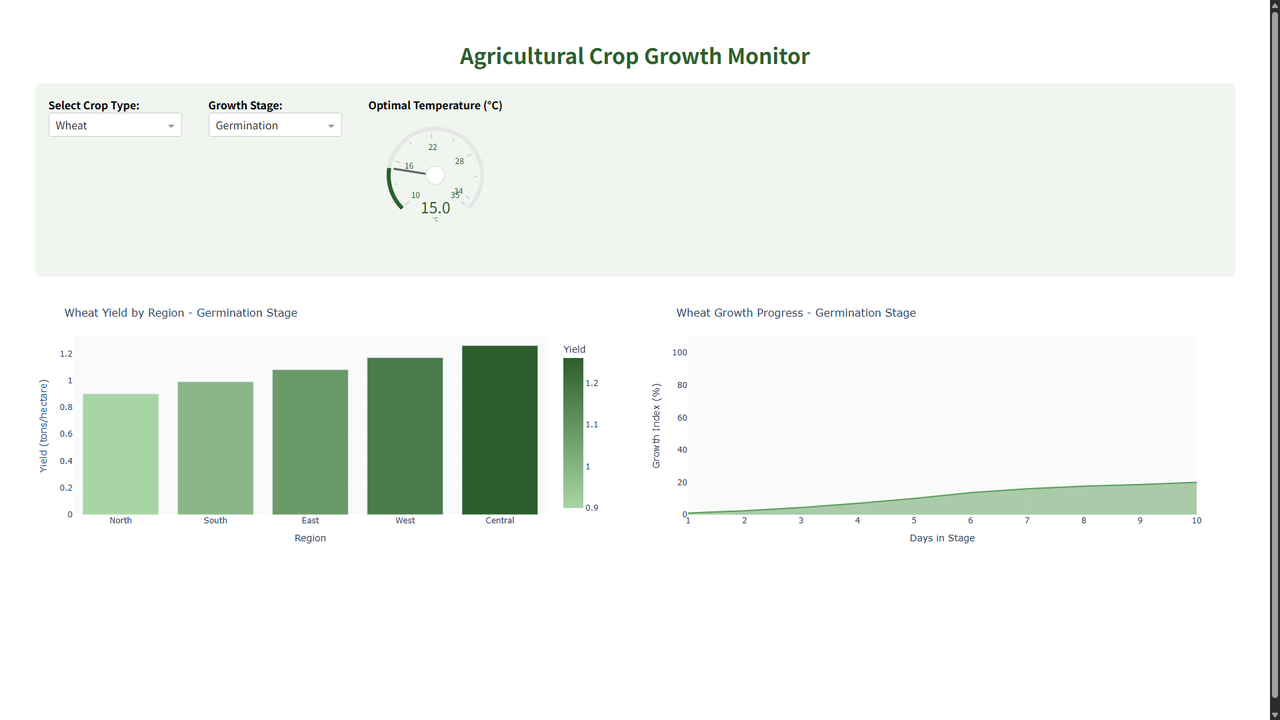}
        \caption{Ground Truth: State A (Wheat)}
        \label{fig:error4_gt_a}
    \end{subfigure}
    \hfill
    \begin{subfigure}[b]{0.48\textwidth}
        \centering
        \includegraphics[width=\textwidth]{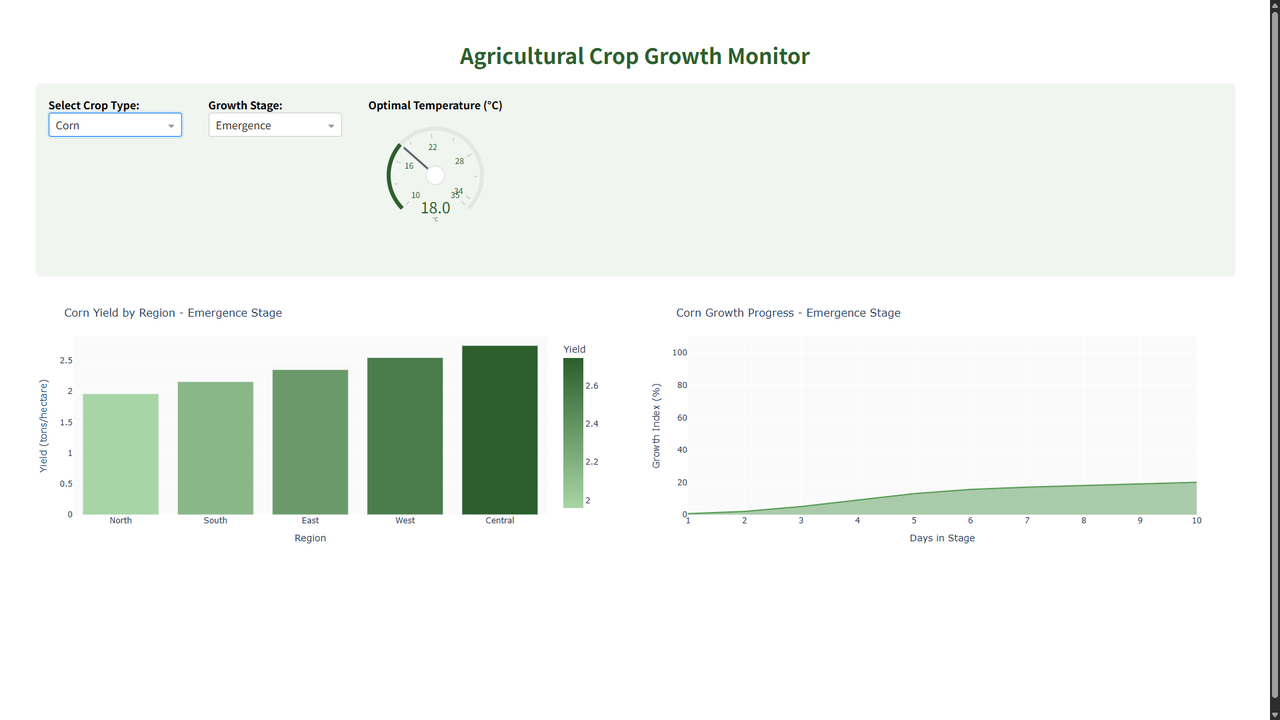}
        \caption{Ground Truth: State B (Corn)}
        \label{fig:error4_gt_b}
    \end{subfigure}

    \vspace{2mm} 

    \begin{subfigure}[b]{0.48\textwidth}
        \centering
        \includegraphics[width=\textwidth]{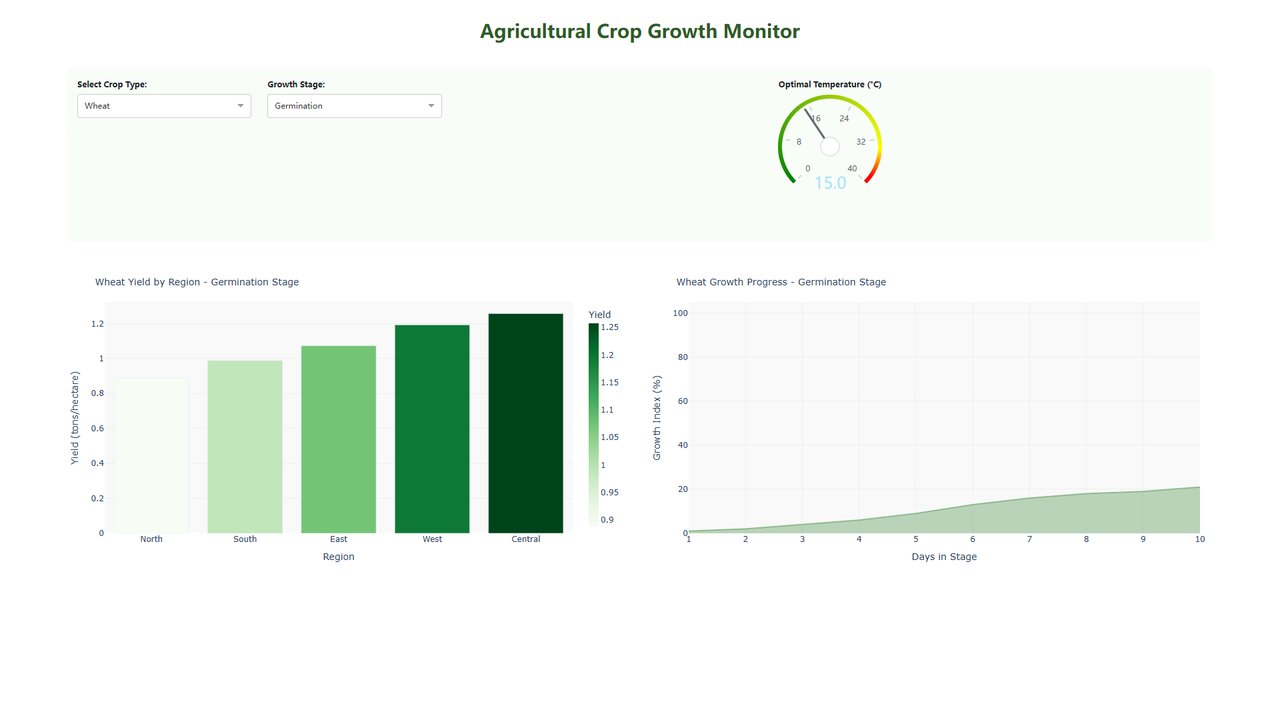}
        \caption{Generated: State A (Wheat)}
        \label{fig:error4_gen_a}
    \end{subfigure}
    \hfill
    \begin{subfigure}[b]{0.48\textwidth}
        \centering
        \includegraphics[width=\textwidth]{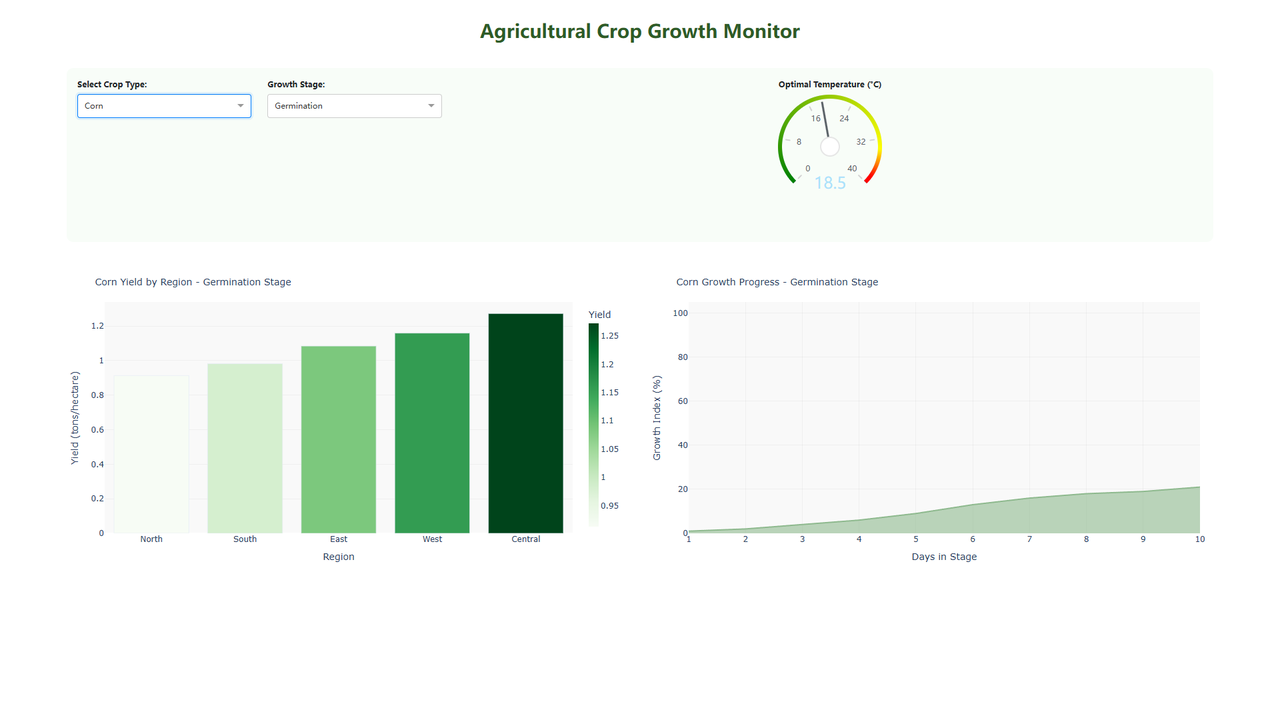}
        \caption{Generated: State B (Broken Dependency)}
        \label{fig:error4_gen_b}
    \end{subfigure}

    \caption{Example of Type IV: State Misinterpretation (Broken Dependency). In the Ground Truth transition (\subref{fig:error4_gt_a} $\rightarrow$ \subref{fig:error4_gt_b}), switching the crop type triggers a cascade update: the chart data shifts significantly, and the secondary dropdown updates to `Emergence'. The generated model (\subref{fig:error4_gen_b}) correctly updates the widget label and chart title to `Corn' but fails to execute the underlying data propagation, leaving the bar chart values and the `Growth Stage' dropdown stuck in the previous `Wheat' state.}
    \label{fig:broken_dependency_error}
\end{figure*}
\subsubsection{Logical dimension - State Misinterpretation (Broken Dependency)}
As illustrated in Figure~\ref{fig:broken_dependency_error}, Type IV errors characterize instances where the model identifies components but misinterprets the causal graph between inputs and outputs. While the generated dashboard successfully captures the primary interaction—updating the "Crop Type" to "Corn" and modifying the chart title—it exhibits an \textbf{edge alignment failure} regarding dependent variables. Specifically, the interaction fails to trigger the necessary downstream updates: the bar chart retains the "Wheat" data distribution (contrast \subref{fig:error4_gen_b} with \subref{fig:error4_gt_b}), and the "Growth Stage" dropdown fails to synchronize with the new crop type. This disconnect suggests the agent treats coupled variables as independent, severing the link between the control trigger and the data visualization layer.

\begin{figure*}[t]
    \centering
    \begin{subfigure}[b]{0.48\textwidth}
        \centering
        \includegraphics[width=\textwidth]{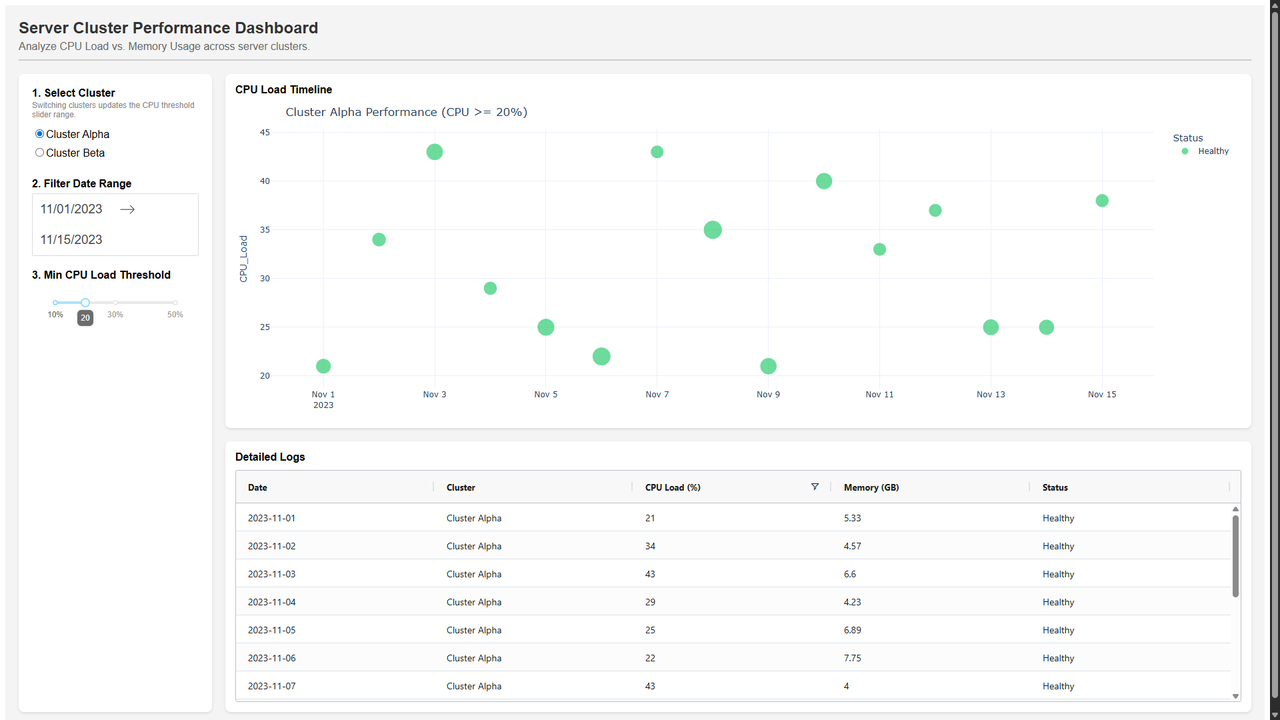}
        \caption{Ground Truth: Cluster Alpha (Normal)}
        \label{fig:error5_gt_a}
    \end{subfigure}
    \hfill
    \begin{subfigure}[b]{0.48\textwidth}
        \centering
        \includegraphics[width=\textwidth]{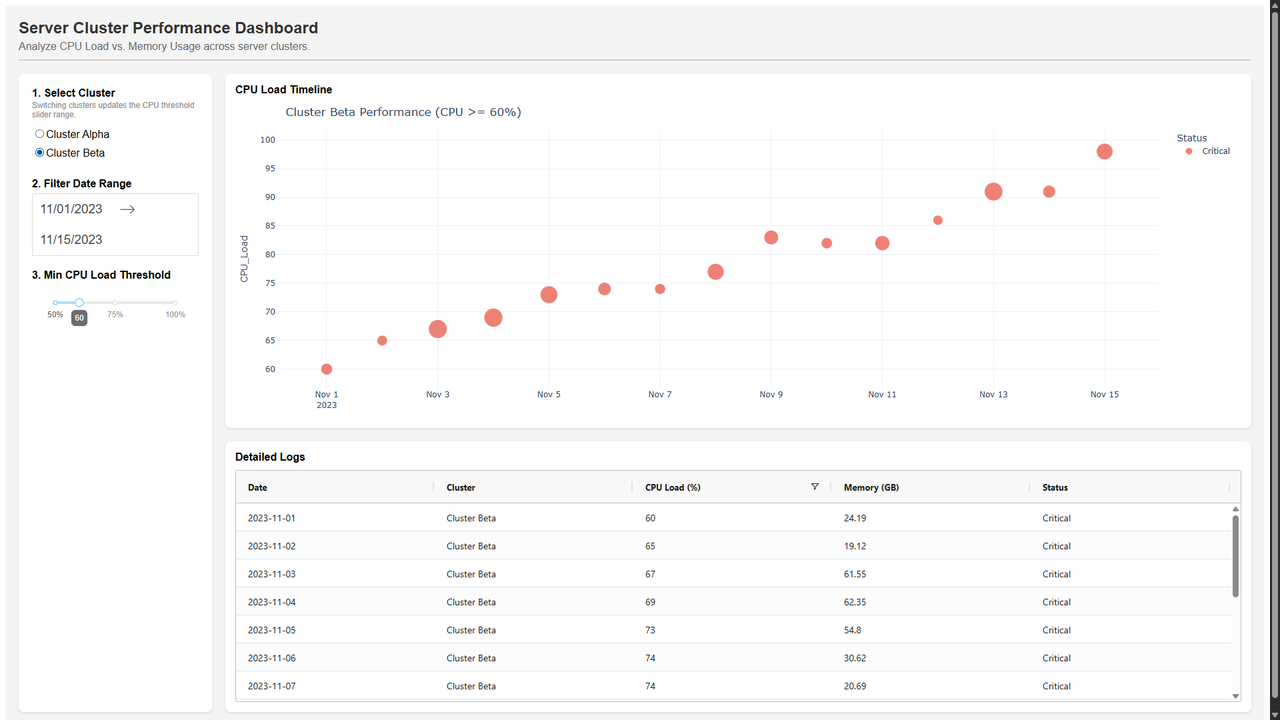}
        \caption{Ground Truth: Cluster Beta (Critical)}
        \label{fig:error5_gt_b}
    \end{subfigure}

    \vspace{2mm} 
    \begin{subfigure}[b]{0.48\textwidth}
        \centering
        \includegraphics[width=\textwidth]{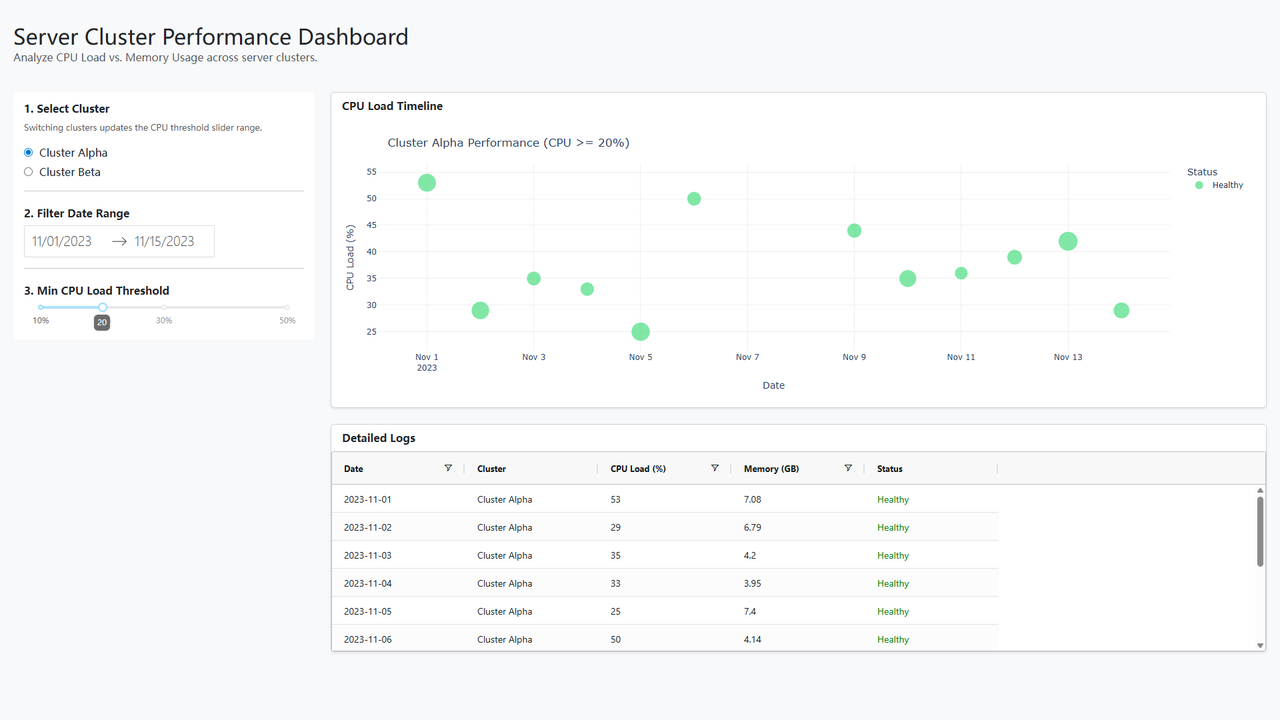}
        \caption{Generated: Cluster Alpha}
        \label{fig:error5_gen_a}
    \end{subfigure}
    \hfill
    \begin{subfigure}[b]{0.48\textwidth}
        \centering
        \includegraphics[width=\textwidth]{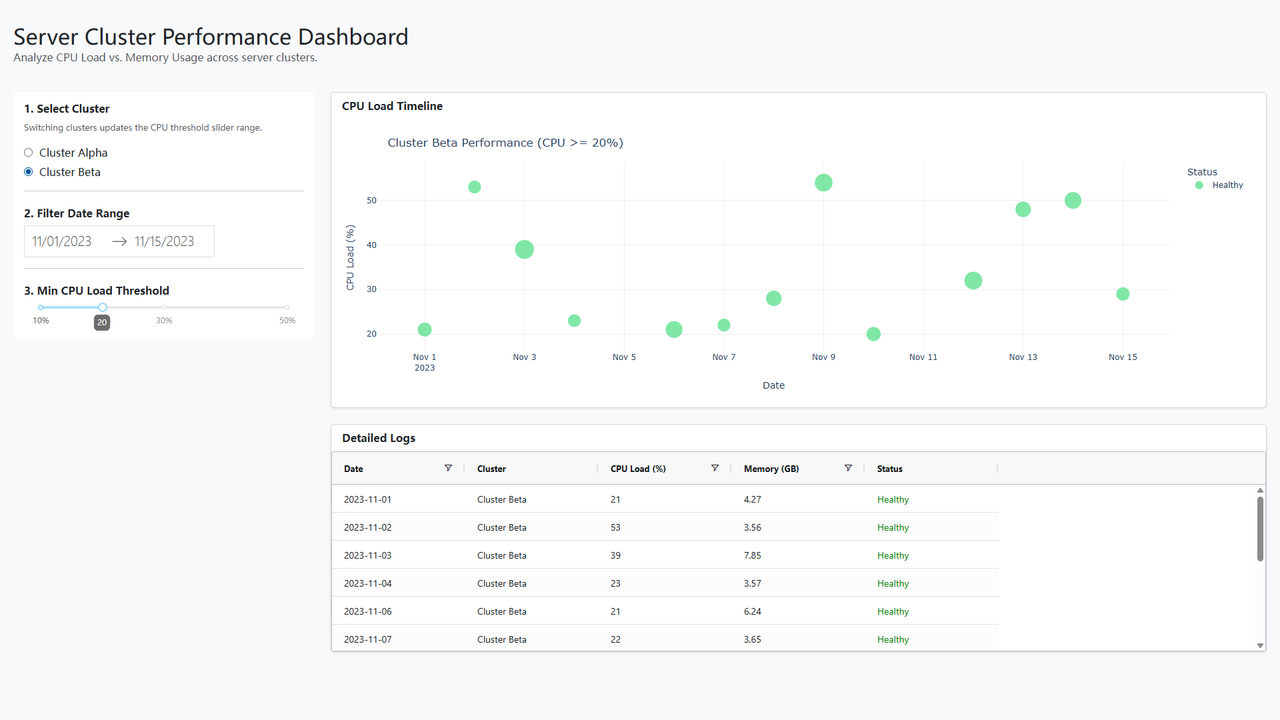}
        \caption{Generated: Transformation Hallucination}
        \label{fig:error5_gen_b}
    \end{subfigure}

    \caption{Example of Type V Error: Data Integrity Failure. Due to the limited size of the \emph{RadioButtons}, the model fails to trigger the state transition to Cluster Beta during exploration. Consequently, while the generated dashboard (\subref{fig:error5_gen_b}) maintains a functional UI skeleton, it fails to reproduce the correct data transformation logic (e.g., color shifting to red and critical threshold scaling), resulting in a logical hallucination where the second state is merely a functional replica of the first.}
    \label{fig:data_integrity_error}
\end{figure*}

\subsubsection{Logical dimension - Transformation Hallucination}
As shown in Figure~\ref{fig:data_integrity_error}, Type V errors (Transformation Hallucination) manifest as a deficit in fine-grained numerical reasoning and logic replication. In this scenario, the model successfully identifies the interactive affordance and establishes the causal link between the "Cluster Beta" radio button and the chart update. However, a semantic gap occurs in the execution of the callback logic.

While the ground truth (\subref{fig:error5_gt_b}) implies a specific data transformation—applying a conditional filter for "Critical" status and updating the visual encoding to red—the generated dashboard (\subref{fig:error5_gen_b}) executes a generic or incorrect update. This results in a "hallucination" where the UI remains interactive and responsive, but the presented data is factually inconsistent with the intended logic, thereby compromising data integrity.

\end{document}